\documentclass[traditabstract]{aa}

\usepackage{caption}

\usepackage{graphicx}
\usepackage{txfonts}
\usepackage{natbib}
\usepackage{enumerate}
\usepackage{bm}
\usepackage[breaklinks]{hyperref}
\hypersetup{
  colorlinks=true,
  linkcolor=blue,
  citecolor=blue,
  urlcolor=blue,
}

\bibpunct{(}{)}{;}{a}{}{,} % to follow A\&A style

\newcommand{\sbs}{SBS\,0335$-$052}
\newcommand{\cgcg}{CGCG\,007$-$025}

\newcommand{\iizw}{II\,Zw\,40}

\newcommand{\twelveco}{$^{12}$CO}
\newcommand{\twelvec}{$^{12}$C}
\newcommand{\thirteenco}{$^{13}$CO}
\newcommand{\coone}{$^{12}$CO(1--0)}
\newcommand{\cotwo}{$^{12}$CO(2--1)}
\newcommand{\cothree}{$^{12}$CO(3--2)}
\newcommand{\cofour}{$^{12}$CO(4--3)}
\newcommand{\thirteencoone}{$^{13}$CO(1--0)}
\newcommand{\thirteencotwo}{$^{13}$CO(2--1)}
\newcommand{\eighteencoone}{C$^{18}$O(1--0)}
\newcommand{\rtwo}{$R_{21}$}
\newcommand{\ncrit}{$n_{\rm crit}$}
\newcommand{\rthree}{$R_{31}$}
\newcommand{\lfir}{$L_{\rm FIR}$}
\newcommand{\lcoone}{$L^\prime_{\rm CO(1-0)}$}
\newcommand{\lcothree}{$L^\prime_{\rm CO(3-2)}$}
\newcommand{\lhcn}{$L^\prime_{\rm HCN(1-0)}$}

\newcommand{\tauco}{$\tau_{\rm CO}$}

\newcommand{\cione}{[C{\sc i}](1--0)}
\newcommand{\citwo}{[C{\sc i}](2--1)}
\newcommand{\ci}{[C{\sc i}]}
\newcommand{\cii}{C{\sc ii}}

\newcommand{\hcn}{HCN(1--0)}
\newcommand{\cn}{CN(1--0)}
\newcommand{\cs}{CS(2--1)}
\newcommand{\nhthree}{NH$_3$}
\newcommand{\htwoco}{H$_2$CO}
\newcommand{\kms}{km\,s$^{-1}$}
\newcommand{\kkms}{K\,km\,s$^{-1}$}
\newcommand{\kkmspc}{K\,km\,s$^{-1}$\,pc$^{2}$}
\newcommand{\htwo}{H$_2$}
\newcommand{\Nhtwo}{$N_{\rm H2}$}
\newcommand{\nhtwo}{$n_{\rm H2}$}
\newcommand{\nco}{$N_{CO}$}
\newcommand{\nhi}{$N_{HI}$}
\newcommand{\ntwelveco}{$N_{12CO}$}
\newcommand{\nthirteenco}{$N_{13CO}$}
\newcommand{\tmb}{$T_{\rm mb}$}
\newcommand{\tkin}{$T_{\rm kin}$}
\newcommand{\tb}{$T_{\rm B}$}
\newcommand{\ta}{$T_{\rm A}^*$}

\newcommand{\dv}{$\Delta V$}

\newcommand{\ropt}{$R_{\rm opt}$}

\newcommand{\av}{$A_V$}
\newcommand{\spit}{{\it Spitzer}}
\newcommand{\hers}{{\it Herschel}}

\newcommand{\logoh}{12$+$log(O/H)}

\newcommand{\mhtwo}{M$_{\rm H2}$}

\newcommand{\mstar}{M$_{\rm star}$}

\newcommand{\aco}{$\alpha_{\rm CO}$}

\newcommand{\cmtwo}{cm$^{-2}$}
\newcommand{\cmthree}{cm$^{-3}$}

\newcommand{\hi}{\rm H{\sc i}}
\newcommand{\hii}{\rm H{\sc ii}}

\newcommand{\micron}{$\mu$m}
\newcommand{\zzsun}{${\mathrm Z/Z}_\odot$}

\def\tex {\ifmmode{{T}_{\rm ex}}\else{$T_{\rm ex}$}\fi}
\def\kms    {\ifmmode{{\rm \ts km\ts s}^{-1}}\else{\ts km\ts s$^{-1}$}\fi}
\def\msun   {\ifmmode{{\rm M}_{\odot}}\else{M$_{\odot}$}\fi}
\def\msunpc   {\ifmmode{{\rm M}_{\odot}\,{\rm pc}^{-2}}\else{M$_{\odot}$\,pc$^{-2}$}\fi}
\def\msunyr   {\ifmmode{{\rm M}_{\odot}\,{\rm yr}^{-1}}\else{M$_{\odot}$\,yr$^{-1}$}\fi}
\def\lsun   {\ifmmode{{\rm L}_{\odot}}\else{L$_{\odot}$}\fi}
\def\zsun   {\ifmmode{{\rm Z}_{\odot}}\else{Z$_{\odot}$}\fi}

\begin{document}

\title{Physical conditions of the molecular gas in metal-poor galaxies
\thanks{Based on observations carried out with the IRAM 30m
and the  Atacama Pathfinder Experiment (APEX). 
IRAM is supported by the INSU/CNRS (France), MPG (Germany), and IGN (Spain),
and
APEX is a collaboration between the Max-Planck-Institut fur Radioastronomie, the European Southern Observatory, and the Onsala Space Observatory.
}
}

\author{L.~K. Hunt \inst{\ref{inst:hunt}}
\and
A. Wei\ss \inst{\ref{inst:weiss}}
\and
C. Henkel \inst{\ref{inst:weiss},\ref{inst:henkelb}}
\and
F. Combes \inst{\ref{inst:combes}}
\and
S. Garc\'{\i}a-Burillo \inst{\ref{inst:garciaburillo}}
\and
V. Casasola \inst{\ref{inst:hunt}}
\and
P. Caselli \inst{\ref{inst:caselli}}
\and
A. Lundgren %\inst{\ref{inst:lundgren}}
\and
R. Maiolino \inst{\ref{inst:maiolino}}
\and
K.~M. Menten \inst{\ref{inst:weiss}}
\and
L. Testi \inst{\ref{inst:hunt},\ref{inst:testi}}
}

\offprints{L. K. Hunt}
\institute{INAF - Osservatorio Astrofisico di Arcetri, Largo E. Fermi, 5, 50125, Firenze, Italy
\label{inst:hunt}
\email{hunt@arcetri.astro.it}
\and
Max-Planck-Institut f\"ur Radioastronomie, Auf dem H\"ugel 69, 53121 Bonn, Germany
\label{inst:weiss}
 \and
Astronomy Department, King Abdulaziz University, P.O. Box 80203, Jeddah, Saudia Arabia
\label{inst:henkelb}
 \and
Observatoire de Paris, LERMA, College de France, CNRS, PSL, Sorbonne University UPMC, F-75014, Paris, France
\label{inst:combes}
 \and
Observatorio Astron\'omico Nacional (OAN)-Observatorio de Madrid,
Alfonso XII, 3, 28014-Madrid, Spain
\label{inst:garciaburillo}
\and
Max-Planck-Institut f\"ur extraterrestrische Physik, Giessenbachstrasse 1, 85748 Garching, Germany
\label{inst:caselli}
% \and
%ALMA JAO, Alonso de Cordova 3107, Vitacura, Casilla 19001, Santiago, Chile
%\label{inst:lundgren}
 \and
Cavendish Laboratory, University of Cambridge, 19 J.J. Thomson Avenue, Cambridge CB3 0HE, UK
\label{inst:maiolino}
\and
ESO, Karl Schwarzschild str. 2, 85748 Garching bei M\"unchen, Germany
\label{inst:testi}
}

   \date{Received  2017/ Accepted  2017}

   \titlerunning{Physical conditions at low metallicity}
   \authorrunning{Hunt et al.}

\abstract{Studying the molecular component of the interstellar medium (ISM) in metal-poor galaxies has been challenging 
%to  measure 
because of the faintness of carbon monoxide emission, the most common proxy of \htwo.
Here we present new detections of molecular gas at low metallicities,
and assess the physical conditions in the gas through
various CO transitions for 8 galaxies.
For one, NGC\,1140 (\zzsun\,$\sim$0.3), 
two detections of \thirteenco\ isotopologues and atomic carbon, \cione\ and an upper limit for \hcn\ are also reported. 
After correcting to a common beam size, we compared \cotwo/\coone\ (\rtwo) and \cothree/\coone\ (\rthree) line ratios 
of our sample with galaxies from the literature and find that only NGC\,1140 shows extreme
values (\rtwo$\sim$\rthree$\sim 2$).
Fitting physical models to the \twelveco\ and \thirteenco\ emission in NGC\,1140
suggests that the molecular gas is cool (kinetic temperature \tkin\,$\la$\,20\,K),
dense (\htwo\ volume density \nhtwo\,$\ga 10^6$\,\cmthree), with
moderate CO column density (\nco\,$\sim 10^{16}$\,\cmtwo) and low
filling factor.
Surprisingly, the [\twelveco]/[\thirteenco] abundance ratio in NGC\,1140 is very low
($\sim 8 - 20$), lower even than the value of 24 found in the Galactic Center.
The young age of the starburst in NGC\,1140 precludes \thirteenco\ enrichment from
evolved intermediate-mass stars; instead 
we attribute the low ratio to charge-exchange reactions and fractionation, 
because of the enhanced efficiency of these processes in cool gas at moderate column densities. 
Fitting physical models to \twelveco\ and \cione\ emission in NGC\,1140
gives an unusually low [\twelveco]/[\twelvec] abundance ratio, suggesting that in this galaxy
atomic carbon is at least 10 times more abundant than \twelveco.
%context
%aims
%methods
%results
%conclusions
\keywords{Galaxies: starburst --- Galaxies: dwarf --- Galaxies: star formation --- Galaxies: ISM --- ISM: molecules
--- Radio lines: ISM} 
}
\maketitle

%---------------------------------------------------------------

\section{Introduction}
\label{sec:intro}

Physical conditions in the molecular component of the interstellar medium (ISM) of 
metal-poor galaxies are difficult to measure.
The main obstacle is the faintness at low metallicities
of the common proxy of \htwo, carbon monoxide emission 
\citep{sage92,taylor98,gondhalekar98,barone00,leroy05,buyle06,leroy07,schruba12,cormier14}.
While the proximity of Local Group galaxies facilitates studies of molecular gas
\citep[e.g.,][]{cohen88,rubio93,fukui99,israel03,bolatto08,pineda12,elmegreen13,paron14,shi15,rubio15,paron16,shi16,schruba17},
beyond the Local Group, CO detections in low-metallicity galaxies are arduous
\citep[e.g.,][]{schruba12,cormier14},
and few in number.
In a previous paper \citep[][hereafter Paper\,I]{hunt15}, we reported \coone\ detections in 8 metal-poor galaxies
outside the Local Group with metallicities ranging from \logoh\,$\sim$7.7 to 8.4,
or equivalently 0.1\,\zsun\ to 0.5\,\zsun\footnote{We assume
the \citet{asplund09} Solar abundance calibration of \logoh\,=\,8.69.}.
This more than doubles the previous number of such detections, and sets the stage for
a more detailed study of the physical conditions in molecular gas in a metal-poor ISM.

\begin{center}
\begin{table*}
      \caption[]{Parameters for observed galaxies} 
\label{tab:params}
\resizebox{\linewidth}{!}{
%\addtolength{\tabcolsep}{7pt}
\addtolength{\tabcolsep}{1pt}
{\small
%\tiny
\begin{tabular}{lrrrrrcccc}
\hline
\multicolumn{1}{c}{Name} &
\multicolumn{2}{c}{Pointed position (J2000)} &
\multicolumn{1}{c}{Redshift} &
\multicolumn{1}{c}{Distance} &
\multicolumn{1}{c}{Distance} &
\multicolumn{1}{c}{Size$^{\mathrm{b}}$} &
\multicolumn{3}{c}{Beam FWHM} \\
&
\multicolumn{1}{c}{RA} &
\multicolumn{1}{c}{Dec.} &
&
\multicolumn{1}{c}{(Mpc)} &
\multicolumn{1}{c}{method$^{\mathrm{a}}$} &
&
\multicolumn{1}{c}{\coone} &
\multicolumn{1}{c}{\cotwo} &
\multicolumn{1}{c}{\cothree} \\
%\multicolumn{1}{c}{\cofour} &
%\multicolumn{1}{c}{\cione} \\
\\
\hline
\\
\cgcg\              & 09:44:01.9 & $-$00:38:32.0 & 0.00483 & 24.5 & CMB   & 27\arcsec\,$\times$\,16\farcs2 & 21\farcs4 & 10\farcs7 & $-$       \\ %& $-$ & $-$ \\
&&&&&& & 2.54\,kpc & 1.27\,kpc & $-$       \\ %& $-$ & $-$ \\
\iizw\              & 05:55:42.6 &    03:23:31.5 & 0.00263 & 11.7 & CMB   & 33\farcs6\,$\times$\,13\farcs2 & $-$       & 27\farcs1 & 18\farcs1 \\ %& $-$ & $-$ \\
&&&&&& & $-$      & 1.537\,kpc & 1.027\,kpc  \\ %& $-$ & $-$ \\
Mrk\,996            & 01:27:35.5 & $-$06:19:36.0 & 0.00541 & 18.1 & CMB   & 36\arcsec\,$\times$\,30\arcsec\   & 21\farcs5 & 10\farcs7 & 18\farcs1 \\ %& $-$ & $-$ \\
&&&&&& & 1.88\,kpc & 0.94\,kpc & 1.59\,kpc  \\ %& $-$ & $-$ \\
NGC\,1140           & 02:54:33.6 & $-$10:01:40.0 & 0.00501 & 19.7 & TF    & 102\arcsec\,$\times$\,54\arcsec\  & 21\farcs4 & 10\farcs7 & 18\farcs1 \\ %& 13\farcs6 & 12\farcs7 \\
&&&&&& & 2.04\,kpc  & 1.02\,kpc   & 1.73\,kpc \\ %& 1.30\,kpc  & 1.21\,kpc \\
NGC\,1156           & 02:59:42.2 &    25:14:14.0 & 0.00125 &  8.1 & Stars & 198\arcsec\,$\times$\,150\arcsec\ & 21\farcs4 & 10\farcs7 & $-$ \\ %& $-$ & $-$ \\
&&&&&& & 0.84\,kpc & 0.42\,kpc \\ %& $-$ & $-$ & $-$ \\
NGC\,3353 (Haro\,3) & 10:45:22.40 &   55:57:37.0 & 0.00315 & 18.1 & TF    & 72\arcsec\,$\times$\,49\farcs8 & 21\farcs4 & 10\farcs7 & $-$ \\ %& $-$ & $-$ \\
&&&&&& & 1.88\,kpc & 0.94\,kpc \\ %&$-$ & $-$ & $-$ \\
NGC\,7077           & 21:29:59.6 &    02:24:51.0 & 0.00384 & 17.2 & TF    & 48\arcsec\,$\times$\,42\arcsec\   & 21\farcs4 & 10\farcs7 & 18\farcs1 \\ %& $-$ & $-$ \\
&&&&&& & 1.78\,kpc & 0.89\,kpc & 1.51\,kpc \\ %& $-$ & $-$ \\
\sbs\               & 03:37:44.0 & $-$05:02:40.0 & 0.01352 & 53.6 & CMB   & 13\farcs8 $\times$\,12\farcs0 & $-$       & 27\farcs4 & 18\farcs3 \\ %& $-$ & $-$ \\
&&&&&& & $-$ & 7.12\,kpc & 4.76\,kpc \\ %& $-$ & $-$ \\
UM\,448             & 11:42:12.4 &    00:20:03.0 & 0.01856 & 81.2 & CMB   & 24\arcsec\,$\times$\,24\arcsec\   & 21\farcs7 & 10\farcs9 & 18\farcs4 \\ %& $-$ & $-$ \\
&&&&&& & 8.54\,kpc & 4.29\,kpc & 7.244,kpc \\ %& $-$ & $-$ \\
UM\,462             & 11:52:37.2 & $-$02:28:10.0 & 0.00353 & 19.5 & CMB   & 36\arcsec\,$\times$\,30\arcsec\   & 21\farcs4 & 10\farcs7 & 18\farcs1 \\ %& $-$ & $-$ \\
&&&&&& & 2.02\,kpc & 1.01\,kpc & 1.71\,kpc \\ %& $-$ & $-$ \\
\\
\hline
\end{tabular}
} 
}
\vspace{0.5\baselineskip}
\begin{description}
\item
[$^{\mathrm{a}}$]~Taken from NED: Cosmic Microwave Background (CMB), Tully Fisher (TF), Stars.
\item
[$^{\mathrm{b}}$]~Taken from NED. 
\end{description}
\end{table*}
\end{center}

In this paper, we present observations of \cotwo\ and \cothree\ for some of the galaxies
in Paper\,I, and for one of them, NGC\,1140, 
also \cofour, \thirteencoone, \thirteencotwo, and \cione\ detections,
as well as upper limits for other molecular transitions including HCN(1-0).
In Sect. \ref{sec:sample}, we briefly discuss the sample, and in 
Sect. \ref{sec:observations}, describe the observations and data reduction. 
Section \ref{sec:beam} presents our approach for correcting the observed temperatures and
fluxes for the different beam sizes.
Sect. \ref{sec:empirical} analyzes the observed line ratios from an empirical
point of view, comparing them with other metal-rich samples, and
Sect. \ref{sec:dense} discusses dense-gas tracers.
Physical models of the line emission of NGC\,1140
%of the CO emission 
are presented in Sect. \ref{sec:models}, 
in order to infer \htwo\ volume density, \nhtwo, kinetic temperature, \tkin,
CO column density, \nco, and the \twelveco/\thirteenco\ 
as well as the \ci/\twelveco\ abundance ratios. %for NGC\,1140;
More limited models are computed for the other galaxies with fewer multiple transitions.
We discuss the results in Sect. \ref{sec:discussion}, and give
our conclusions in Sect. \ref{sec:conclusions}.

\section{The targets\label{sec:sample}}

A complete description of source selection and the individual galaxies
is given in Paper\,I.
Here we briefly outline the main selection criteria and general
characteristics of our targets.

The majority of the galaxies in the observing sample already had \htwo\
detections, either of the ro-vibrational transitions at 2\,\micron\
or the rotational transitions in the mid-infrared with \spit/IRS
\citep{hunt10}.
Additional galaxies were included
with previous CO observations lacking a clear detection.
%The metallicities of the targets range of \logoh\,=\,7.74 ($\sim$0.1\,\zsun)
%to \logoh\,=\,8.37 ($\sim$0.5\,\zsun).
Stellar masses, estimated 
from 3.6\,\micron\ luminosities after the subtraction of nebular emission,
range from $\sim 10^8$ to $10^{10}$\,\msun\ (UM\,448 is more massive than
this, and is the clear result of an interaction with another galaxy, see Paper\,I for more details).
The relatively high specific star-formation rates (ratio of SFR and stellar mass, SFR/\mstar\,=\,sSFR) of the sample make them starburst galaxies
%(SFR\,$\sim 0.1$ to $\sim 2$\,\msunyr; again
%UM\,448 is an outlier with SFR $\sim$ 11\,\msunyr). 
(sSFR\,$\ga 10^{-10}$\,yr$^{-1}$). 
They are also gas rich, with gas-mass fractions (relative to total baryonic
mass) as high as $\sim$0.8.
%More details are given in Paper\,I.

\section{The observations \label{sec:observations}}

We have observed \coone\ and \cotwo\ 
in our sample galaxies over a three-year period from 2008 to 2010
with the IRAM 30-m telescope (Pico Veleta, Spain). 
The measurements of \cothree\ and for a few galaxies also of \cotwo\ were acquired with the APEX 12-m telescope 
(Chile)\footnote{This publication is based on data acquired with the Atacama Pathfinder Experiment (APEX). APEX is a collaboration between the Max-Planck-Institut fur Radioastronomie, the European Southern Observatory, and the Onsala Space Observatory.}.
NGC\,1140 was singled out for a more intense observing campaign in which
we observed \thirteencoone, \hcn, other 3-mm transitions, %at the 30m,
and \thirteencotwo\ at the 30m;
\cofour\ and \cione\ measurements were obtained with APEX.
The \coone\ observations of this sample were presented in Paper\,I.
Table \ref{tab:params} reports selected observational parameters for the
sample galaxies.
Our single-dish observations sample fairly large regions within each galaxy,
ranging from 420\,pc for the smallest beam in the closest galaxy
(\cotwo, NGC\,1156)
to 8.5\,kpc for the largest beam in the most distant one
(\coone, UM\,448).

\subsection{IRAM \label{sec:iram}}

We observed eight galaxies at the 30m 
%\footnote{\iizw\ and \sbs\ were observed only in the higher-J lines with APEX.} 
in \coone\ and \cotwo, first with the older ABCD receivers (proposals 036-08, 227-09), 
then with the Eight Mixer Receiver (EMIR, proposal 097-10)
using the Wideband Line Multiple Autocorrelator (WILMA) backend. 
Earlier spectra were acquired at an intrinsic resolution
of 4\,MHz, while later ones using EMIR at 2\,MHz.
%Observations of NGC\,1140 and NGC\,1156
%were performed with the AB receivers in November, 2008, using the 1 and
%4\,MHz backends at 3 and 1.3\,mm (88-115\,GHz, 219-229\,GHz, respectively).
%During the 2008 run,
%the precipitable water vapor was $\la$4\,mm, 
%giving system temperatures of T$_{\rm sys}$ of $\la$370\,K (270--330\,K) 
%at 3\,mm (1.3\,mm);  
%in 2009, the conditions were worse ($\la$11\,mm), but
%the EMIR receivers had comparable T$_{\rm sys}$, $\sim$335\,K (330\,K) 
%at 3\,mm (1.3\,mm).
%For our most extensive run (2010), T$_{\rm sys}$ varied from $\sim$230\,K
%to $\ga$400\,K, with precipitable water vapor ranging from $\sim$ 4\,mm to 10\,mm.
For NGC\,1140, the EMIR E090 receiver was tuned to an intermediate frequency, $\sim$113\,GHz,
in order to cover other 3\,mm transitions simultaneously together with \coone\
in the lower-inner sideband in two polarizations.
We adopted wobbler switching with a throw of 90\arcsec, and
used the standard intensity calibration with two absorbers at different temperatures. 
Pointing was checked every hour or two on nearby planets %(Uranus)
or bright quasars, and focus tests
were performed every 4\,hrs during the night and every 3\,hrs 
during the day.
Typical pointing errors were $\la$2\arcsec, and
never exceeded $\sim$3\arcsec.

With the EMIR setup for NGC\,1140, we were able to also observe additional transitions, including
\thirteencoone, \thirteencotwo, \eighteencoone, and the dense-gas tracers \hcn, 
\cn, and \cs.
% Christian says remove because mixing up observations and results
%Only \thirteencoone\ and \thirteencotwo\ were detected.
%The constraints on dense gas imposed by \hcn\ upper limit
%will be discussed in Sects. \ref{sec:dense} and \ref{sec:hcn}.

%In 2009, we also observed \thirteencoone\ in NGC\,3353, without using extra time
%because of a convenient frequency setting, but did not detect the transition.

\begin{figure*}[ht!]
%\begin{minipage}[t]{\textwidth}
\vspace{\baselineskip}
\hbox{
\centerline{
\includegraphics[angle=0,width=0.3\linewidth]{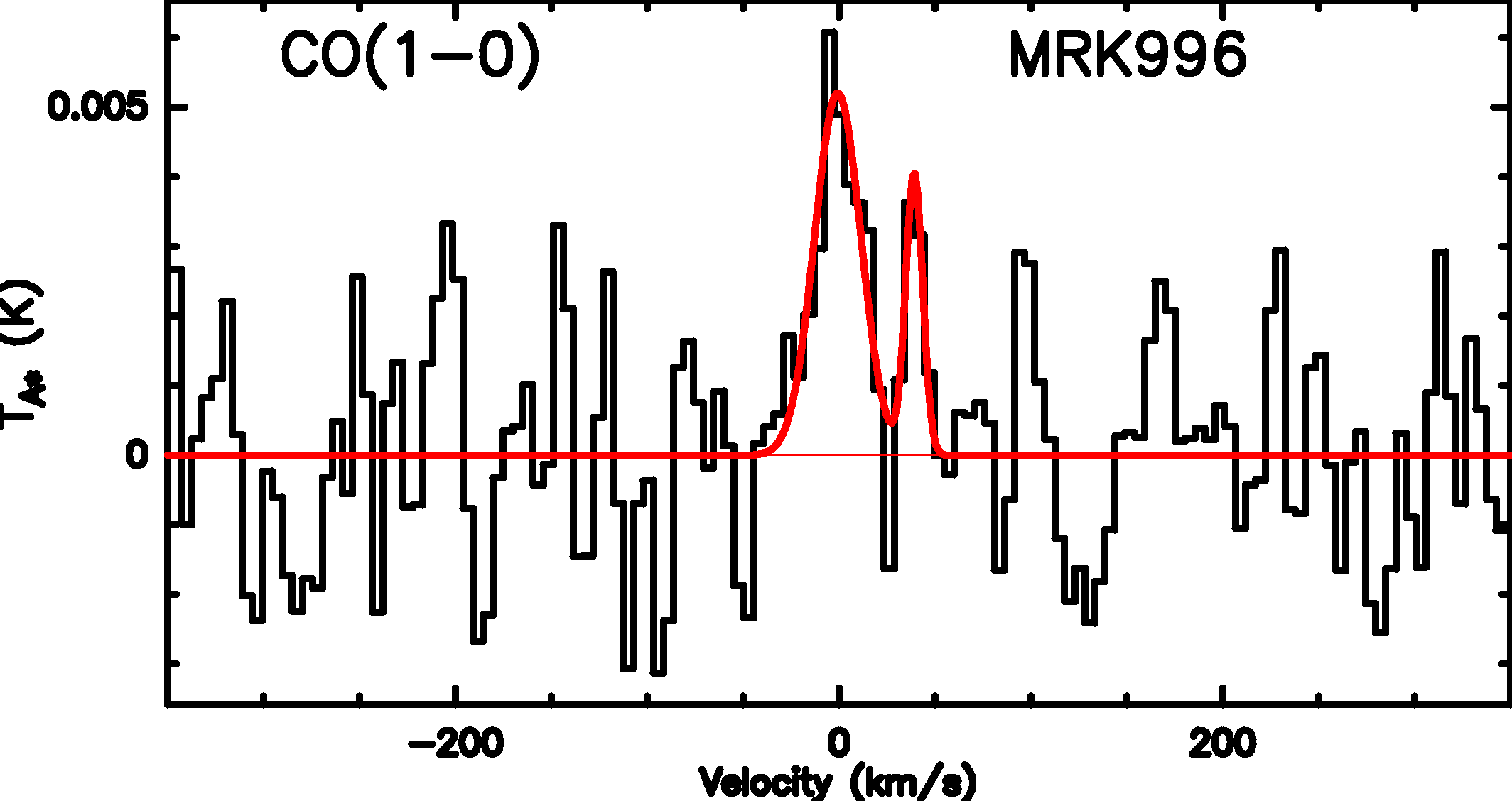}
\hspace{0.2cm}
\includegraphics[angle=0,width=0.3\linewidth]{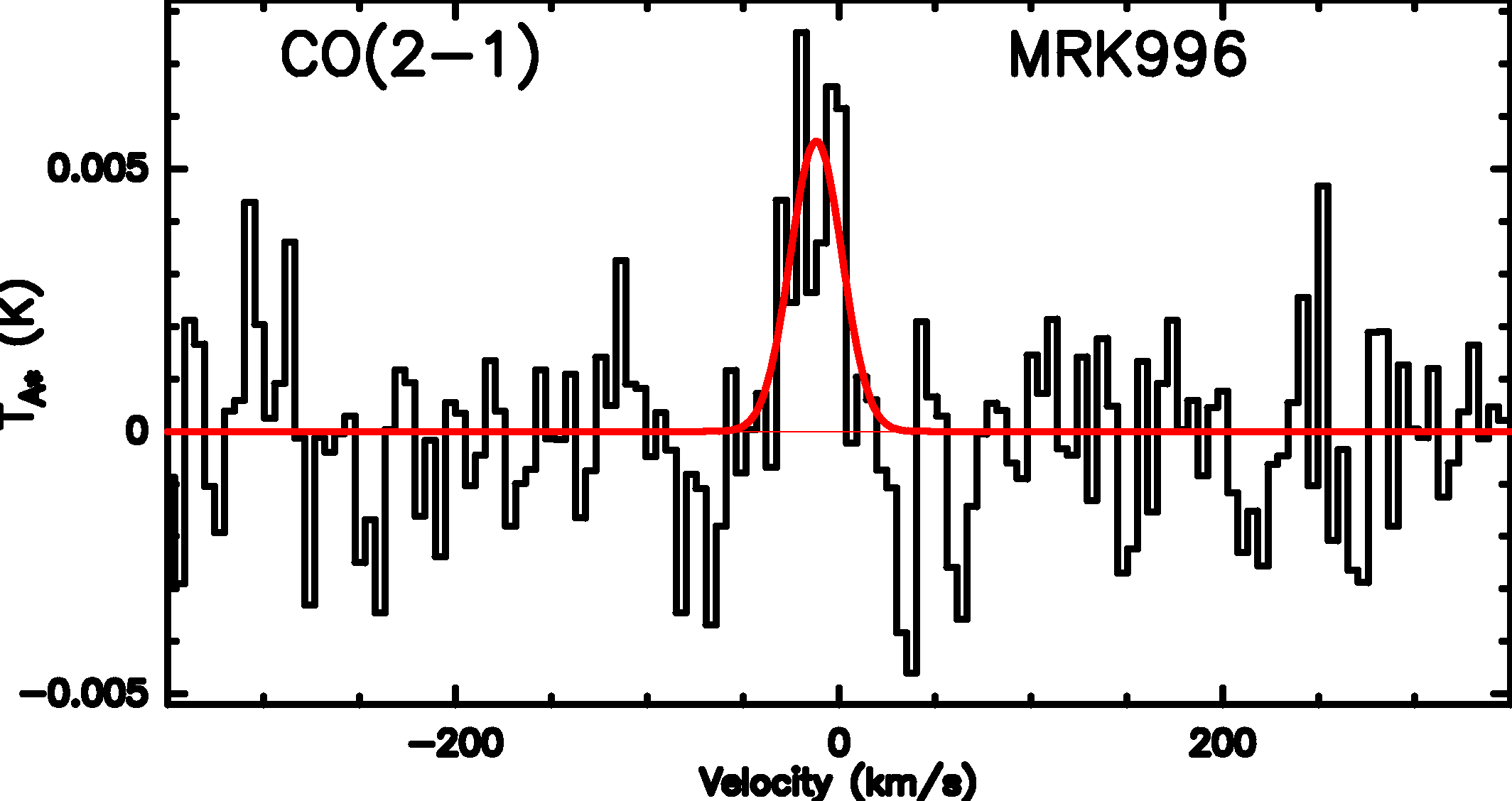}
\hspace{0.2cm}
\includegraphics[angle=0,width=0.3\linewidth]{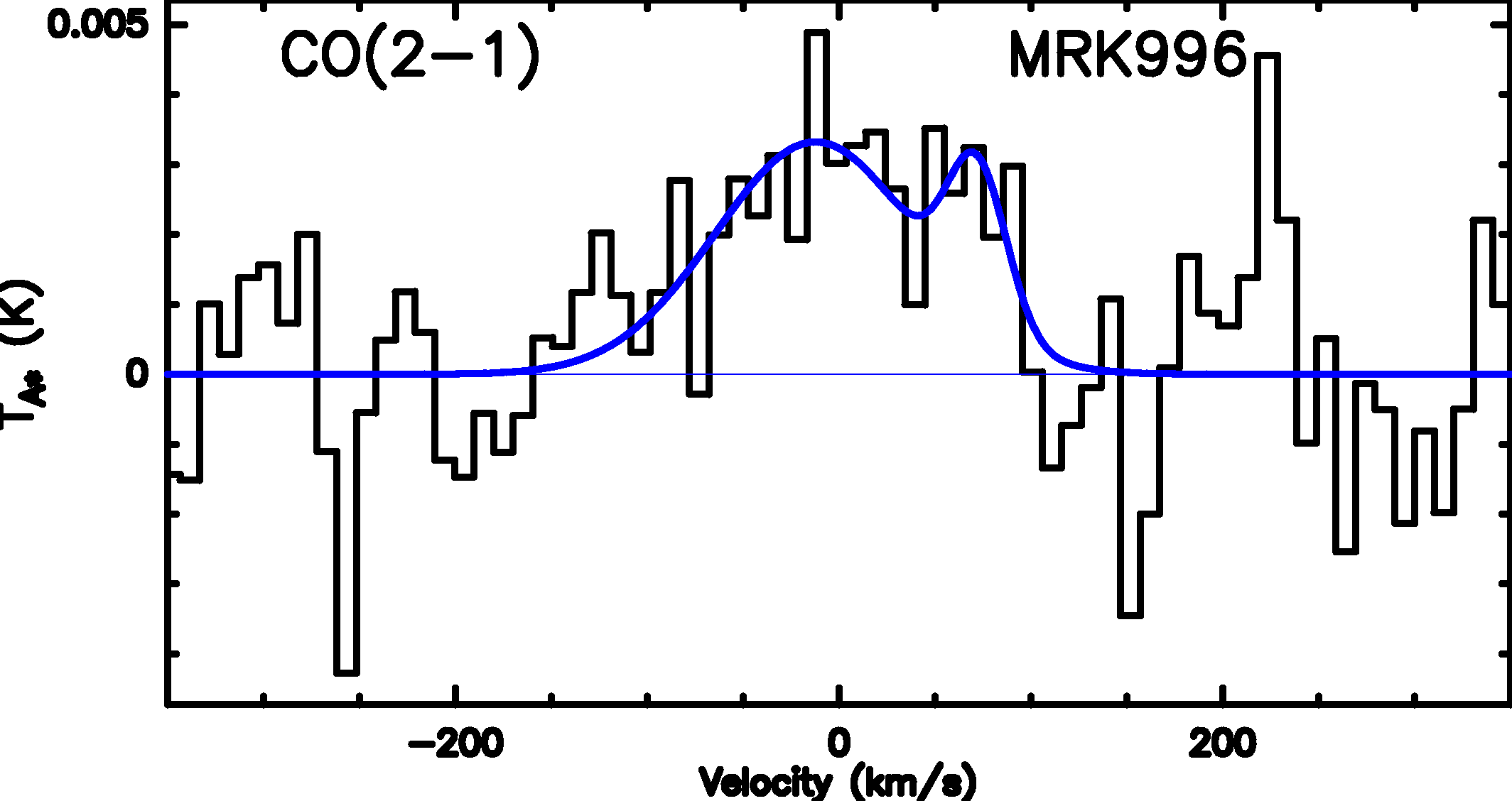}
}
}
\vspace{\baselineskip}
\hbox{
\centerline{
\includegraphics[angle=0,width=0.3\linewidth]{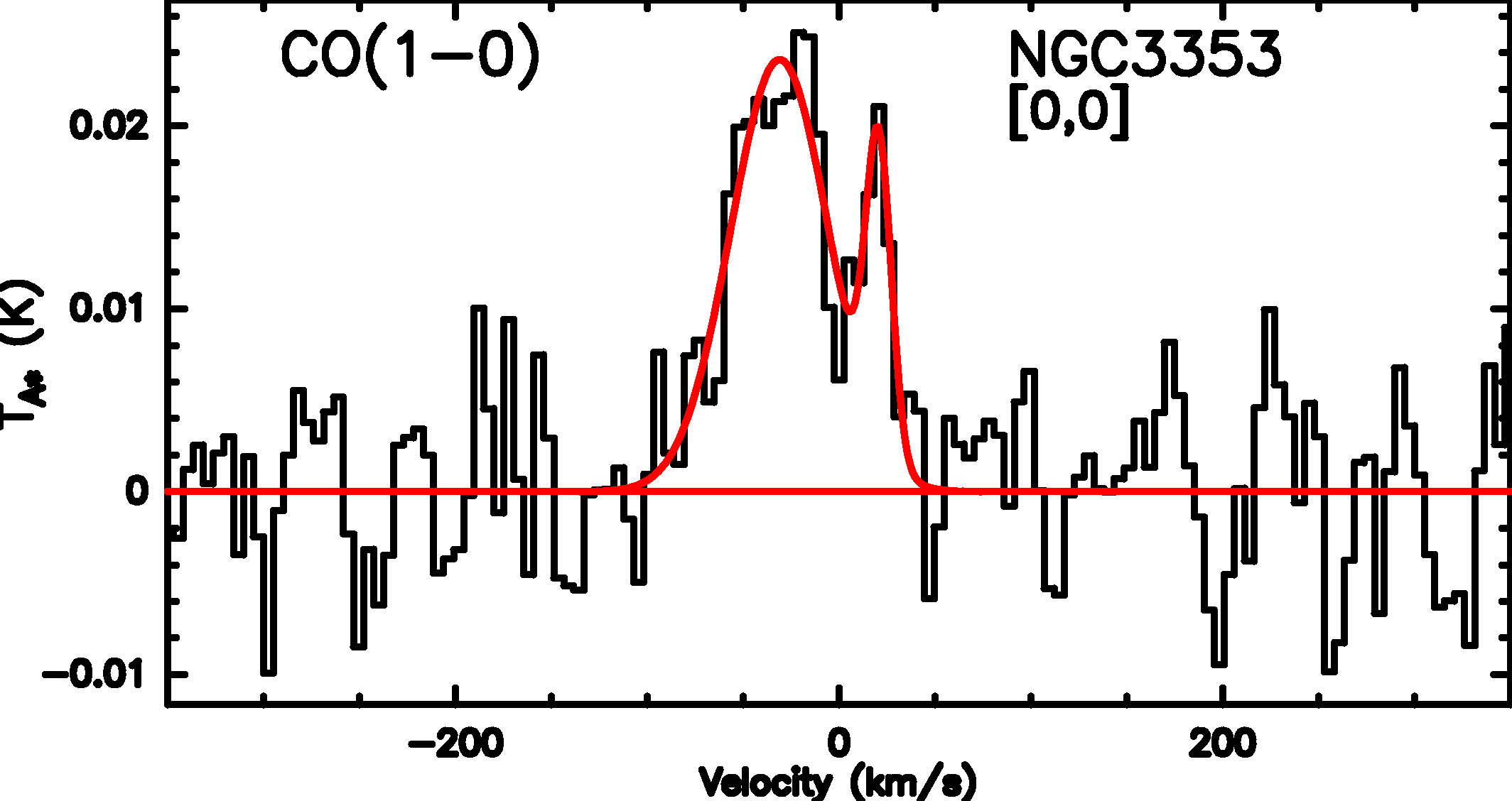}
\hspace{0.2cm}
\includegraphics[angle=0,width=0.3\linewidth]{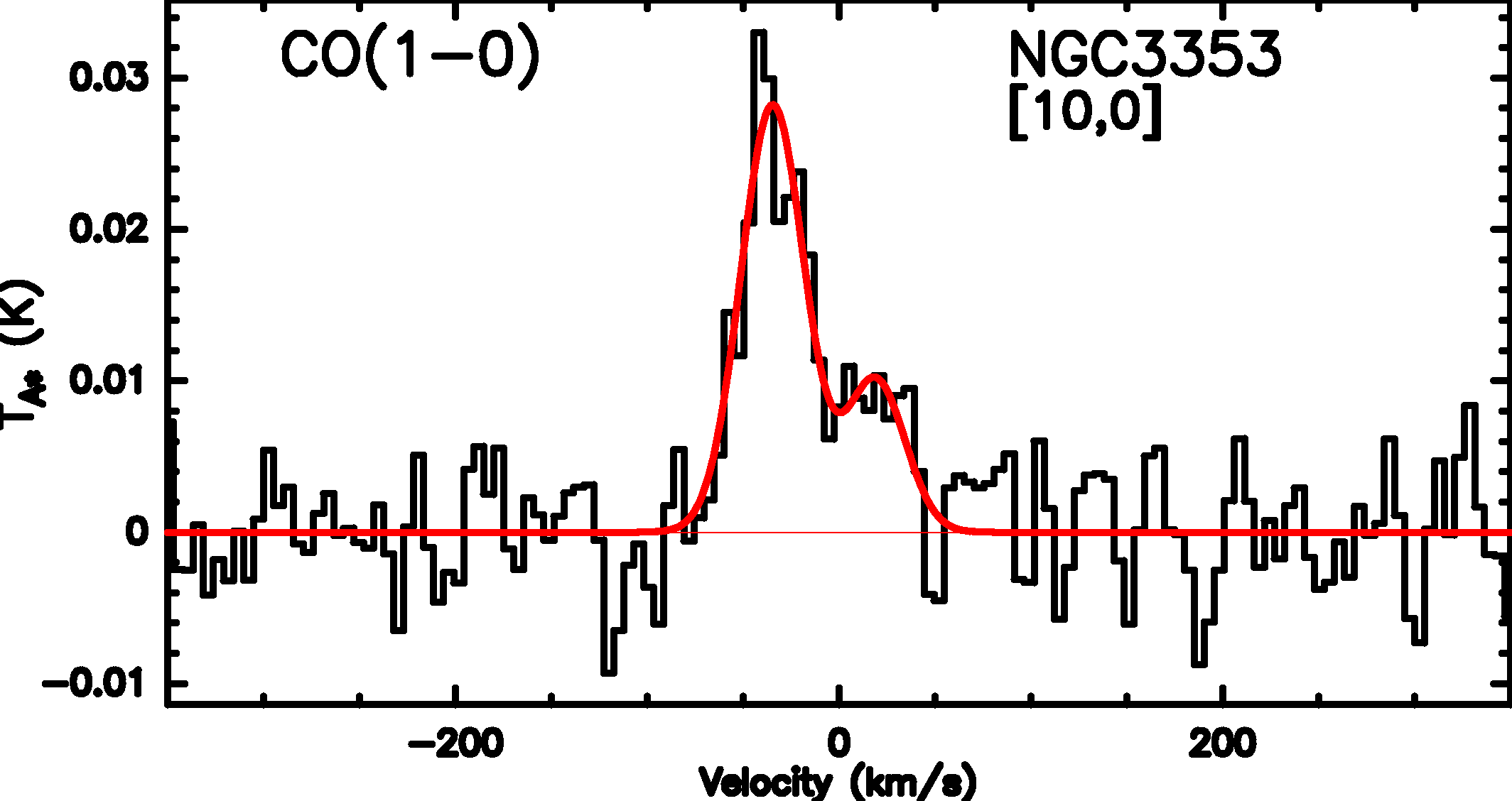}
\hspace{0.2cm}
\includegraphics[angle=0,width=0.3\linewidth]{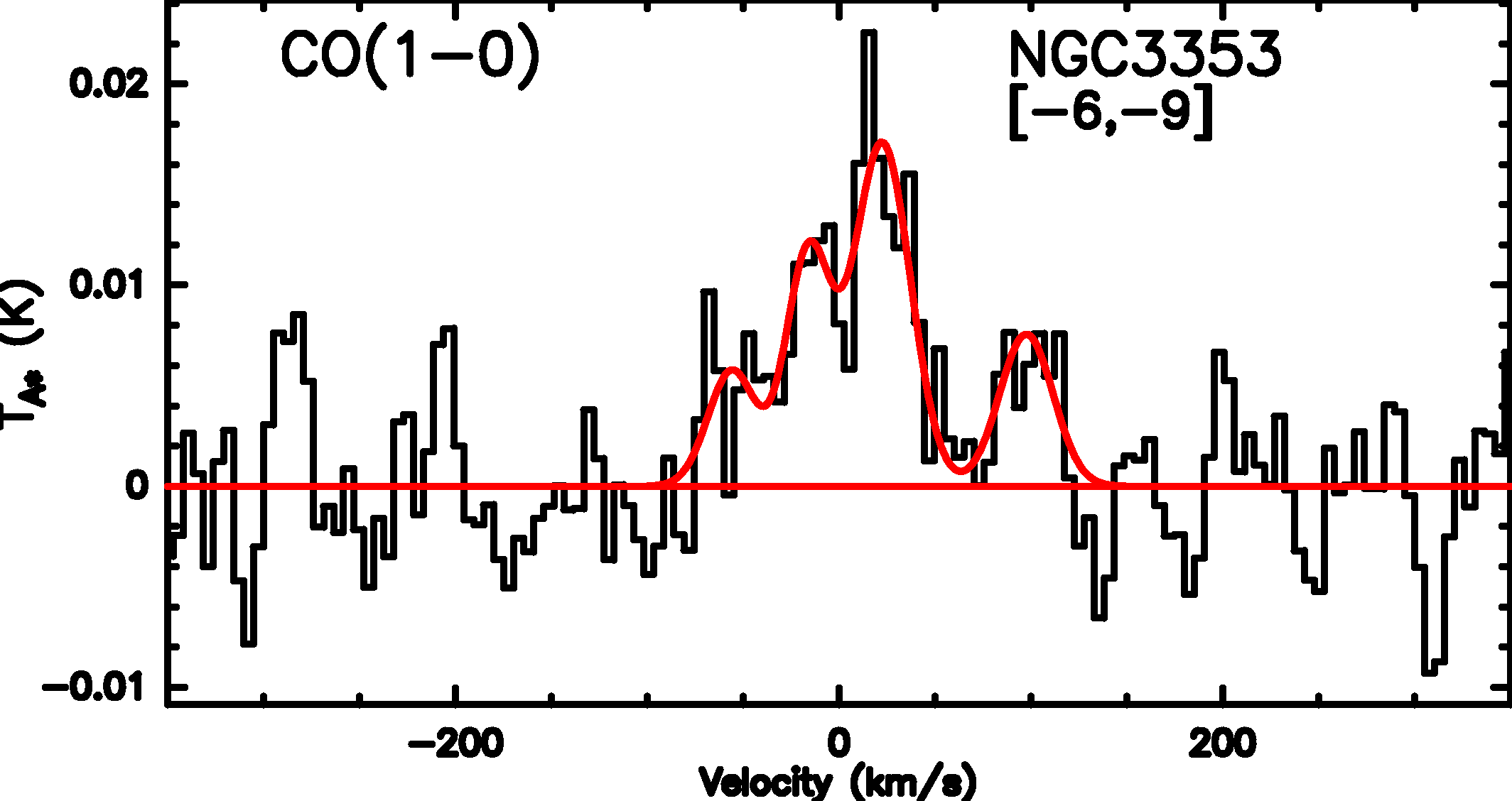}
}
}
\vspace{\baselineskip}
\hbox{
\centerline{
\includegraphics[angle=0,width=0.3\linewidth]{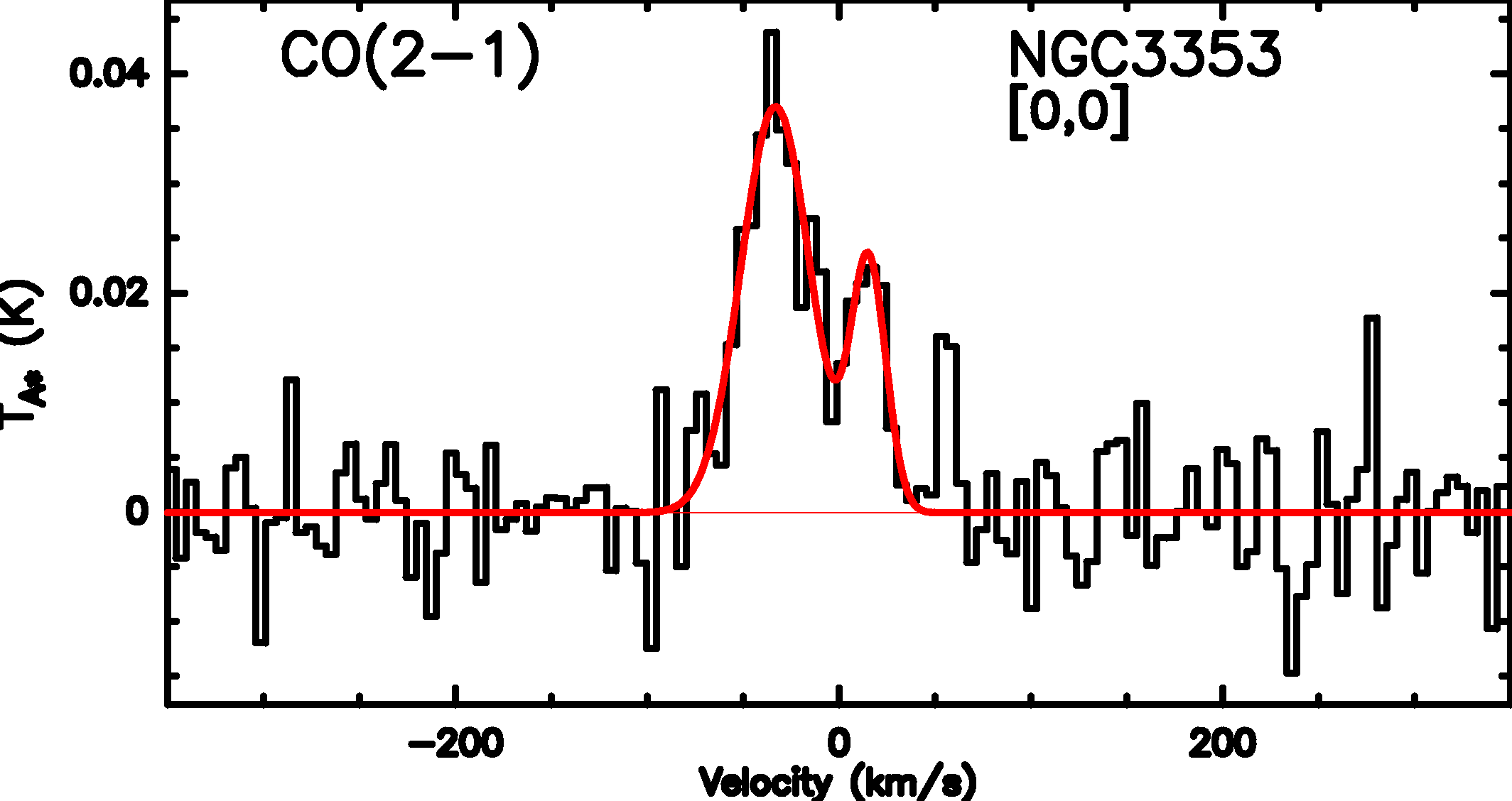}
\hspace{0.2cm}
\includegraphics[angle=0,width=0.3\linewidth]{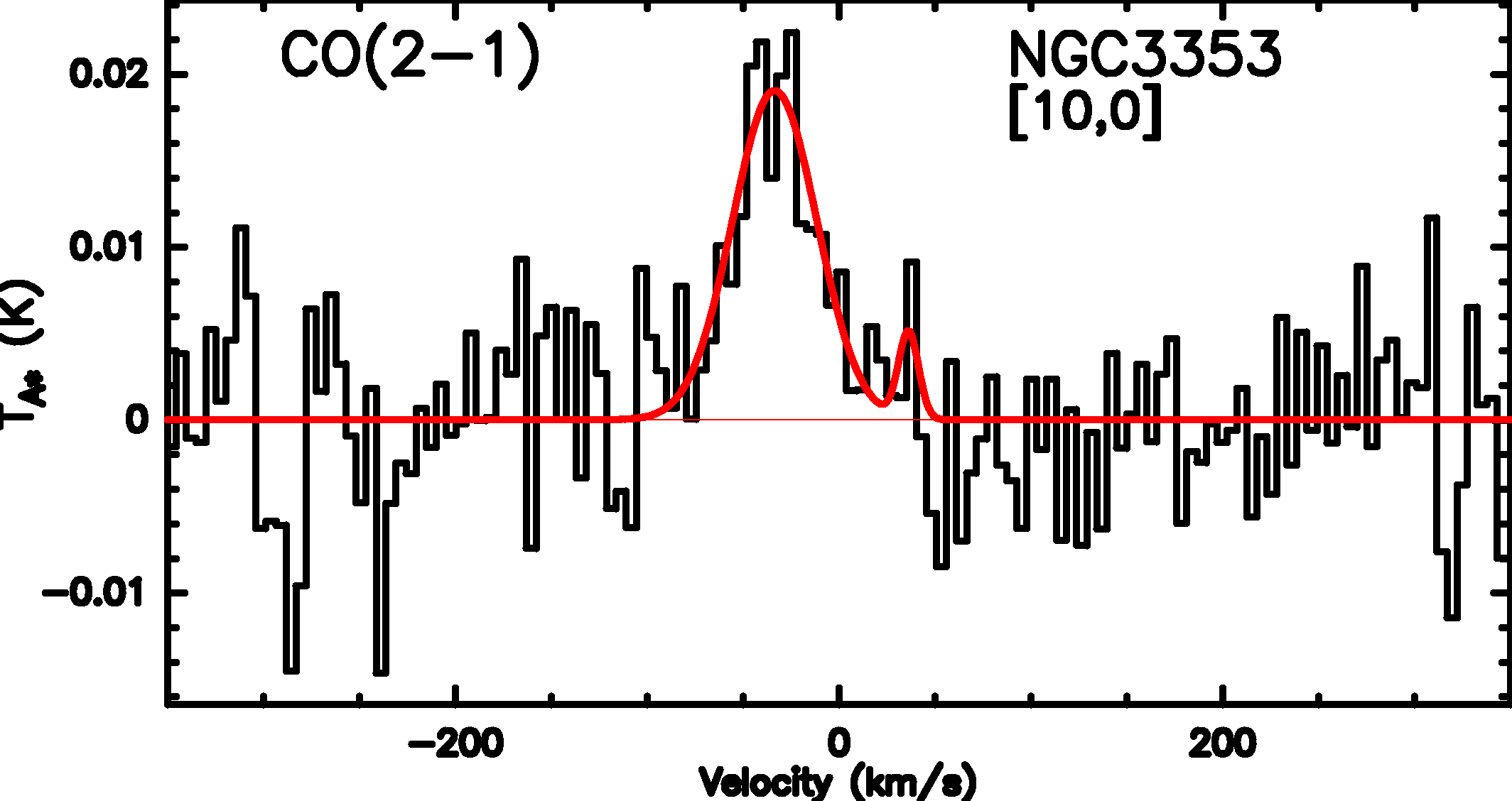}
\hspace{0.2cm}
\includegraphics[angle=0,width=0.3\linewidth]{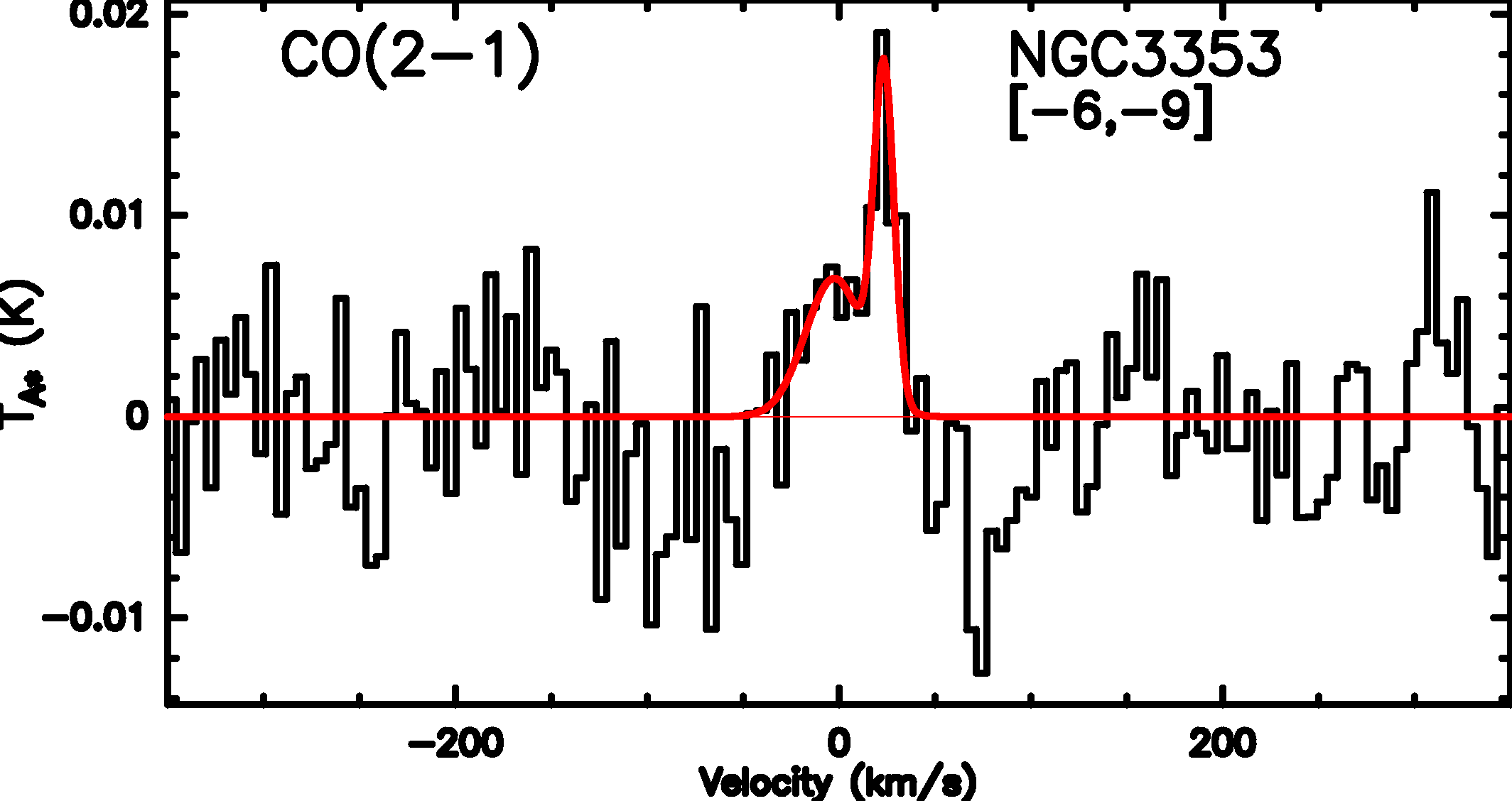}
}
}
\vspace{\baselineskip}
\hbox{
\centerline{
\includegraphics[angle=0,width=0.3\linewidth]{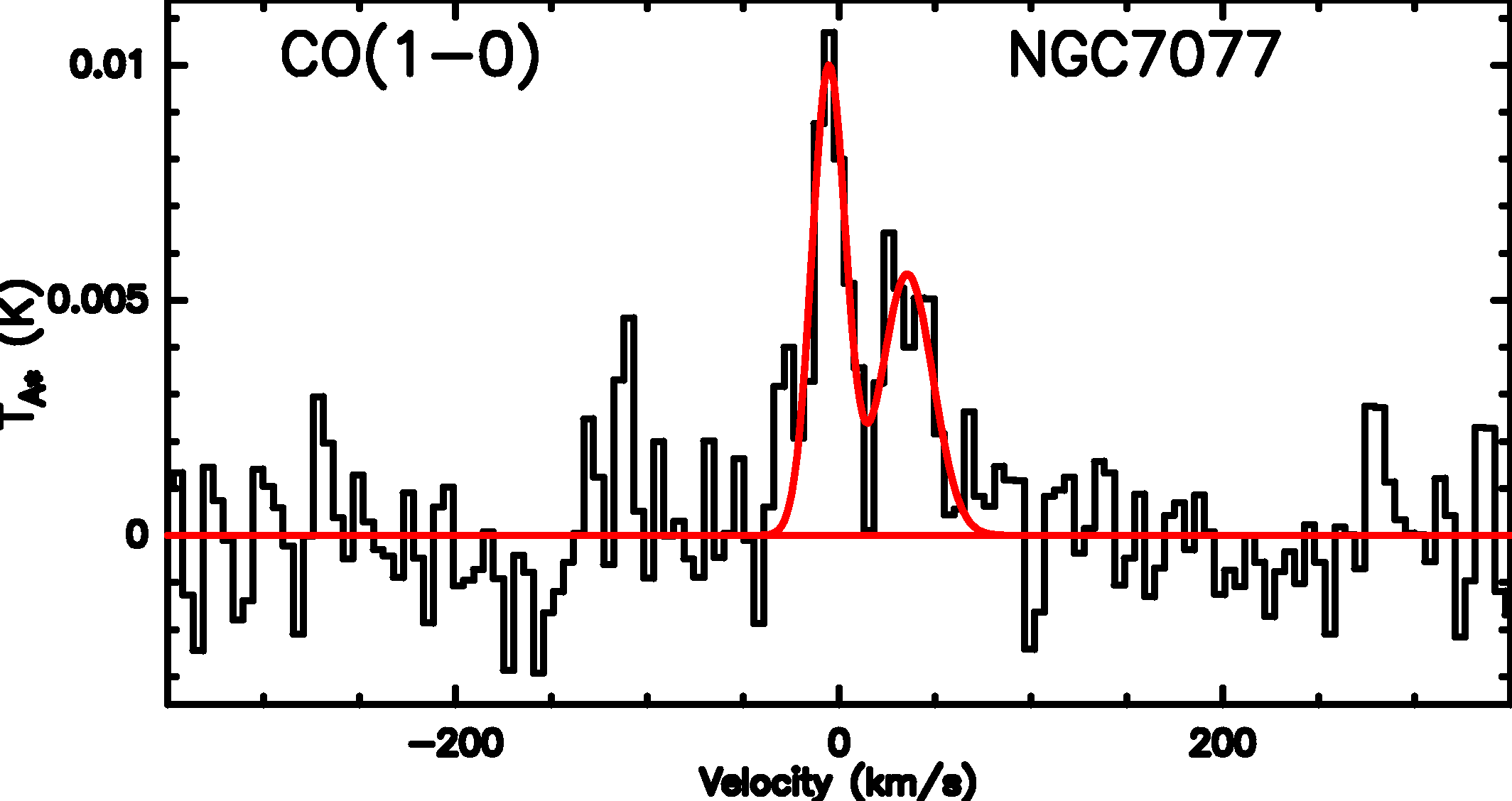}
\hspace{0.2cm}
\includegraphics[angle=0,width=0.3\linewidth]{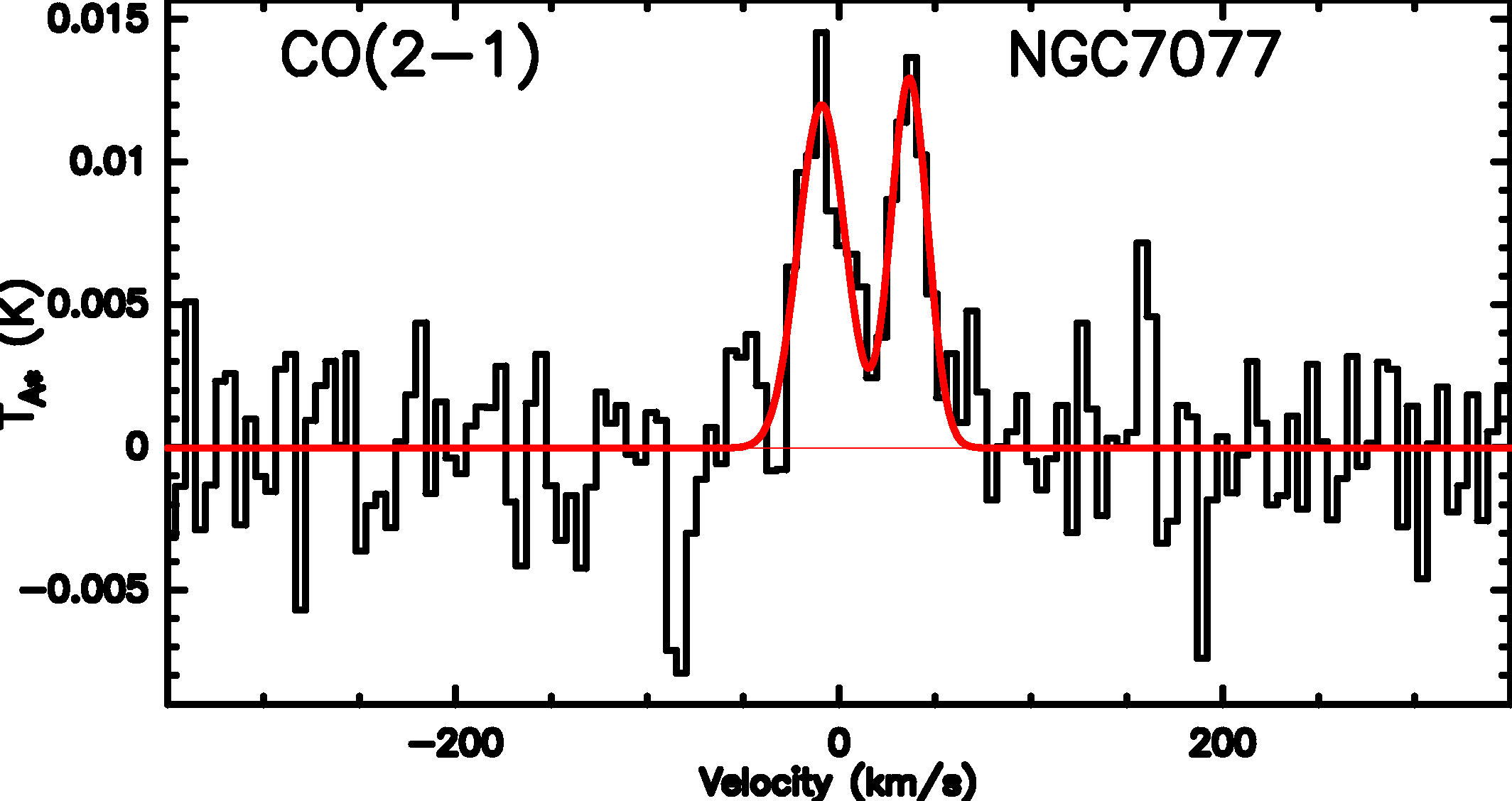}
\hspace{0.2cm}
\includegraphics[angle=0,width=0.3\linewidth]{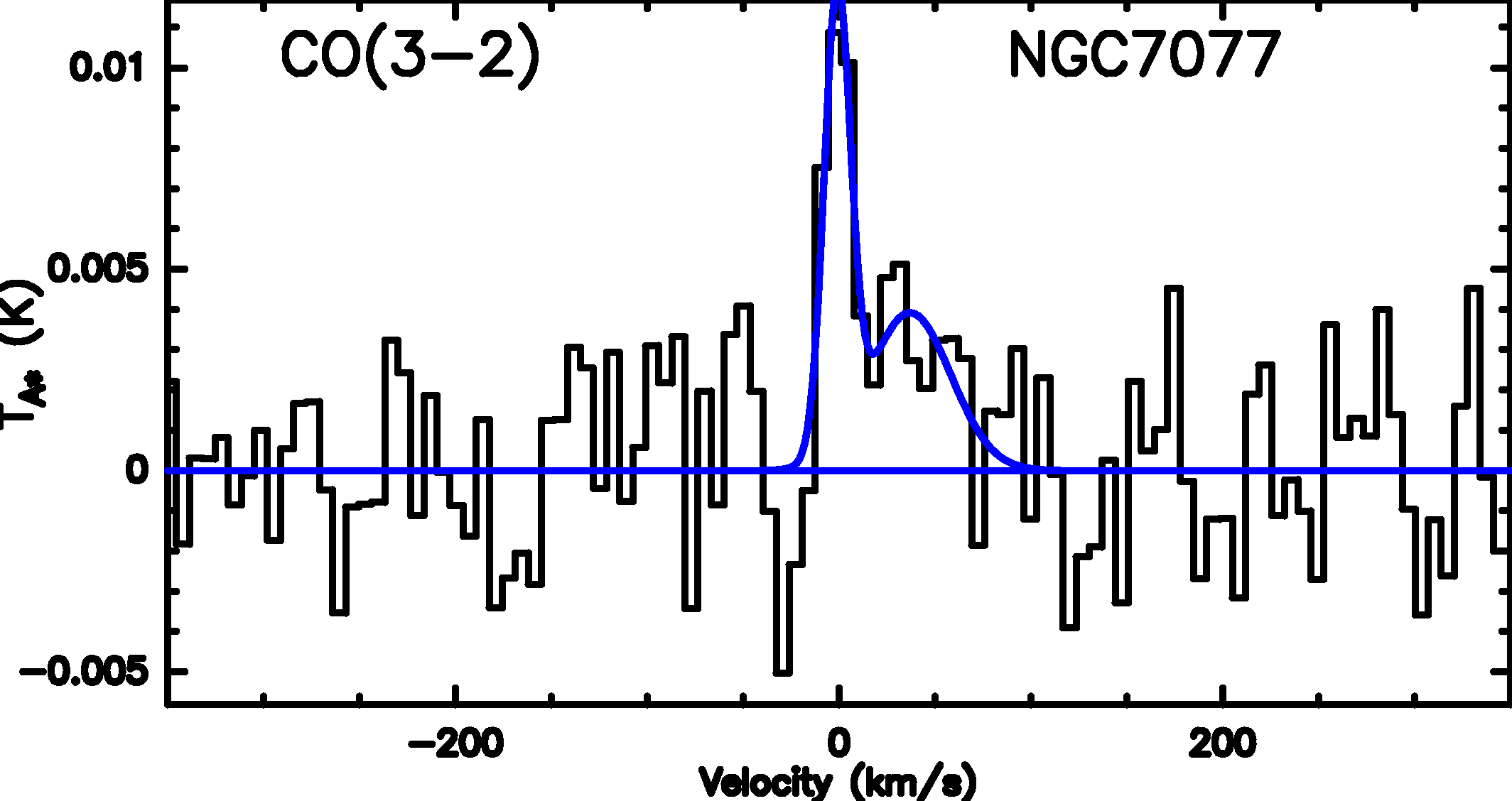}
}
}
\vspace{\baselineskip}
\hbox{
\centerline{
\includegraphics[angle=0,width=0.3\linewidth]{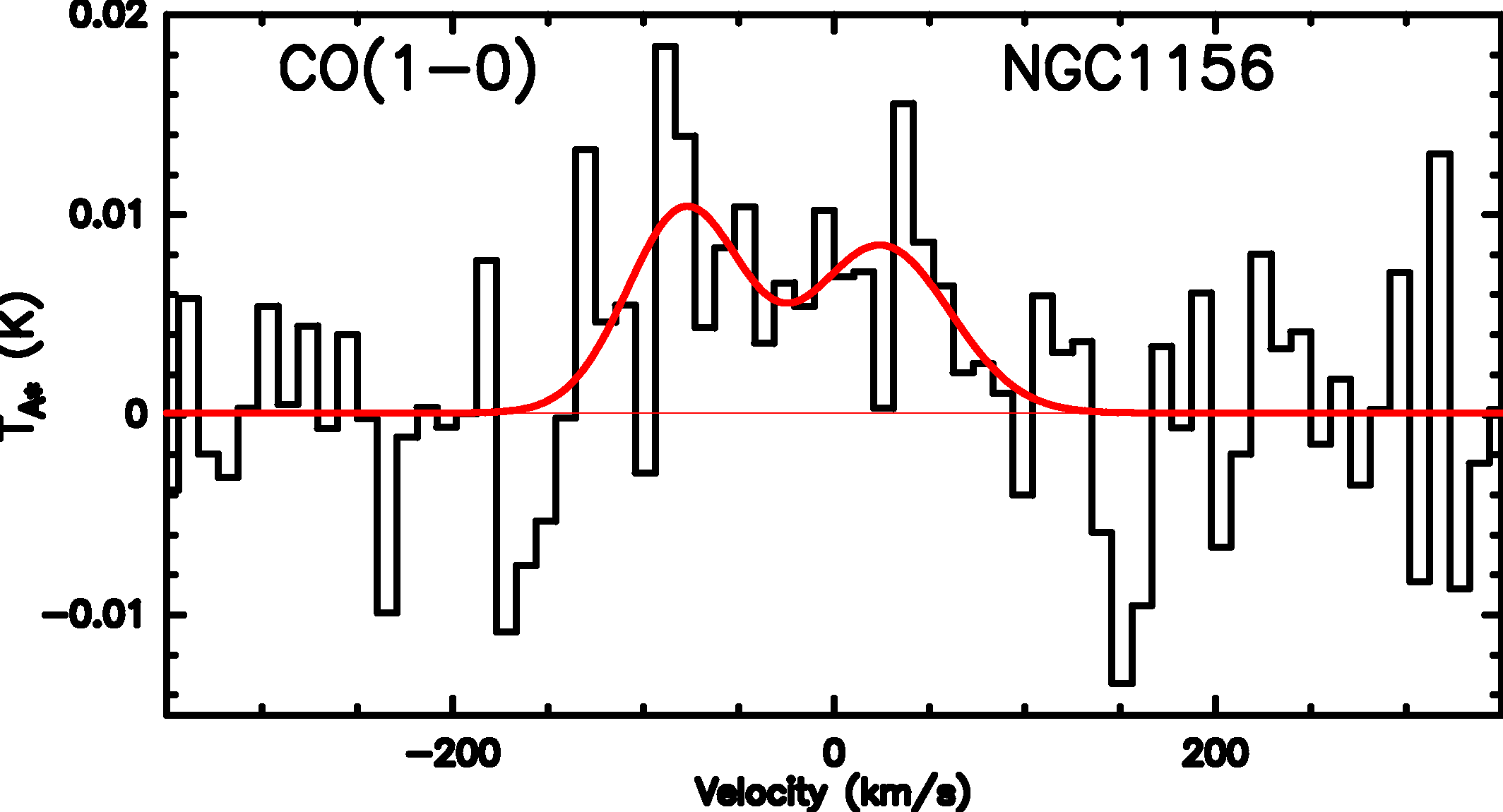}
\hspace{0.2cm}
\includegraphics[angle=0,width=0.3\linewidth]{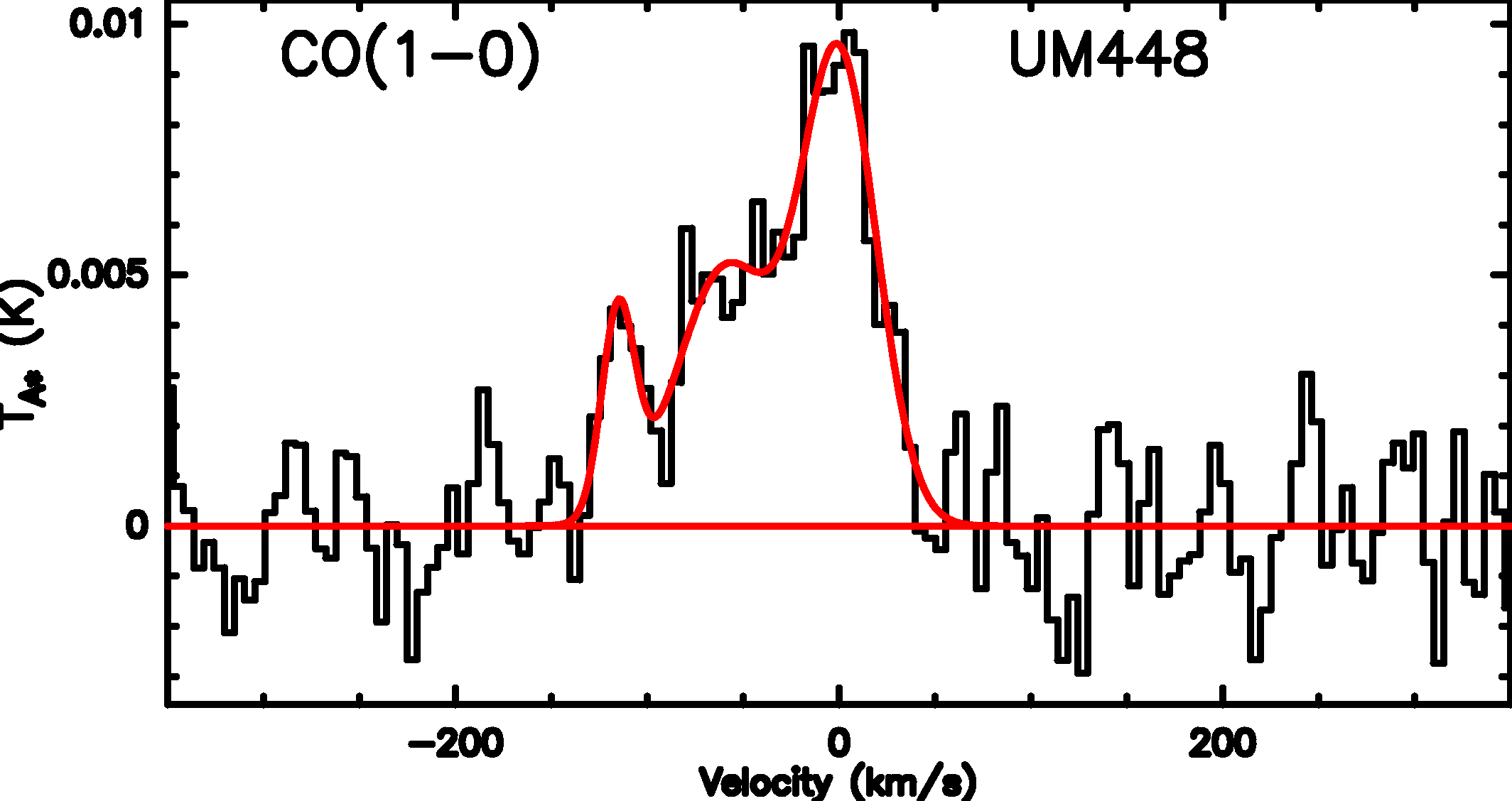}
\hspace{0.2cm}
\includegraphics[angle=0,width=0.3\linewidth]{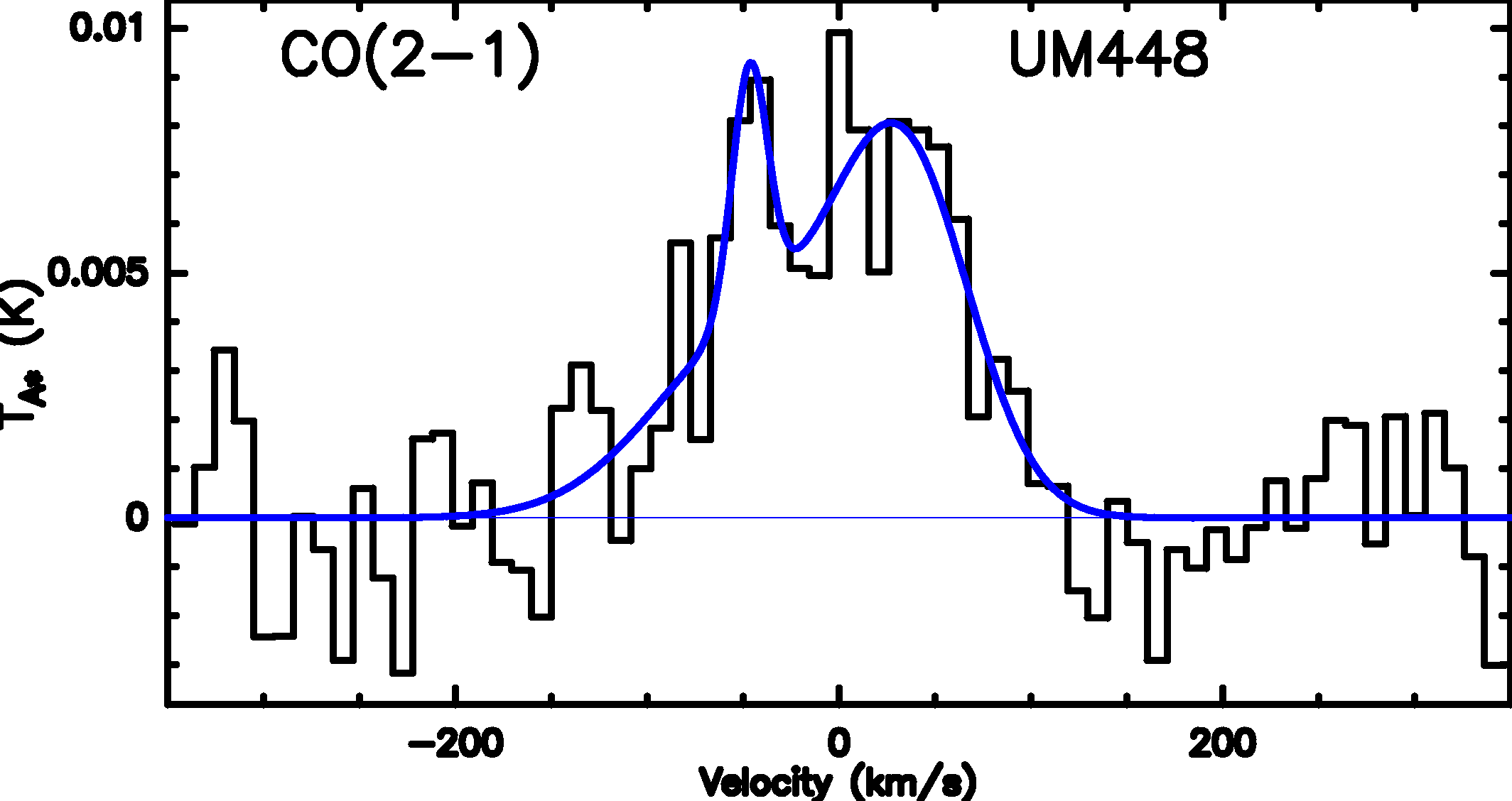}
}
}
\vspace{\baselineskip}
\hbox{
\centerline{
\includegraphics[angle=0,width=0.3\linewidth]{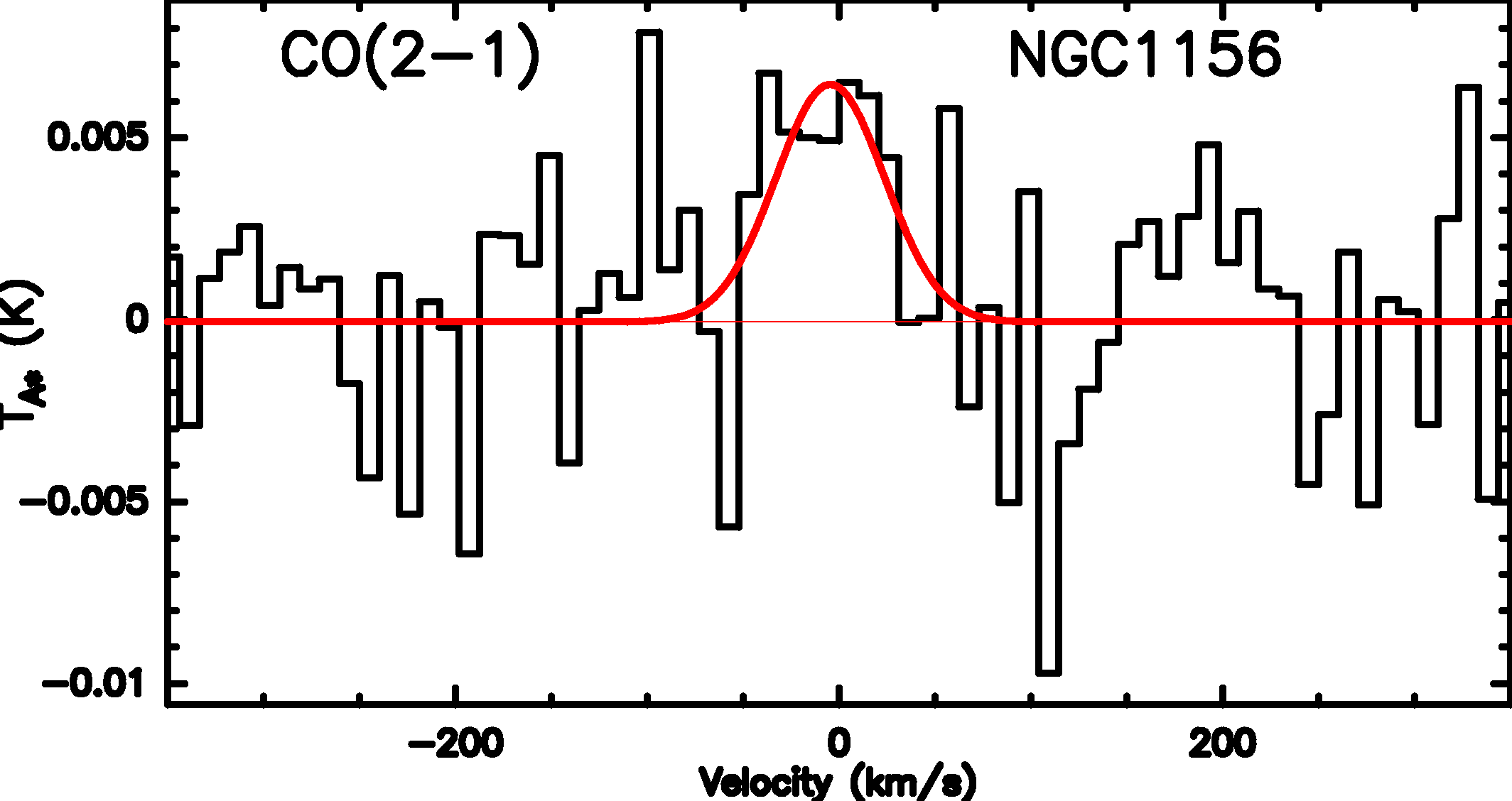}
\hspace{0.2cm}
\includegraphics[angle=0,width=0.3\linewidth]{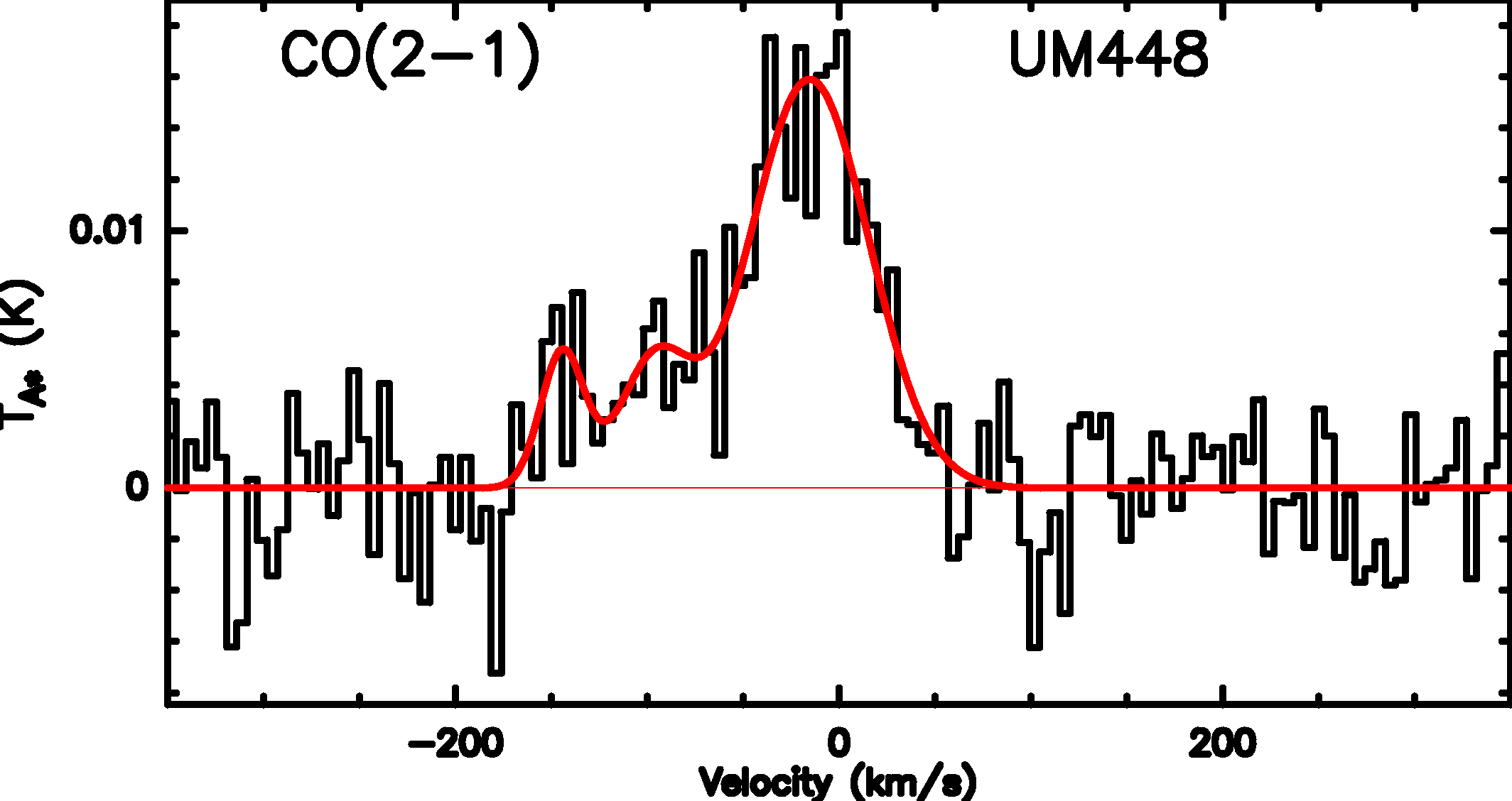}
\hspace{0.2cm}
\includegraphics[angle=0,width=0.3\linewidth]{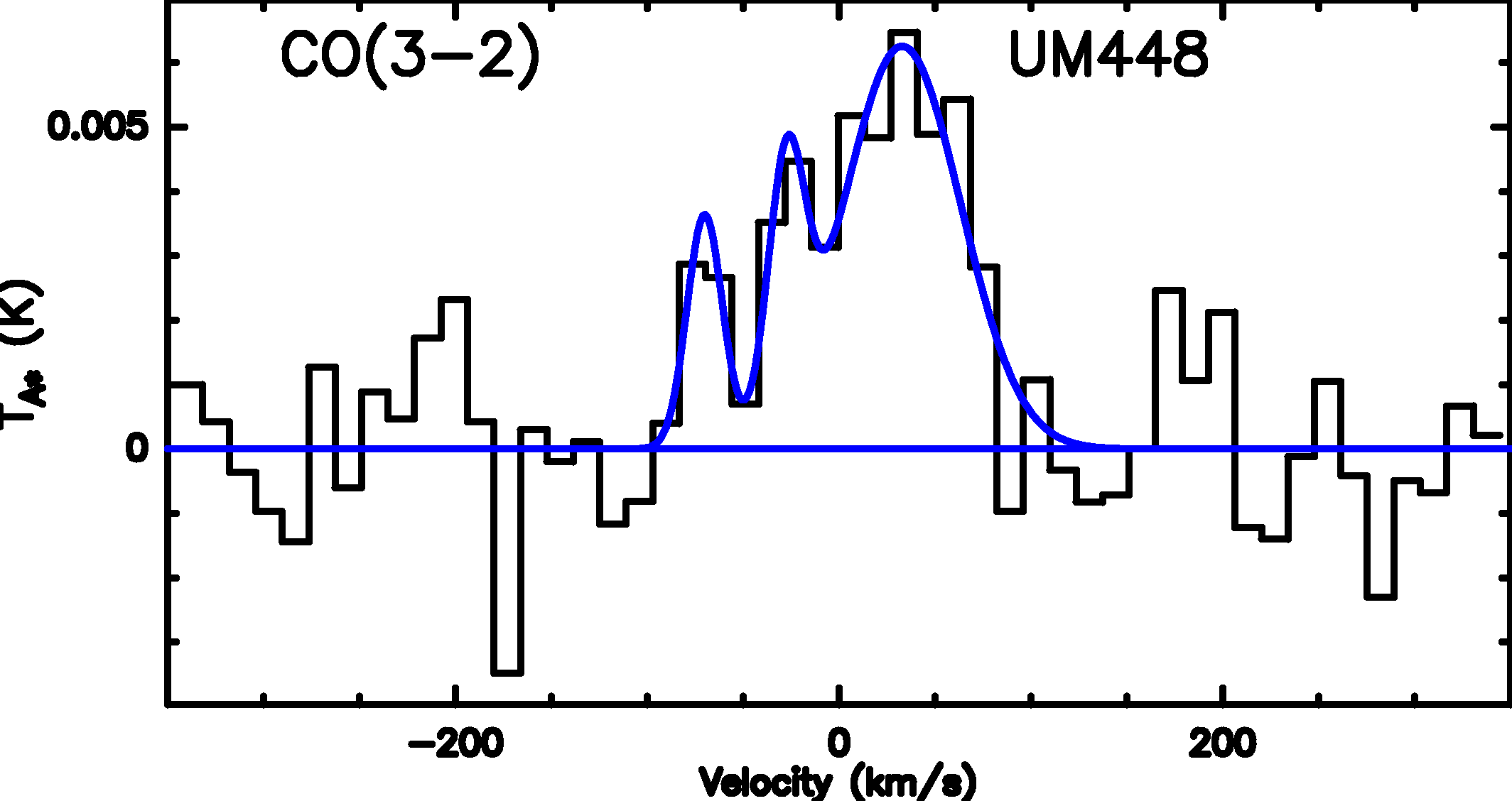}
}
}

%\end{minipage}
%\hbox{
%\centerline{
%\begin{minipage}[c]{0.3\textwidth}
%\vspace{\baselineskip}
%\includegraphics[angle=0,width=\linewidth]{UM462-co10-1gauss_15kmschannels_TA_pretty}
%\end{minipage}
%\hspace{0.8cm}
%\begin{minipage}[c]{0.585\textwidth}
%\vspace{-0.8cm}
\caption{Various \twelveco\ transitions for Mrk\,996, NGC\,3353, NGC\,7077, NGC\,1156,
and UM\,448.
Top row: \coone, \cotwo\ (IRAM), and \cotwo\ (APEX) for Mrk\,996; 
second row: \coone\ for NGC\,3353 (Haro\,3) for three pointings
([0,0], [0,+10\arcsec], [-6\arcsec, -9\arcsec]);
third row: \cotwo\ for NGC\,3353 (Haro\,3) for three pointings;
fourth row: \coone, \cotwo\ (IRAM), and \cothree\ (APEX) for NGC\,7077 (Mrk\,900);
fifth row: \coone\ for NGC\,1156 and UM\,448, and \cotwo\ (APEX) for UM\,448; 
bottom row:  \cotwo\ (IRAM) for NGC\,1156 and UM\,448 and \cothree\ (APEX) for UM\,448.
The baselines are shown as a horizontal solid line, together with the multiple-component
Gaussian fits as described in Sect. \ref{sec:reduction}.
The vertical axes are in \ta\ units, while in Tables \ref{tab:lines} 
and \ref{tab:linesagain} the units are \tmb,
converted from \ta\ as described in Sect. \ref{sec:reduction}.
Velocity channels are as reported in Table \ref{tab:lines}.
The red curves and lines refer to 30-m spectra, while the blue ones to APEX. 
\label{fig:lines}
}
%\end{minipage}
%}
%}
\end{figure*}

\begin{figure*}[ht]
\begin{minipage}[t]{\textwidth}
\vspace{\baselineskip}
\hbox{
\centerline{
\includegraphics[angle=0,width=0.3\linewidth]{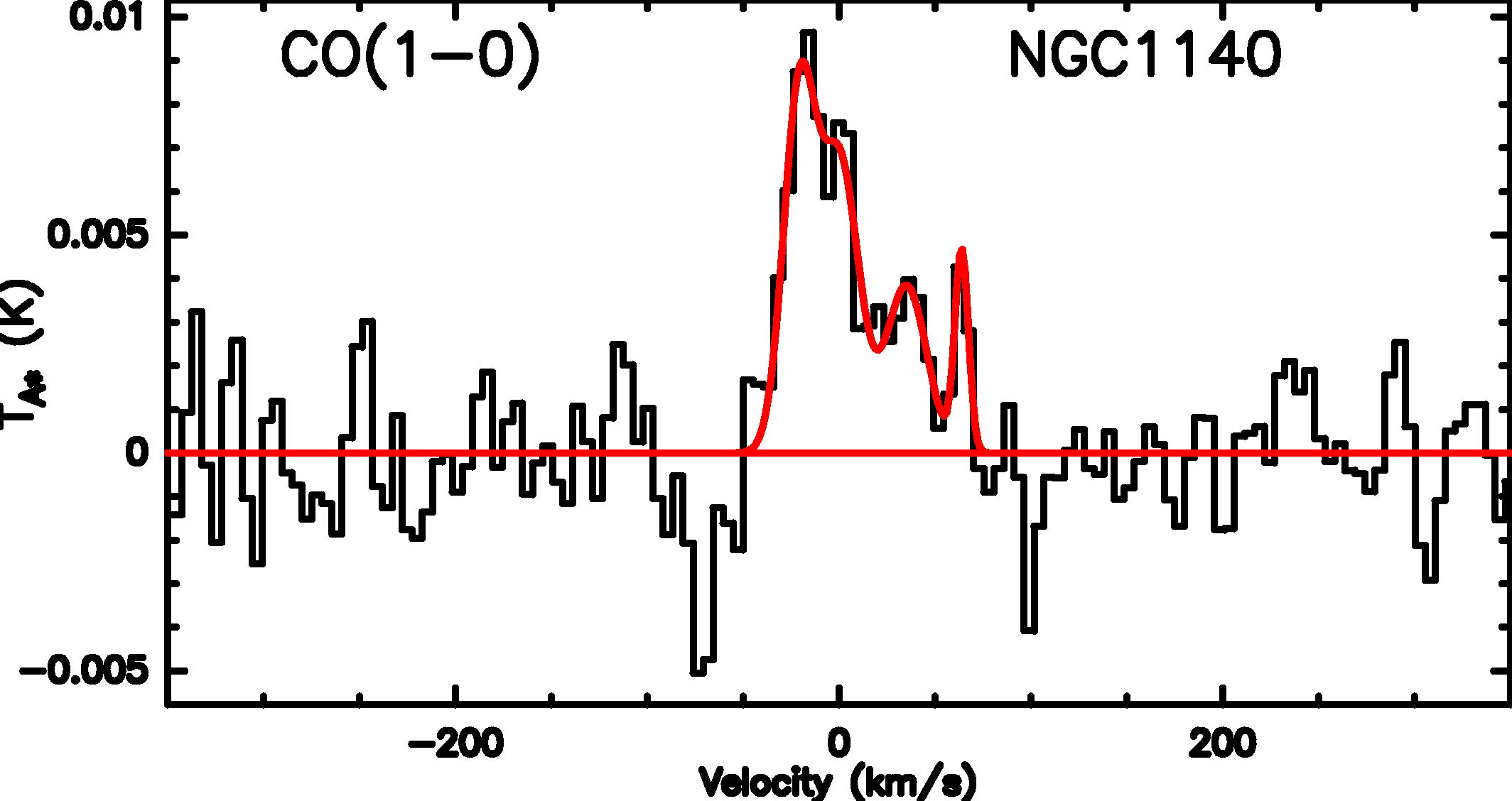}
\hspace{0.2cm}
\includegraphics[angle=0,width=0.3\linewidth]{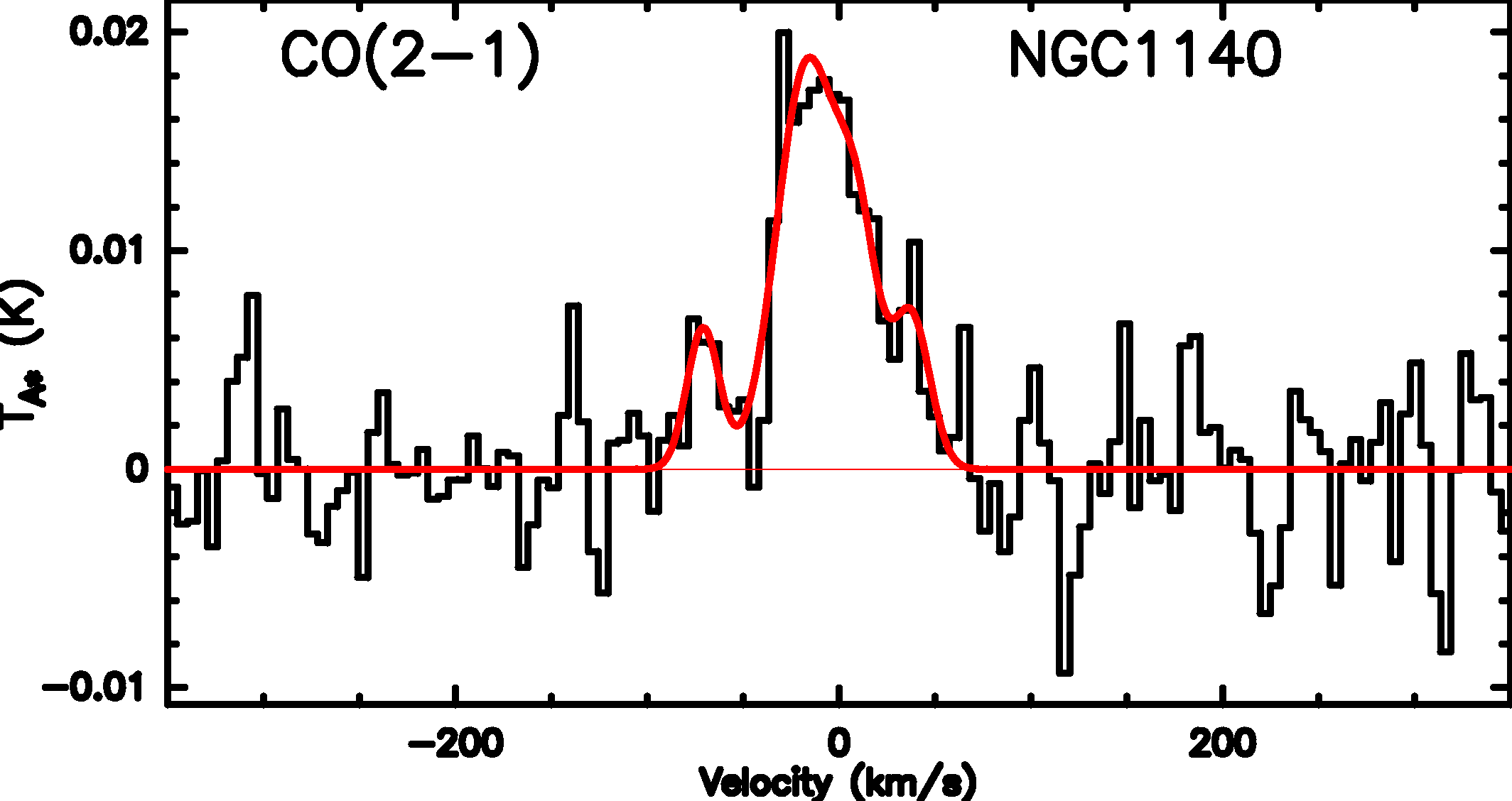}
\hspace{0.2cm}
\includegraphics[angle=0,width=0.3\linewidth]{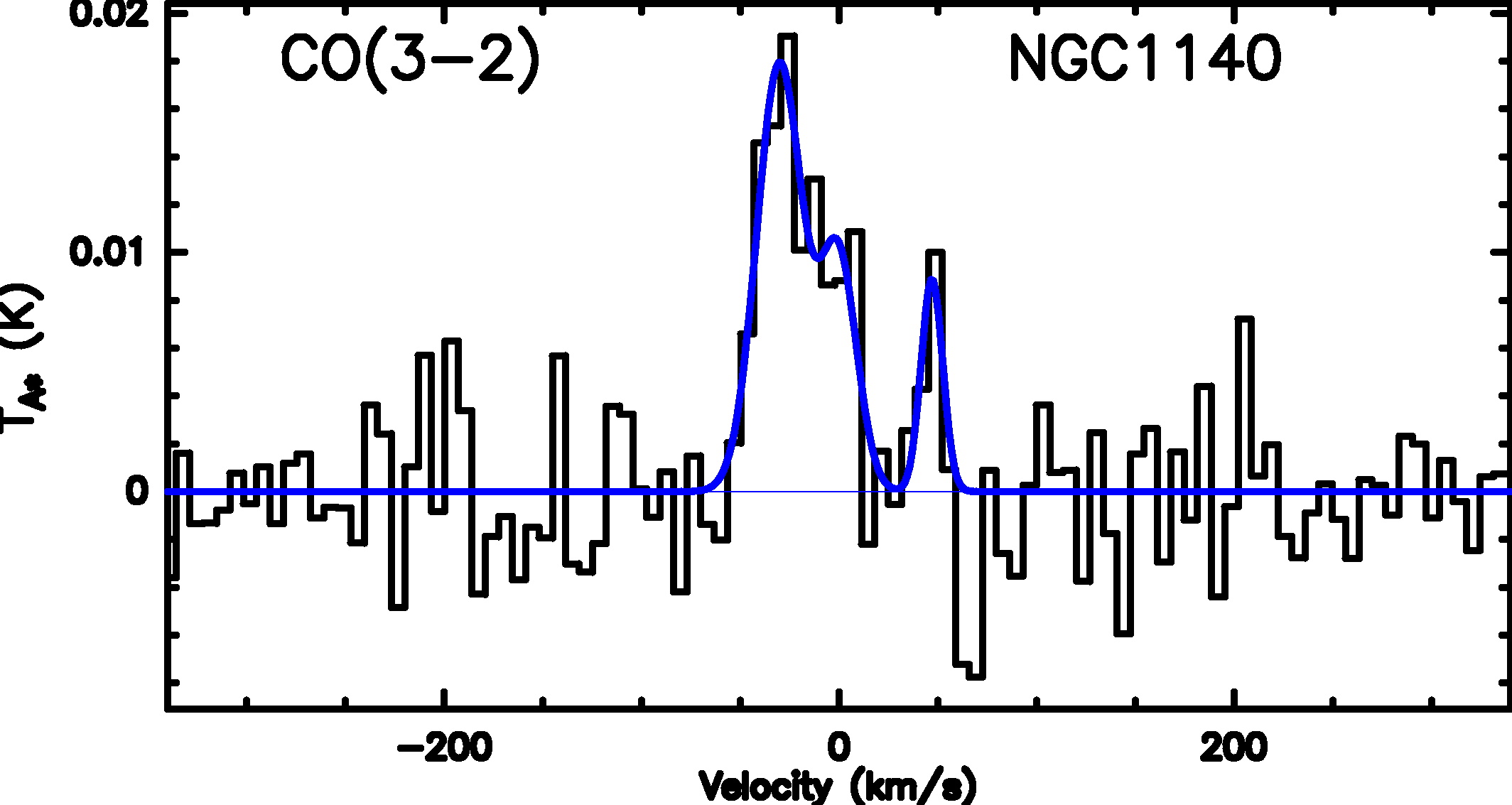}
}
}
\vspace{\baselineskip}
\hbox{
\centerline{
\includegraphics[angle=0,width=0.3\linewidth]{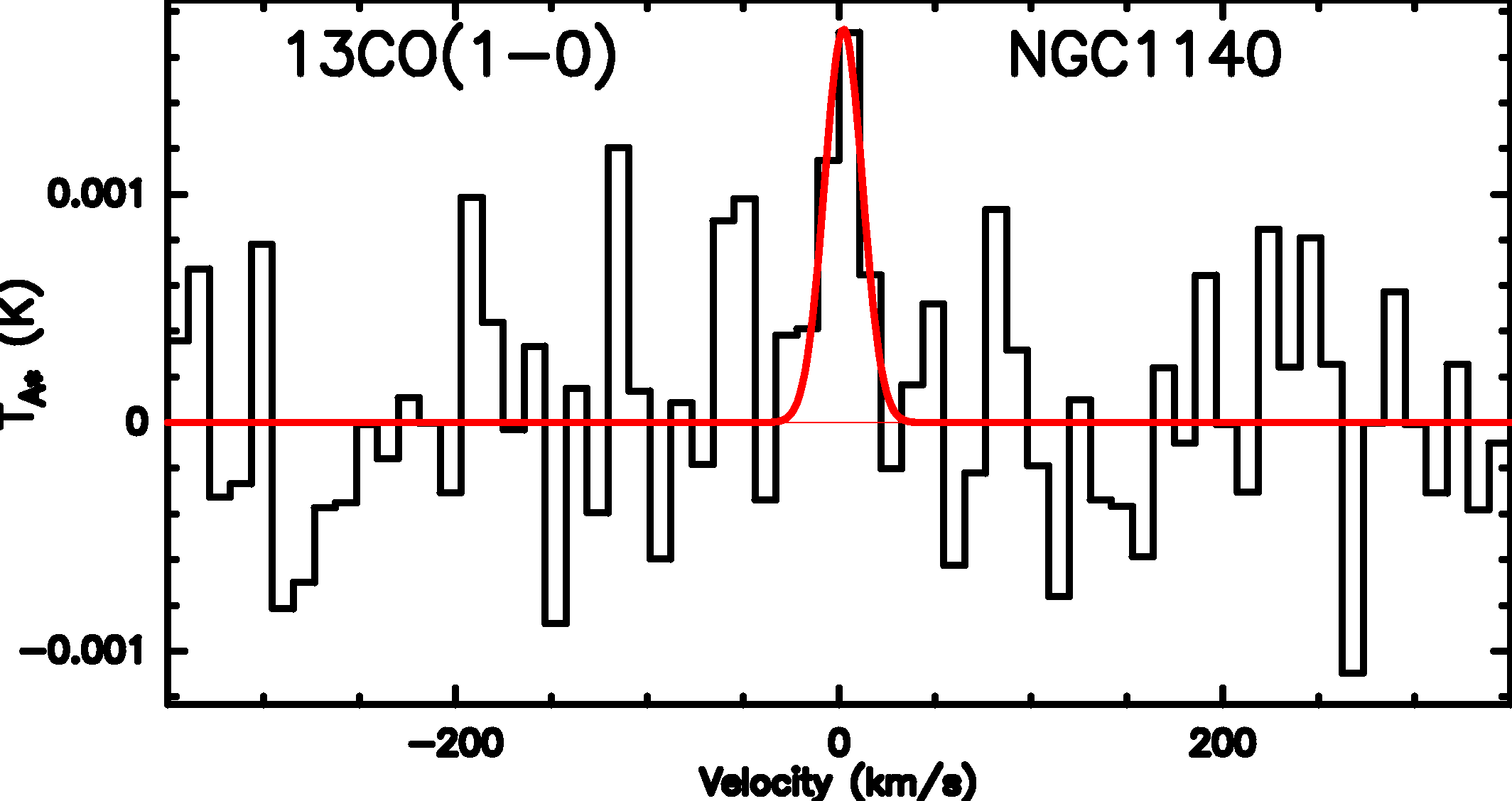}
\hspace{0.2cm}
\includegraphics[angle=0,width=0.3\linewidth]{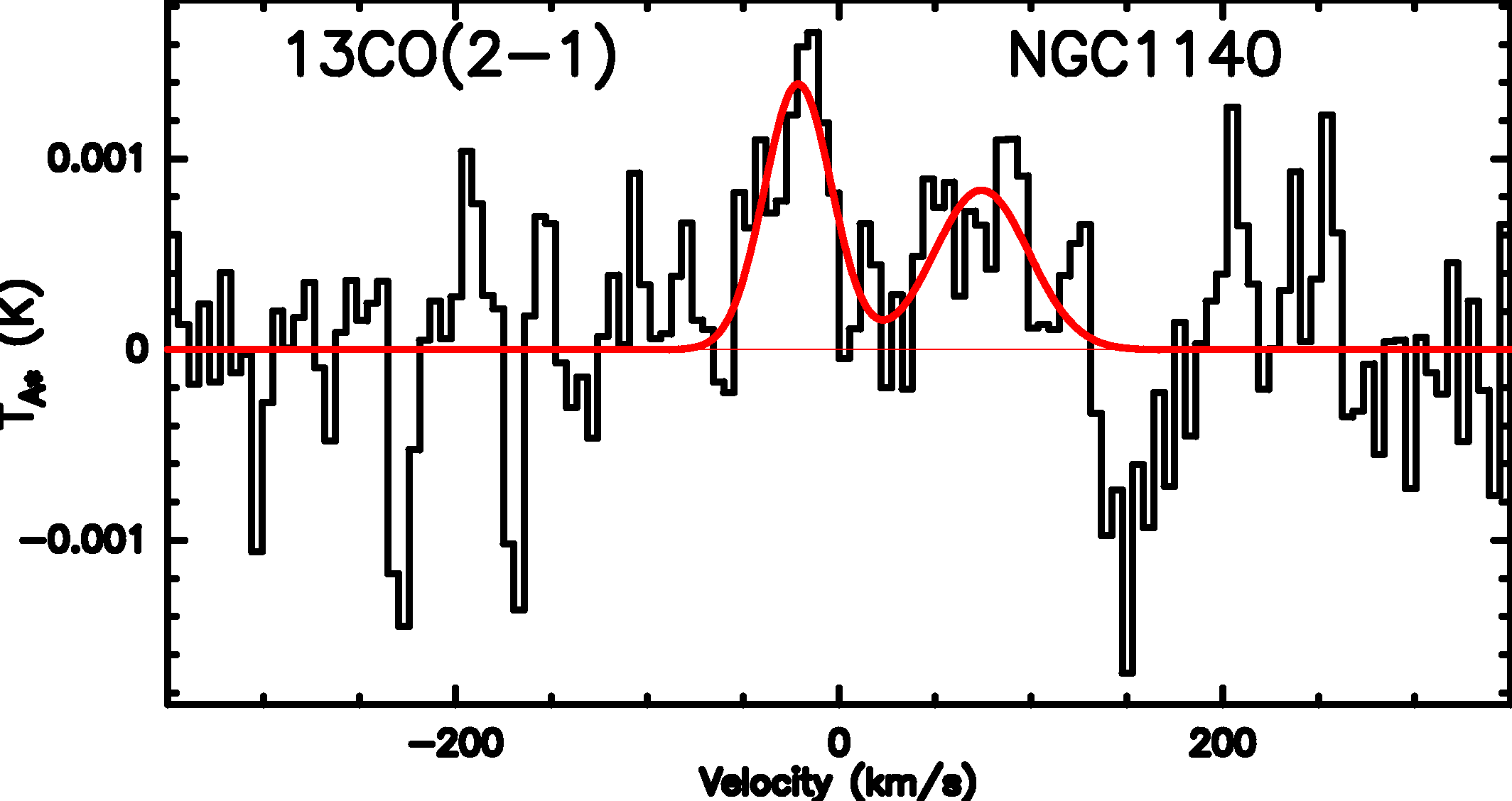}
\hspace{0.2cm}
\includegraphics[angle=0,width=0.3\linewidth]{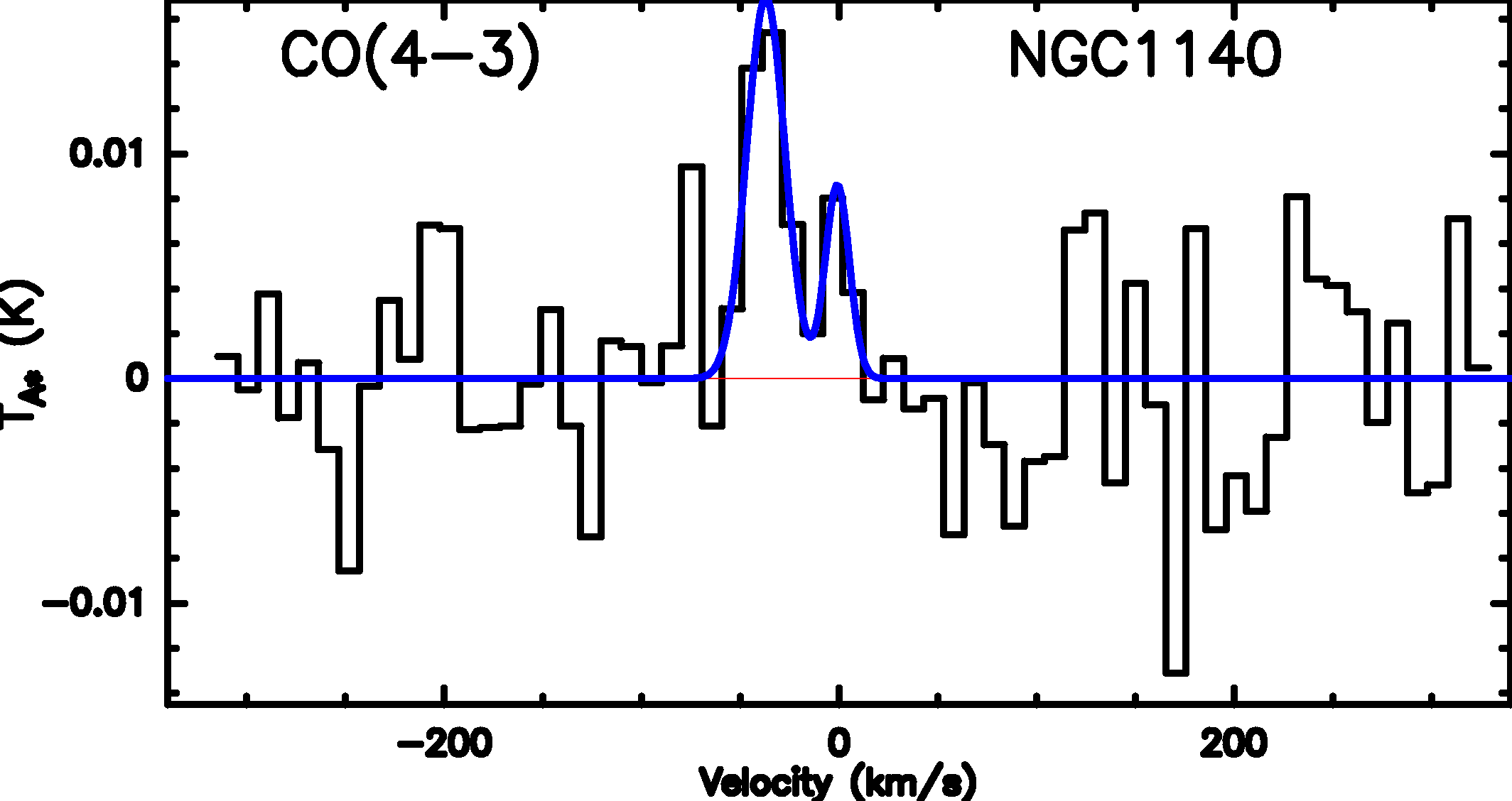}
}
}
\end{minipage}
\hbox{
\centerline{
\begin{minipage}[c]{0.3\textwidth}
\vspace{\baselineskip}
\includegraphics[angle=0,width=\linewidth]{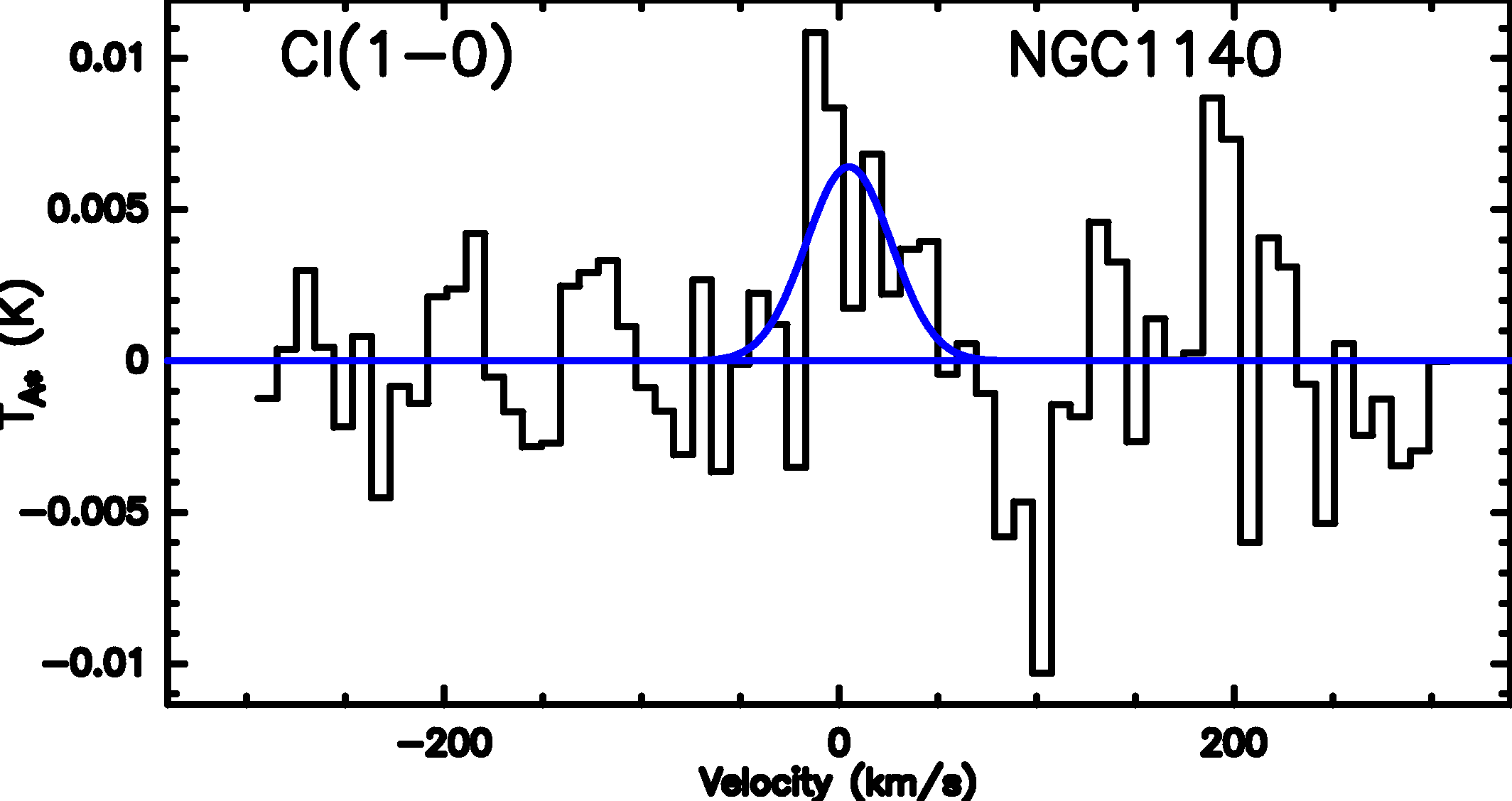}
\end{minipage}
\hspace{0.8cm}
\begin{minipage}[c]{0.585\textwidth}
%\vspace{-0.8cm}
\caption{\twelveco, \thirteenco, and \cione\ transitions for NGC\,1140.
Top row: \coone, \cotwo\ (IRAM), and \cothree\ (APEX);
second row: \thirteencoone, \thirteencotwo\ (IRAM), and \cofour\ (APEX);
third row: \cione\ (APEX).
As in Fig. \ref{fig:lines},
the baselines are shown as a horizontal solid line, together with the multiple-component
Gaussian fits as described in Sect. \ref{sec:reduction}.
The vertical axes are in \ta\ units, while in Table \ref{tab:n1140} the units are \tmb,
converted from \ta\ as described in Sect. \ref{sec:reduction}.
Velocity channels are as reported in Table \ref{tab:lines}.
The red curves and lines refer to 30-m spectra, while the blue ones to APEX. 
\label{fig:n1140}
}
\end{minipage}
}
}
\end{figure*}

% -----------------------------------------------------------------
% --- Table 1: observations
%
\begin{center}
\begin{table*}
      \caption[]{CO emission line parameters for CGCG\,007$-$025, Mrk\,996, NGC\,1156, and NGC\,3353$^{\rm a}$} 
\label{tab:lines}
\resizebox{\linewidth}{!}{
%\addtolength{\tabcolsep}{7pt}
\addtolength{\tabcolsep}{1pt}
{\small
%\tiny
\begin{tabular}{lrrcrrrrrrrrrr}
\hline
\multicolumn{1}{c}{Name} &
\multicolumn{1}{c}{Systemic} &
\multicolumn{1}{c}{Transition} &
\multicolumn{1}{c}{Component} &
\multicolumn{1}{c}{Beam} &
\multicolumn{1}{c}{Channel} &
\multicolumn{1}{c}{T$_{\rm rms}$} &
\multicolumn{1}{c}{T$_{\rm peak}$} &
\multicolumn{1}{c}{Offset$^{\rm b}$} &
\multicolumn{1}{c}{FWHM} &
\multicolumn{1}{c}{$I_{CO}$} &
\multicolumn{1}{c}{$I_{CO}^{\rm cor}$ $^{\mathrm c}$} &
\multicolumn{1}{c}{$S_{CO}^{\rm cor}$ $^{\mathrm c}$} &
\multicolumn{1}{c}{Log($L^\prime_{\rm cor}$) $^{\mathrm c}$} \\
&
\multicolumn{1}{c}{velocity} &
&& \multicolumn{1}{c}{size} & 
\multicolumn{1}{c}{width} &
\multicolumn{1}{c}{(mK)} &
\multicolumn{1}{c}{(mK)} &
\multicolumn{1}{c}{(\kms)} &
\multicolumn{1}{c}{(\kms)} &
\multicolumn{1}{c}{(\kkms)} &
\multicolumn{1}{c}{(\kkms)} &
\multicolumn{1}{c}{(Jy\,\kms)} &
\multicolumn{1}{c}{(\kkmspc)} \\
& \multicolumn{1}{c}{(\kms)}
&&& 
\multicolumn{1}{c}{(\arcsec)}
& \multicolumn{1}{c}{(\kms)} & & & & & & \\
\hline
% redone 19/8/2016
% redone 1/3/2017 the luminosities after Axel's comment
\\ 
CGCG\,007-025 & 1449 & \coone\    & Total & 21.4 & 15.7 &  1.3 &   2.3 &  -34 &  129 &    0.314   (  0.092) &    0.314   (  0.092) &    1.60    &    6.368   (  0.126)\\
CGCG\,007-025 & & \cotwo\    & Total      & 10.7 & 10.5 &  3.9 & $-$ & $-$ & $-$ & $<$0.31    & $-$ & $-$ & $<$5.83   \\
CGCG\,007-025 & & \cotwo\    & Total      & 10.7 & 15.7 &  3.4 & $-$ & $-$ & $-$ & $<$0.31    & $-$ & $-$ & $<$5.95   \\
\\ 
Mrk\,996     & 1622 & \coone\    & 1     & 21.5 &  5.2 &  1.8 &   6.0 &   -1 &   28 &    0.189   (  0.034) &    0.182   (  0.033) &    0.97    &    5.890   (  0.077)\\
Mrk\,996     & & \coone\    & 2     & 21.5 &  5.2 &  1.8 &   4.7 &   39 &   10 &    0.054   (  0.019) &    0.052   (  0.019) &    0.28    &    5.347   (  0.154)\\
Mrk\,996     & & \coone\    & Total & 21.5 &  5.2 &  1.8 &   6.0 &    0 &   40 &    0.243   (  0.039) &    0.235   (  0.037) &    1.25    &    5.999   (  0.069)\\
Mrk\,996     & & \cotwo\    & Total & 10.7 &  5.2 &  2.5 &   3.8 &  -12 &   32 &    0.291   (  0.047) &    0.128   (  0.020) &    2.69    &    5.729   (  0.070)\\
Mrk\,996     & & \cotwo\    & 1     & 27.2 & 10.2 &  2.1 &   6.0 &  -12 &  123 &    0.563   (  0.131) &    0.782   (  0.182) &    15.34   &    6.485   (  0.101)\\
Mrk\,996     & & \cotwo\    & 2     & 27.2 & 10.2 &  2.1 &   4.0 &   72 &   36 &    0.113   (  0.076) &    0.156   (  0.106) &    3.07    &    5.786   (  0.295)\\
Mrk\,996     & & \cotwo\    & Total & 27.2 & 10.2 &  2.1 &   6.0 &    0 &   84 &    0.675   (  0.151) &    0.938   (  0.210) &    18.40   &    6.564   (  0.097)\\
Mrk\,996     & & \cothree\  & Total      & 18.1 & 13.6 &  1.7 & $-$ & $-$ & $-$ & $<$0.21    & $-$ & $-$ & $<$5.74   \\
\\ 
NGC\,1156    & 375 & \coone\    & 1     & 21.4 & 10.4 &  7.1 &  12.2 &  -78 &   71 &    0.933   (  0.406) &    0.929   (  0.404) &    5.00    &    5.904   (  0.189)\\
NGC\,1156    & & \coone\    & 2     & 21.4 & 10.4 &  7.1 &  10.0 &   25 &   85 &    0.916   (  0.433) &    0.912   (  0.431) &    4.91    &    5.896   (  0.205)\\
NGC\,1156    & & \coone\    & Total & 21.4 & 10.4 &  7.1 &  12.2 &    0 &  103 &    1.849   (  0.593) &    1.841   (  0.590) &    9.91    &    6.201   (  0.139)\\
NGC\,1156    & & \cotwo\    & Total & 10.7 & 10.4 &  4.8 &   9.4 &   -5 &   67 &    0.725   (  0.206) &    0.668   (  0.190) &    14.19   &    5.755   (  0.123)\\
\\
NGC\,3353 [0\arcsec, 0\arcsec] & 944  & \coone\  & 1     & 21.4 &  5.2 &  5.5 &  27.6 &  -31 &   60 &    1.824   (  0.177) &    1.771   (  0.172) &    9.50    &    6.880   (  0.042)\\
NGC\,3353 [0\arcsec, 0\arcsec] & & \coone\  & 2     & 21.4 &  5.2 &  5.5 &  19.7 &   21 &   16 &    0.339   (  0.112) &    0.329   (  0.109) &    1.76    &    6.149   (  0.144)\\
NGC\,3353 [0\arcsec, 0\arcsec] & & \coone\  & Total & 21.4 &  5.2 &  5.5 &  27.6 &    0 &   52 &    2.163   (  0.210) &    2.100   (  0.204) &    11.26   &    6.954   (  0.042)\\
NGC\,3353 [0\arcsec, 0\arcsec] & & \thirteencoone\  & Total      & 22.4 & 10.9 &  2.0 & $-$ & $-$ & $-$ & $<$2.10    & $-$ & $-$ & $<$5.95   \\
NGC\,3353 [0\arcsec, 0\arcsec] & & \cotwo\  & 1     & 10.7 &  5.2 &  7.9 &  31.7 &  -33 &   42 &    2.611   (  0.232) &    1.430   (  0.127) &    30.27   &    6.781   (  0.039)\\
NGC\,3353 [0\arcsec, 0\arcsec] & & \cotwo\  & 2     & 10.7 &  5.2 &  7.9 &  19.5 &   15 &   22 &    0.814   (  0.175) &    0.446   (  0.096) &    9.43    &    6.275   (  0.093)\\
NGC\,3353 [0\arcsec, 0\arcsec] & & \cotwo\  & Total & 10.7 &  5.2 &  7.9 &  31.7 &    0 &   48 &    3.424   (  0.290) &    1.876   (  0.159) &    39.70   &    6.899   (  0.037)\\
\\ 
NGC\,3353 [0\arcsec, $+$10\arcsec] & 944 & \coone\  & 1     & 21.4 &  5.2 &  4.5 &  33.0 &  -35 &   39 &    1.412   (  0.123) &    1.371   (  0.119) &    7.35    &    6.769   (  0.038)\\
NGC\,3353 [0\arcsec, $+$10\arcsec] & & \coone\  & 2     & 21.4 &  5.2 &  4.5 &  11.8 &   19 &   36 &    0.468   (  0.123) &    0.454   (  0.119) &    2.43    &    6.289   (  0.114)\\
NGC\,3353 [0\arcsec, $+$10\arcsec] & & \coone\  & Total & 21.4 &  5.2 &  4.5 &  33.0 &    0 &   54 &    1.880   (  0.174) &    1.825   (  0.168) &    9.79    &    6.893   (  0.040)\\
NGC\,3353 [0\arcsec, $+$10\arcsec] & & \cotwo\  & 1     & 10.7 &  5.2 &  7.9 &  16.2 &  -34 &   52 &    1.635   (  0.207) &    0.896   (  0.114) &    18.96   &    6.578   (  0.055)\\
NGC\,3353 [0\arcsec, $+$10\arcsec] & & \cotwo\  & 2     & 10.7 &  5.2 &  7.9 &   4.3 &   36 &   12 &    0.103   (  0.079) &    0.056   (  0.044) &    1.19    &    5.377   (  0.336)\\
NGC\,3353 [0\arcsec, $+$10\arcsec] & & \cotwo\  & Total & 10.7 &  5.2 &  7.9 &  16.2 &    0 &   70 &    1.738   (  0.221) &    0.952   (  0.121) &    20.15   &    6.605   (  0.055)\\
\\ 
NGC\,3353 [$-$6\arcsec, $-$9\arcsec] & 944 & \coone\  & 1     & 21.4 &  5.2 &  4.6 &   6.7 &  -56 &   29 &    0.217   (  0.037) &    0.211   (  0.036) &    1.13    &    5.955   (  0.075)\\
NGC\,3353 [$-$6\arcsec, $-$9\arcsec] & & \coone\  & 2     & 21.4 &  5.2 &  4.6 &  13.6 &  -16 &   28 &    0.407   (  0.037) &    0.395   (  0.036) &    2.12    &    6.229   (  0.040)\\
NGC\,3353 [$-$6\arcsec, $-$9\arcsec] & & \coone\  & 3     & 21.4 &  5.2 &  4.6 &  20.0 &   22 &   35 &    0.758   (  0.037) &    0.736   (  0.036) &    3.95    &    6.499   (  0.021)\\
NGC\,3353 [$-$6\arcsec, $-$9\arcsec] & & \coone\  & 4     & 21.4 &  5.2 &  4.6 &   8.8 &   98 &   33 &    0.319   (  0.037) &    0.310   (  0.036) &    1.66    &    6.123   (  0.051)\\
NGC\,3353 [$-$6\arcsec, $-$9\arcsec] & & \coone\  & Total & 21.4 &  5.2 &  4.6 &  20.0 &    0 &   78 &    1.701   (  0.075) &    1.652   (  0.073) &    8.86    &    6.850   (  0.019)\\
NGC\,3353 [$-$6\arcsec, $-$9\arcsec] & & \cotwo\  & 1     & 10.7 &  5.2 &  6.1 &   5.9 &   -2 &   36 &    0.405   (  0.190) &    0.222   (  0.104) &    4.70    &    5.972   (  0.204)\\
NGC\,3353 [$-$6\arcsec, $-$9\arcsec] & & \cotwo\  & 2     & 10.7 &  5.2 &  6.1 &  13.8 &   23 &   12 &    0.332   (  0.115) &    0.182   (  0.063) &    3.85    &    5.886   (  0.151)\\
NGC\,3353 [$-$6\arcsec, $-$9\arcsec] & & \cotwo\  & Total & 10.7 &  5.2 &  6.1 &  13.8 &    0 &   25 &    0.737   (  0.223) &    0.404   (  0.122) &    8.55    &    6.232   (  0.131)\\
\\
\hline
\end{tabular}
} 
}
\vspace{0.5\baselineskip}
\begin{description}
\item
[$^{\mathrm{a}}$] All temperature units in this table are main-beam, \tmb.
Values in parentheses are the $1\sigma$ uncertainties.
The total line flux is the sum of the individual components, and the
total uncertainty is calculated by adding in quadrature
the individual uncertainties. 
Systemic velocities are heliocentric, taken from NED, and used 
as the (optical convention) reference in Fig. \ref{fig:lines}.
%The total line width is the quadrature sum of the individual widths. 
\item
[$^{\mathrm{b}}$] Offsets of the component's central velocity relative to the
systemic one.
\item
[$^{\mathrm{c}}$] These have been corrected for beam dilution
to a common beam size of 22\arcsec\ as described in Sect. \ref{sec:beam}
and Appendix \ref{sec:appendix_exponential}.
\end{description}
\end{table*}
\end{center}
% -----------------------------------------------------------------

 \begin{center}
\begin{table*}
      \caption[]{CO emission line parameters for NGC\,7077, UM\,448 and UM\,462$^{\rm a}$} 
\label{tab:linesagain}
\resizebox{\linewidth}{!}{
%\addtolength{\tabcolsep}{7pt}
\addtolength{\tabcolsep}{3pt}
{\small
%\tiny
\begin{tabular}{lrrcrrrrrrrrrr}
\hline
\multicolumn{1}{c}{Name} &
\multicolumn{1}{c}{Systemic} &
\multicolumn{1}{c}{Transition} &
\multicolumn{1}{c}{Component} &
\multicolumn{1}{c}{Beam} &
\multicolumn{1}{c}{Channel} &
\multicolumn{1}{c}{T$_{\rm rms}$} &
\multicolumn{1}{c}{T$_{\rm peak}$} &
\multicolumn{1}{c}{Offset$^{\rm b}$} &
\multicolumn{1}{c}{FWHM} &
\multicolumn{1}{c}{$I_{CO}$} &
\multicolumn{1}{c}{$I_{CO}^{\rm cor}$ $^{\mathrm c}$} &
\multicolumn{1}{c}{$S_{CO}^{\rm cor}$ $^{\mathrm c}$} &
\multicolumn{1}{c}{Log($L^\prime_{\rm cor}$) $^{\mathrm c}$} \\
&
\multicolumn{1}{c}{velocity} &
&& \multicolumn{1}{c}{size} & 
\multicolumn{1}{c}{width} &
\multicolumn{1}{c}{(mK)} &
\multicolumn{1}{c}{(mK)} &
\multicolumn{1}{c}{(\kms)} &
\multicolumn{1}{c}{(\kms)} &
\multicolumn{1}{c}{(\kkms)} &
\multicolumn{1}{c}{(\kkms)} &
\multicolumn{1}{c}{(Jy\,\kms)} &
\multicolumn{1}{c}{(\kkmspc)} \\
& \multicolumn{1}{c}{(\kms)}
&&& 
\multicolumn{1}{c}{(\arcsec)}
& \multicolumn{1}{c}{(\kms)} & & & & & & \\
\hline
\\ 
NGC\,7077    & 1152 & \coone\    & 1     & 21.4 &  5.2 &  1.7 &  11.6 &   -5 &   21 &    0.265   (  0.034) &    0.257   (  0.033) &    1.38    &    5.996   (  0.055)\\
NGC\,7077    & & \coone\    & 2     & 21.4 &  5.2 &  1.7 &   6.5 &   36 &   31 &    0.224   (  0.040) &    0.217   (  0.039) &    1.16    &    5.923   (  0.077)\\
NGC\,7077    & & \coone\    & Total & 21.4 &  5.2 &  1.7 &  11.6 &    0 &   41 &    0.489   (  0.052) &    0.474   (  0.050) &    2.54    &    6.262   (  0.046)\\
NGC\,7077    & & \cotwo\    & 1     & 10.7 &  5.2 &  3.9 &   9.6 &   -9 &   29 &    0.578   (  0.086) &    0.298   (  0.044) &    6.30    &    6.055   (  0.064)\\
NGC\,7077    & & \cotwo\    & 2     & 10.7 &  5.2 &  3.9 &  10.4 &   37 &   23 &    0.485   (  0.084) &    0.250   (  0.043) &    5.28    &    5.979   (  0.075)\\
NGC\,7077    & & \cotwo\    & Total & 10.7 &  5.2 &  3.9 &  10.4 &    0 &   46 &    1.063   (  0.120) &    0.548   (  0.062) &    11.58   &    6.320   (  0.049)\\
NGC\,7077    & & \cothree\  & 1     & 18.1 &  6.8 &  2.8 &  12.1 &   -1 &   16 &    0.259   (  0.057) &    0.208   (  0.046) &    9.12    &    5.864   (  0.096)\\
NGC\,7077    & & \cothree\  & 2     & 18.1 &  6.8 &  2.8 &   4.2 &   37 &   50 &    0.277   (  0.092) &    0.223   (  0.074) &    9.78    &    5.894   (  0.143)\\
NGC\,7077    & & \cothree\  & Total & 18.1 &  6.8 &  2.8 &  12.1 &    0 &   38 &    0.536   (  0.085) &    0.431   (  0.068) &    18.89   &    6.180   (  0.069)\\
\\ 
UM\,448      & 5564 & \coone\    & 1     & 21.7 &  5.3 &  1.6 &   4.8 & -115 &   46 &    0.106   (  0.034) &    0.104   (  0.033) &    0.54    &    6.934   (  0.138)\\
UM\,448      & & \coone\    & 2     & 21.7 &  5.3 &  1.6 &   6.1 &  -59 &   61 &    0.398   (  0.140) &    0.392   (  0.137) &    2.04    &    7.509   (  0.152)\\
UM\,448      & & \coone\    & 3     & 21.7 &  5.3 &  1.6 &  10.9 &    0 &   46 &    0.547   (  0.125) &    0.539   (  0.123) &    2.80    &    7.647   (  0.099)\\
UM\,448      & & \coone\    & Total & 21.7 &  5.3 &  1.6 &  10.9 &    0 &  115 &    1.051   (  0.190) &    1.035   (  0.187) &    5.38    &    7.931   (  0.079)\\
UM\,448      & & \cotwo\    & 1     & 10.9 &  5.3 &  4.0 &   3.8 & -145 &   71 &    0.212   (  0.093) &    0.105   (  0.046) &    2.16    &    6.932   (  0.190)\\
UM\,448      & & \cotwo\    & 2     & 10.9 &  5.3 &  4.0 &   3.7 &  -95 &   52 &    0.407   (  0.118) &    0.202   (  0.058) &    4.15    &    7.216   (  0.126)\\
UM\,448      & & \cotwo\    & 3     & 10.9 &  5.3 &  4.0 &  12.3 &  -15 &   71 &    1.871   (  0.099) &    0.930   (  0.049) &    19.10   &    7.879   (  0.023)\\
UM\,448      & & \cotwo\    & Total & 10.9 &  5.3 &  4.0 &  12.3 &    0 &  130 &    2.490   (  0.179) &    1.237   (  0.089) &    25.41   &    8.003   (  0.031)\\
UM\,448      & & \cotwo\    & 1     & 27.6 & 10.4 &  2.1 &   8.9 &  -47 &   22 &    0.153   (  0.081) &    0.208   (  0.111) &    3.98    &    7.198   (  0.232)\\
UM\,448      & & \cotwo\    & 2     & 27.6 & 10.4 &  2.1 &   6.7 &  -45 &  118 &    0.614   (  0.212) &    0.839   (  0.290) &    16.03   &    7.803   (  0.150)\\
UM\,448      & & \cotwo\    & 3     & 27.6 & 10.4 &  2.1 &  12.1 &   34 &   82 &    0.776   (  0.191) &    1.059   (  0.261) &    20.25   &    7.904   (  0.107)\\
UM\,448      & & \cotwo\    & Total & 27.6 & 10.4 &  2.1 &  12.1 &    0 &   81 &    1.543   (  0.297) &    2.106   (  0.406) &    40.27   &    8.202   (  0.084)\\
UM\,448      & & \cothree\  & 1     & 18.4 & 13.8 &  1.6 &   3.9 &  -70 &   22 &    0.111   (  0.042) &    0.090   (  0.034) &    3.82    &    6.827   (  0.165)\\
UM\,448      & & \cothree\  & 2     & 18.4 & 13.8 &  1.6 &   4.2 &  -27 &   23 &    0.126   (  0.069) &    0.101   (  0.056) &    4.32    &    6.881   (  0.238)\\
UM\,448      & & \cothree\  & 3     & 18.4 & 13.8 &  1.6 &   6.7 &   33 &   72 &    0.637   (  0.049) &    0.513   (  0.039) &    21.88   &    7.585   (  0.033)\\
UM\,448      & & \cothree\  & Total & 18.4 & 13.8 &  1.6 &   6.7 &    0 &  103 &    0.875   (  0.094) &    0.705   (  0.076) &    30.02   &    7.723   (  0.047)\\
\\ 
UM\,462      & 1057 & \coone\    & Total & 21.4 & 15.7 &  2.5 &   6.7 &   -8 &  123 &    0.872   (  0.157) &    0.872   (  0.157) &    4.43    &    6.613   (  0.078)\\
UM\,462      & & \cotwo\    & Total      & 10.7 & 15.7 &  4.8 & $-$ & $-$ & $-$ & $<$0.87    & $-$ & $-$ & $<$5.72   \\
UM\,462      & & \cothree\  & Total      & 18.1 & 13.6 &  1.6 & $-$ & $-$ & $-$ & $<$0.87    & $-$ & $-$ & $<$5.78   \\
\\
\hline
\end{tabular}
} 
}
\vspace{0.5\baselineskip}
\begin{description}
\item
[$^{\mathrm{a}}$] All temperature units in this table are main-beam, \tmb.
Values in parentheses are the $1\sigma$ uncertainties.
The total line flux is the sum of the individual components, and the
total uncertainty is calculated by adding in quadrature
the individual uncertainties.
%The total line width is the quadrature sum of the individual widths. 
Systemic velocities are heliocentric, taken from NED, and used 
as the (optical convention) reference in Fig. \ref{fig:lines}.
\item
[$^{\mathrm{b}}$] Offsets of the component's central velocity relative to the systemic one.
\item
[$^{\mathrm{c}}$] These have been corrected for beam dilution
to a common beam size of 22\arcsec\ as described in Sect. \ref{sec:beam}
and Appendix \ref{sec:appendix_exponential}.
\end{description}
\end{table*}
\end{center}
% -----------------------------------------------------------------

% --- Table 3: NGC1140
%
\begin{center}
\begin{table*}
      \caption[]{Emission line parameters for NGC\,1140$^{\rm a}$} 
\label{tab:n1140}
\resizebox{\linewidth}{!}{
%\addtolength{\tabcolsep}{7pt}
\addtolength{\tabcolsep}{3pt}
{\small
%\tiny
\begin{tabular}{lrcrrrrrrrrrr}
\hline
\multicolumn{1}{c}{Name} &
\multicolumn{1}{c}{Transition} &
\multicolumn{1}{c}{Component} &
\multicolumn{1}{c}{Beam} &
\multicolumn{1}{c}{Channel} &
\multicolumn{1}{c}{T$_{\rm rms}$} &
\multicolumn{1}{c}{T$_{\rm peak}$} &
\multicolumn{1}{c}{Offset$^{\rm b}$} &
\multicolumn{1}{c}{FWHM} &
\multicolumn{1}{c}{$I_{CO}$} &
\multicolumn{1}{c}{$I_{CO}^{\rm cor}$ $^{\mathrm c}$} &
\multicolumn{1}{c}{$S_{CO}^{\rm cor}$ $^{\mathrm c}$} &
\multicolumn{1}{c}{Log($L^\prime_{\rm cor}$) $^{\mathrm c}$} \\
&&& \multicolumn{1}{c}{size} & 
\multicolumn{1}{c}{width} &
\multicolumn{1}{c}{(mK)} &
\multicolumn{1}{c}{(mK)} &
\multicolumn{1}{c}{(\kms)} &
\multicolumn{1}{c}{(\kms)} &
\multicolumn{1}{c}{(\kkms)} &
\multicolumn{1}{c}{(\kkms)} &
\multicolumn{1}{c}{(Jy\,\kms)} &
\multicolumn{1}{c}{(\kkmspc)} \\
& && 
\multicolumn{1}{c}{(\arcsec)}
& \multicolumn{1}{c}{(\kms)} & & & & & & \\
\hline
\\
NGC\,1140    & \coone\    & 1     & 21.4 &  5.2 &  1.6 &   9.5 &  -21 &   20 &    0.205   (  0.016) &    0.200   (  0.015) &    1.07    &    6.004   (  0.033)\\
NGC\,1140    & \coone\    & 2     & 21.4 &  5.2 &  1.6 &   7.8 &    0 &   24 &    0.201   (  0.016) &    0.196   (  0.015) &    1.05    &    5.996   (  0.034)\\
NGC\,1140    & \coone\    & 3     & 21.4 &  5.2 &  1.6 &   4.5 &   35 &   25 &    0.123   (  0.016) &    0.120   (  0.015) &    0.64    &    5.782   (  0.055)\\
NGC\,1140    & \coone\    & 4     & 21.4 &  5.2 &  1.6 &   5.4 &   64 &    8 &    0.049   (  0.016) &    0.048   (  0.015) &    0.26    &    5.386   (  0.138)\\
NGC\,1140    & \coone\    & Total & 21.4 &  5.2 &  1.6 &   9.5 &    0 &   85 &    0.578   (  0.031) &    0.564   (  0.031) &    3.02    &    6.455   (  0.024)\\
\\
NGC\,1140    & \cotwo\    & 1     & 10.7 &  5.2 &  5.0 &   6.0 &  -71 &   20 &    0.212   (  0.047) &    0.126   (  0.028) &    2.65    &    5.797   (  0.096)\\
NGC\,1140    & \cotwo\    & 2     & 10.7 &  5.2 &  5.0 &  17.0 &  -17 &   37 &    1.124   (  0.047) &    0.668   (  0.028) &    14.10   &    6.522   (  0.018)\\
NGC\,1140    & \cotwo\    & 3     & 10.7 &  5.2 &  5.0 &   8.5 &   10 &   25 &    0.389   (  0.047) &    0.231   (  0.028) &    4.88    &    6.061   (  0.052)\\
NGC\,1140    & \cotwo\    & 4     & 10.7 &  5.2 &  5.0 &   6.4 &   38 &   23 &    0.260   (  0.047) &    0.155   (  0.028) &    3.26    &    5.886   (  0.078)\\
NGC\,1140    & \cotwo\    & Total & 10.7 &  5.2 &  5.0 &  17.0 &    0 &  109 &    1.985   (  0.093) &    1.180   (  0.056) &    24.89   &    6.769   (  0.020)\\
\\
NGC\,1140    & \cothree\  & 1     & 18.1 &  6.8 &  3.6 &  20.0 &  -30 &   27 &    0.686   (  0.182) &    0.577   (  0.153) &    25.24   &    6.423   (  0.115)\\
NGC\,1140    & \cothree\  & 2     & 18.1 &  6.8 &  3.6 &  11.0 &    0 &   22 &    0.308   (  0.171) &    0.259   (  0.144) &    11.33   &    6.075   (  0.241)\\
NGC\,1140    & \cothree\  & 3     & 18.1 &  6.8 &  3.6 &   9.9 &   47 &   13 &    0.162   (  0.044) &    0.136   (  0.037) &    5.96    &    5.796   (  0.117)\\
NGC\,1140    & \cothree\  & Total & 18.1 &  6.8 &  3.6 &  20.0 &    0 &   77 &    1.155   (  0.252) &    0.972   (  0.212) &    42.52   &    6.650   (  0.095)\\
\\
NGC\,1140    & \cofour\   & 1     & 13.6 & 10.2 &  7.5 &  18.2 &  -37 &   23 &    0.645   (  0.168) &    0.440   (  0.114) &    33.11   &    6.291   (  0.113)\\
NGC\,1140    & \cofour\   & 2     & 13.6 & 10.2 &  7.5 &   9.2 &   -1 &   14 &    0.209   (  0.132) &    0.142   (  0.090) &    10.71   &    5.801   (  0.274)\\
NGC\,1140    & \cofour\   & Total & 13.6 & 10.2 &  7.5 &  18.2 &    0 &   38 &    0.854   (  0.213) &    0.582   (  0.146) &    43.82   &    6.413   (  0.109)\\
\\
NGC\,1140    & \thirteencoone\ & Total & 22.4 & 10.9 &  0.6 &   2.1 &    2 &   23 &    0.052   (  0.016) &    0.053   (  0.016) &    0.26    &    5.423   (  0.131)\\
NGC\,1140    & \thirteencotwo\ & 1     & 11.2 &  5.5 &  0.9 &   1.3 &  -21 &   41 &    0.093   (  0.020) &    0.057   (  0.012) &    1.09    &    5.450   (  0.093)\\
NGC\,1140    & \thirteencotwo\ & 2     & 11.2 &  5.5 &  0.9 &   0.8 &   74 &   58 &    0.079   (  0.021) &    0.048   (  0.013) &    0.93    &    5.381   (  0.117)\\
NGC\,1140    & \thirteencotwo\ & Total & 11.2 &  5.5 &  0.9 &   1.3 &    0 &   95 &    0.172   (  0.029) &    0.105   (  0.018) &    2.02    &    5.718   (  0.073)\\
\\
NGC\,1140    & C18O(1$-$0) & Total      & 22.5 & 11.0 &  0.5 & $-$ & $-$ & $-$ & $<$0.10    & $-$ & $-$ & $<$5.40   \\
NGC\,1140    & HCN(1$-$0) & Total      & 21.8 & 12.8 &  0.5 & $-$ & $-$ & $-$ & $<$0.10    & $-$ & $-$ & $<$5.44   \\
NGC\,1140    & CN(1$-$0)  & Total      & 21.8 & 10.6 &  1.2 & $-$ & $-$ & $-$ & $<$0.10    & $-$ & $-$ & $<$5.75   \\
NGC\,1140    & CS(2$-$1)  & Total      & 25.2 & 12.3 &  0.6 & $-$ & $-$ & $-$ & $<$0.10    & $-$ & $-$ & $<$5.64   \\
\\
NGC\,1140    & \cione\    & Total & 12.7 &  9.6 &  5.7 &   6.8 &    5 &   51 &    0.563   (  0.188) &    0.369   (  0.123) &    31.63   &    6.215   (  0.145)\\
\\
\hline
\end{tabular}
}
}
\vspace{0.5\baselineskip}
\begin{description}
\item
[$^{\mathrm{a}}$] All temperature units in this table are main-beam, \tmb.
Values in parentheses are the $1\sigma$ uncertainties.
The total line flux is the sum of the individual components, and the
total uncertainty is calculated by adding in quadrature
the individual uncertainties. 
We assumed a (heliocentric) systemic velocity of 1501\,\kms\ (from NED) for NGC\,1140, and used this as the (optical convention) reference velocity
in Fig. \ref{fig:n1140}. 
\item
[$^{\mathrm{b}}$] Offsets of the component's central velocity relative to the systemic one.
\item
[$^{\mathrm{c}}$] These have been corrected for beam dilution
to a common beam size of 22\arcsec\ as described in Sect. \ref{sec:beam}
and Appendix \ref{sec:appendix_exponential}.
%\item
%[$^{\mathrm{d}}$] With aperture corrections for total galaxy flux {\bf as described in Paper\,I (a factor of 1.64 for NGC\,1140)}.
\end{description}
\end{table*}
\end{center}
% -----------------------------------------------------------------

% --- Table 4: ULs
%
\begin{center}
\begin{table*}
      \caption[]{3$\sigma$ upper limits of undetected emission lines for II\,Zw\,40 and \sbs$^{\rm a}$} 
\label{tab:UL}
%\resizebox{\linewidth}{!}{
%\addtolength{\tabcolsep}{7pt}
{\small
%\tiny
\begin{tabular}{lrcrrrr}
\hline
\multicolumn{1}{c}{Name} &
\multicolumn{1}{c}{Transition} &
\multicolumn{1}{c}{Component} &
\multicolumn{1}{c}{Beam} &
\multicolumn{1}{c}{Channel} &
\multicolumn{1}{c}{T$_{\rm rms}$} &
%\multicolumn{1}{c}{Offset} &
%\multicolumn{1}{c}{FWHM} &
%\multicolumn{1}{c}{$I_{CO}^{\rm b}$} \\
\multicolumn{1}{c}{$I_{CO}$} \\
&&& \multicolumn{1}{c}{size (\arcsec)} & 
\multicolumn{1}{c}{width (\kms)} &
\multicolumn{1}{c}{(mK)} &
%\multicolumn{1}{c}{(\kms)} &
%\multicolumn{1}{c}{(\kms)} &
\multicolumn{1}{c}{(\kkms)} \\
\hline
\\
%II\,Zw\,40       & \cotwo\    & Total & 27.1 & 10.2 &  0.0025 & $<$0.230   \\
%II\,Zw\,40       & \cothree\  & Total & 18.1 & 13.6 &  0.0019 & $<$0.230   \\
II\,Zw\,40   & \cotwo\    & Total & 27.1 & 10.2 &  2.5 & $<$0.23     \\ %& $<$5.75   \\
II\,Zw\,40   & \cothree\  & Total & 18.1 & 13.6 &  1.9 & $<$0.23     \\ %& $<$5.40   \\
\\
%\sbs\  & \cotwo\    & Total & 27.4 & 10.3 &  0.0022 & $<$0.200   \\
%\sbs\  & \cothree\  & Total & 18.3 & 13.7 &  0.0027 & $<$0.330   \\
SBS0335-052  & \cotwo\    & Total & 27.4 & 10.3 &  2.2 &  $<$0.20     \\ %& $<$7.03   \\
SBS0335-052  & \cothree\  & Total & 18.3 & 13.7 &  2.7 &  $<$0.33     \\ %& $<$6.87   \\
\\ 
\hline
\end{tabular}
%}
}
\vspace{0.5\baselineskip}
\begin{description}
\item
[$^{\mathrm{a}}$] All temperature units in this table are main-beam, \tmb.
%\item
%[$^{\mathrm{b}}$] Values in parentheses are the uncertainties.
%The total line flux is the sum of the individual components, and the
%total uncertainty is calculated by adding in quadrature
%the individual uncertainties. 
\end{description}
\end{table*}
\end{center}
% -----------------------------------------------------------------

\subsection{APEX \label{sec:apex}}

The \cothree\ observations of six of our targets were performed with the Swedish Heterodyne 
Facility Instrument (SHeFI)
at the 12-m Atacama Pathfinder Experiment telescope (APEX).
The observations were acquired over three observing periods from 2008 to 2010 as part of the
proposals 082.B-0795A, 085.F-9320A, and 087.F-9316A. 
\cotwo\ measurements were also obtained for some sources, %although detected in only
as reported in Tables \ref{tab:lines}, \ref{tab:linesagain}, and \ref{tab:UL}.
We used the ``standard'' receiver$+$backend configurations,
APEX-1 (HET230$+$FTTS1) at 230\,GHz 
and APEX-2 (HET345$+$FTTS1) at 345\,GHz, with an intrinsic resolution of 0.12\,MHz at both frequencies. 
The sources were small enough to be observed in one pointing with wobbler
switching (in ON/OFF mode); pointing was checked every 2 hrs, and calibrations
were performed every 10 minutes.
Pointing accuracy was typically $\la$2\arcsec.
System temperatures for the \cothree\ (\cotwo) observations ranged from 
$\sim$180\,K to 670\,K ($\sim$160\,K to 270\,K)
with precipitable water vapor from 0.2\,mm to 1.1\,mm (1.2\,mm to 3\,mm).

Only NGC\,1140 was observed in the \cofour\ and \cione\ transitions (Period 82),
following the same observing protocols as for the lower frequencies but with APEX-3 (HET460$+$FTTS1) at 460\,GHz
(intrinsic resolution 0.98\,MHz).
System temperatures for these higher frequency observations ranged
from $\sim$380\,K to 740\,K ($\sim$370\,K to 550\,K) for \cione\ (\cofour). 
Water vapor columns ranged from 0.2\,mm to 0.6\,mm for \ci, and
from 0.3\,mm to 0.4\,mm for \cofour.

\subsection{Data reduction \label{sec:reduction}}

We adopted the GILDAS/CLASS data reduction package 
(http://www.iram.fr/IRAMFR/GILDAS) 
to obtain averaged spectra for all transitions.
To remove the baseline from each spectrum, a polynomial % baseline 
%of order 3-4 %order of 0 or 1 
was fitted to the line-free regions of each scan (defined a priori) and 
subtracted; the scans were thereafter Hanning smoothed and averaged, and a constant baseline subtracted. 
We measured the peak
intensities, central velocities, full width half-maximum (FWHM) and
velocity-integrated fluxes of the detected lines by fitting Gaussian profiles
to the data. 
In most cases, there is significant velocity structure in the emission, so
we fit the line profiles to multiple Gaussians.
In these cases we use the sum of the integrated line intensities 
from the multiple Gaussian fits in the analysis. % (see Table \ref{tab:lines}).

Antenna temperatures (\ta) have been converted to main-beam brightness temperatures (\tmb)
by dividing the antenna temperatures by $\eta\,\equiv\,B_{\rm eff}/F_{\rm eff}$, 
where $B_{\rm eff}$ and $F_{\rm eff}$ are
the beam and forward hemisphere efficiencies, respectively 
(i.e., \tmb\,=\,\ta/$\eta$).  
To convert the measured \coone\ antenna temperatures,
[\tmb\ (K)] to fluxes [$S$ (Jy)], we used the standard conversion factors for both
telescopes.
The \coone, \cotwo, and \cothree\ line profiles for the galaxies in which CO was detected
are shown in Fig. \ref{fig:lines}, together with the
Gaussian fits obtained using the GILDAS/CLASS package as
described above and reported in Tables \ref{tab:lines}, \ref{tab:linesagain}, and \ref{tab:n1140}. 
The upper limits
for \iizw\ and \sbs\ are given in Table \ref{tab:UL}.
Limits for non-detections are 3$\sigma$, taking the width of three contiguous velocity channels and 
the rms noise in line-free regions of the spectrum. 

For the total flux in each transition,
we have included the different velocity components in each galaxy into a global, spectrally-integrated
sum.
While some of the spectra show slightly different profiles in the different transitions,
the signal-to-noise (S/N) is insufficient to allow an analysis of the separate components.
%Indeed, there is no compelling evidence on the basis of the low S/N ratios of the {\it individual velocity components},
%that the ISM in these galaxies is inhomogeneous. 
Although a few of the individual components may be too narrow to be significant
(because their width is smaller than three velocity channels, e.g., Mrk\,996, NGC\,3353, NGC\,1140, see Figs. \ref{fig:lines}, \ref{fig:n1140}), 
we have checked that the total intensities 
are robust to the details of the fitting
by repeating the fits with different numbers of components. 
Some of the measured values in the tables differ slightly from those given
in Paper\,I; this is because we have redone the data reduction for all transitions
in order to render the results as consistent as possible.
The main difference is the newer version of CLASS used here that seems to give integrated 
fluxes which can differ by a few percent (but within the error limits)
with respect to those in Paper\,I (performed with an older version of CLASS).

In Paper\,I, 
for the galaxies with the largest apparent sizes, 
to estimate total \coone\ flux beyond even the largest beam 
we applied aperture corrections.
%as reported in Tables \ref{tab:lines} and \ref{tab:n1140}.
%Although they are reported in the tables,
Here they are not adopted because we prefer to correct the observations to a common beam size,
rather than also extrapolating to an uncertain total flux. 

\subsection{Comparison with previous observations \label{sec:cocomparison}}

A detailed comparison of our \coone\ measured intensities relative to previous work is given
in Paper\,I.
Here we compare only the higher-$J$ CO transitions with published literature values.
\citet{sage92} observed \iizw, NGC\,3353 (Haro\,3), NGC\,7077, and UM\,462 in \cotwo,
and securely detected only the first three.
Like the \coone\ measurements, their results from the IRAM 30m for \cotwo\ in NGC\,3353
are in good agreement with the line parameters deduced from our central pointing.
For NGC\,7077, our \cotwo\ flux of 1.06\,\kkms\ is $\ga$50\% larger than the value of 0.68\,\kkms\
reported by \citet{sage92}; given the similar uncertainties, it is difficult to understand
the discrepancy, especially since our \coone\ flux is smaller than theirs by $\sim$30\%.
Our \cotwo\ non-detection for UM\,462 is consistent with the non-detection by \citet{sage92}, while our
upper limit is smaller.
Instead, our non-detection with APEX of \cotwo\ in \iizw\ is inconsistent with \citet{sage92};
at the flux level found by them, we should have detected this galaxy, as our 3$\sigma$ upper limit is
several times smaller.

NGC\,1140 and \iizw\ were also observed previously in \cotwo\ 
with the IRAM 30m by \citet{albrecht04}. 
Although the 2--1/1--0 line ratios for NGC\,1140 are similar, they find 
$\sim$40-60\% larger intensities in both lines. 
The line intensity they report for the IRAM 30-m spectrum of \iizw\ (UGCA\,116) is larger than our APEX upper limit,
although less than half of the 30m detection reported by \citet{sage92}.
The peak \cothree\ temperatures observed with ALMA in the main knots in \iizw\ by
\citet{kepley16} exceed those of our upper limits; but beam dilution
% xxx see Karl's comment on v2, April, 2017, no reply
is probably playing an important role since the ALMA beam is $\sim$0\farcs5
compared to the APEX beam of $\sim$18\arcsec.

One of our targets, NGC\,7077, was observed by \citet{mao10}.
The ratio of \cothree\ to \coone\ we find for this galaxy of 1.1 is in excellent
agreement with their study.

In conclusion, there is no clear systematic difference between our results and previous work.
Nevertheless, the formal uncertainties given in Tables \ref{tab:lines}--\ref{tab:UL}
should probably be considered as lower limits to the true uncertainties.

\section{Correction to a common beam size\label{sec:beam}}

The main difficulty in the interpretation of line ratios from single-dish
observations is the correction for different beam sizes.
The physical models %(e.g., the RADEX models 
we discuss in Sect. \ref{sec:models} rely
on brightness temperatures \tb\ within a common beam size, while we observe %beam-averaged 
main-beam brightness temperatures, \tmb, within different beams.
To correct for this,
we have pursued two independent approaches:
\begin{enumerate}[(1)]
\item
used maps of the cool dust emission ($\sim$160\,\micron) to scale the
different transitions to a common beam size of 22\arcsec,
with the underlying assumption that the molecular emission is distributed like the cool dust.
The PACS 160 \micron\ maps are particularly well suited to such an endeavor because
their intrinsic FWHM is $\sim$11\arcsec, similar to the smallest beam size
in our observations (\cotwo, 10\farcs7).
\item
assumed that the molecular emission follows a Gaussian distribution with a FWHM size to be
determined by letting it vary in the fit to the line intensities.
This can be done in a robust way only for NGC\,1140 where we have a sufficient number
of line intensities to constrain the fit. 
\end{enumerate}

Observations at high spatial resolution ($\la 2$\,pc) of metal-poor galaxies
in the Local Group suggest that the CO clouds are more concentrated than the dust emission
\citep[e.g.,][]{rubio15,schruba17}; if this were true also on the spatial scales probed
by our observations ($\ga$500 times larger), then our assumption that the gas follows the
dust would be incorrect.
Thus, analyzing beam correction factors under the assumption of a Gaussian distribution 
enables an independent assessment of the validity of the inferred beam corrections.

Appendix \ref{sec:appendix_exponential} describes the formulation of
the correction to a common beam size assuming that the molecular emission follows the dust.
In brief, we have established that the distribution of the cool dust emission in our targets
is well described by an exponential distribution
by extracting azimuthally-averaged radial brightness profiles; they are all well fit by
an exponential. 
Moreover, for our targets, 160\,\micron\ emission is longward of the spectral energy distribution
peak, thus tracing as well as possible the cool dust component.
We have checked that free-free emission is unimportant at 160\,\micron\ by inferring its amplitude
from the observed SFR (Paper\,I); for all galaxies for which we have dust images, the expected free-free component
is $\la$ 0.1\% of the observed emission.

We performed aperture photometry to obtain growth curves, in order to approximate
the cumulative flux distribution we expect for the molecular gas in the different beams.
Finally, we have followed an analytical approach to fit the growth curves, in order
to derive the corrections to the expected flux in different apertures.
Table \ref{tab:corrections} in Appendix \ref{sec:appendix_exponential} gives the multiplicative
correction factors, and the corrected velocity-integrated \tmb\ values that we will
use throughout the paper unless otherwise noted;
corrected intensities are also given in the last three columns in Tables \ref{tab:lines},
\ref{tab:linesagain}, and \ref{tab:n1140}.
While high-$J$ CO emission may be more compact than the gas traced by the lower-$J$ lines,
the different radial distributions are expected to be a second-order effect;
our proposed correction provides the bulk of the ``first-order" correction.

Appendix \ref{sec:appendix_gaussian} discusses the form of the growth curves for a Gaussian source distribution.
The main difficulty with applying the Gaussian assumption is the unknown source size;
while the source extent can be measured from the analysis of the dust emission,
there is no similar way to assess the size of a Gaussian distribution.
Thus, when we estimate the physical conditions in the molecular gas by fitting the line
intensities (for NGC\,1140, the only source for which this is possible),
we use the fitting procedure itself to define the source size. 
The expectation
is that because of the similarity of some of the beams in our observations
(e.g., \cotwo\ with IRAM, \cofour\ with APEX), the fit will be sensitive to the beam
correction.
This turns out to be the case as will be described in Sect. \ref{sec:radexn1140}.

Mrk\,996 shows a markedly broader \cotwo\ profile in the larger (27\arcsec) APEX beam,
relative to the IRAM one (11\arcsec).
This could be an illustration of the spatial extension of the \cotwo\ emitting gas,
because of the broader velocity range in the bigger beam;
on the other hand, the APEX spectrum is rather noisy, so its reliability is uncertain.
Nevertheless,
the optical diameter of Mrk\,996 is 36\arcsec\ so it is conceivable that there is more CO emission
in the larger APEX beam.
%Interestingly, the observed \cotwo/\coone\ ratio (in \tmb\ units) using the APEX observation
%is quite high, $\sim 2.8$.
UM\,448 was also observed in \cotwo\ with both the 30m and APEX, but in this galaxy
the shape and breadth of the two profiles are more similar,
possibly because the galaxy is smaller, 24\arcsec\ in diameter.

\section{Empirical line ratios\label{sec:ratios}}
\label{sec:empirical}

Before describing our models of the molecular gas, here we discuss three commonly used diagnostic
ratios, \cotwo/\coone, \cothree/\coone, and \cione/\coone.
These ratios are calculated using velocity-integrated main-beam temperature,
$\int T_{\rm mb}\ d\varv$ (see Tables \ref{tab:lines}, \ref{tab:linesagain}, and \ref{tab:n1140}).

\begin{figure}[ht!]
\vspace{\baselineskip}
\hbox{
\centerline{
\includegraphics[angle=0,width=0.98\linewidth]{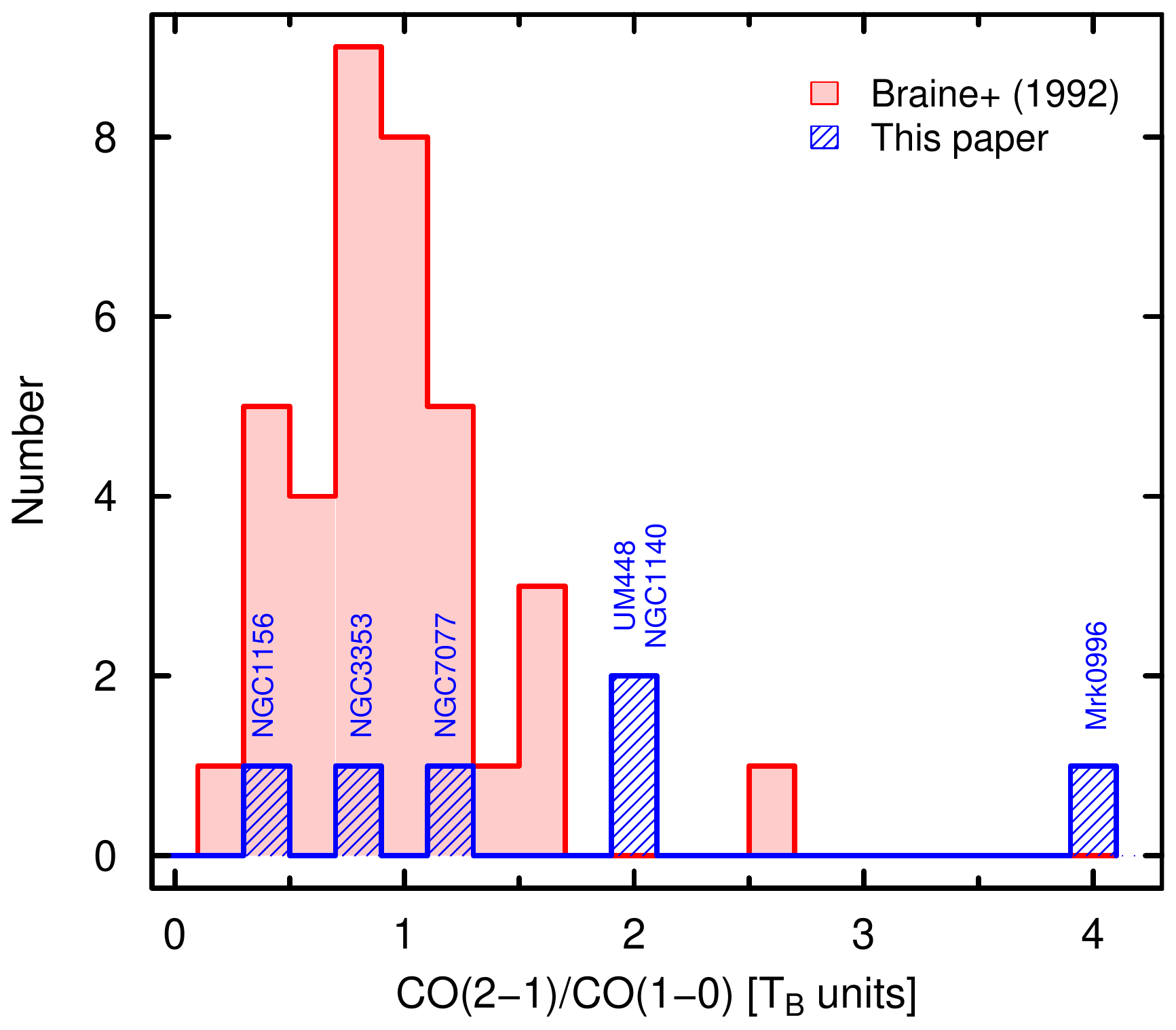}
}
}
\caption{Histogram of observed \cotwo/\coone\ line ratios for our targets corrected
for beam dilution as described in Sect. \ref{sec:beam}. 
Because of the larger beam, for Mrk\,996 and UM\,448, we show the APEX \cotwo\ measurements rather than the ones from IRAM. 
Also plotted are the line ratios corrected for beam dilution of a large sample of spiral galaxies observed by
\citet{braine92}. 
%Our measurements have been corrected for beam diltuassuming the distribution of 160\,\micron\ dust emission also applies to the molecular gas
%as described in Sect. \ref{sec:beam}.
}
\label{fig:lineratios21}
\end{figure}

\begin{figure}[ht!]
\vspace{\baselineskip}
\hbox{
\centerline{
\includegraphics[angle=0,width=0.98\linewidth]{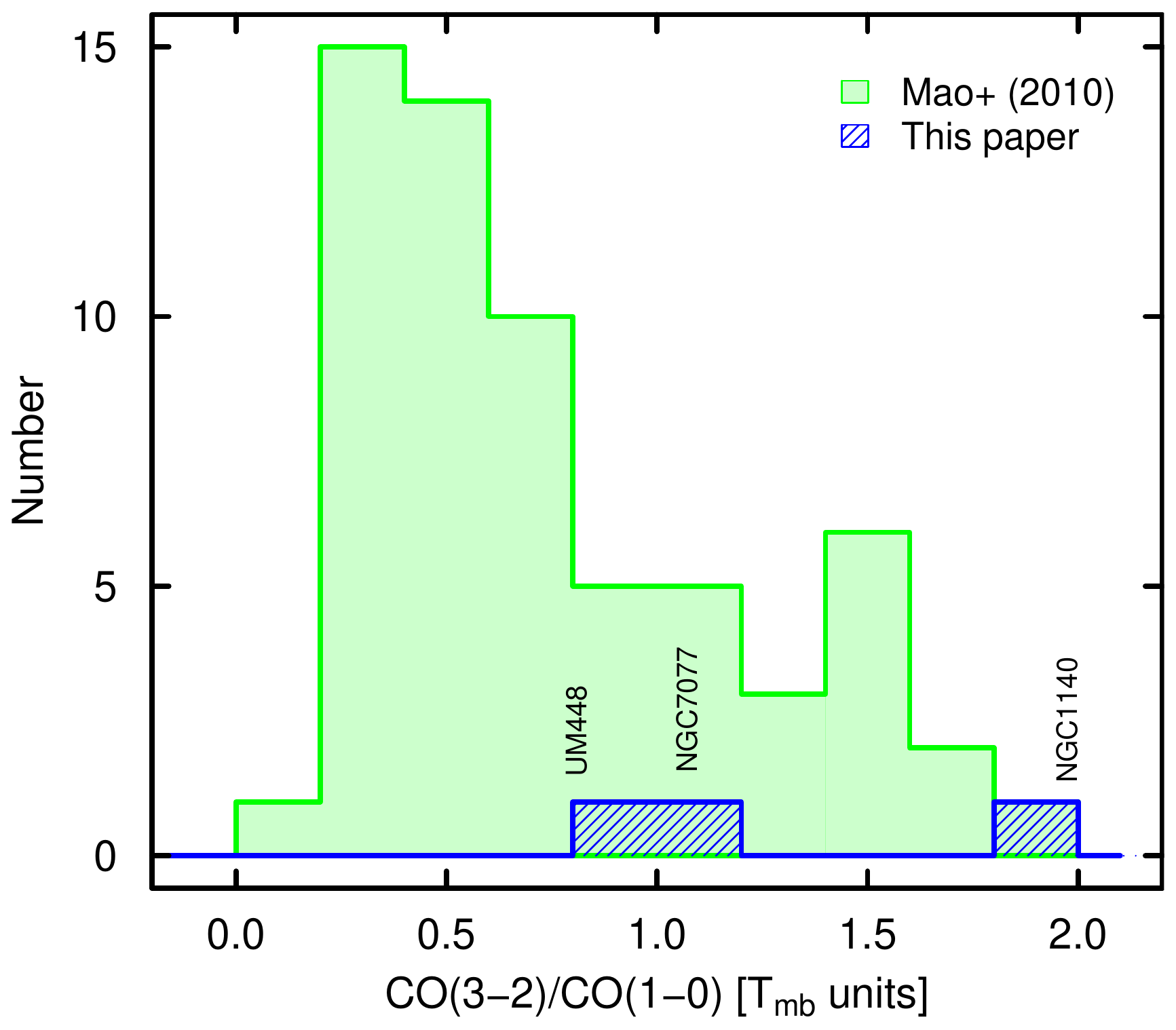}
}
}
\caption{Histogram of observed \cothree/\coone\ line ratios for our targets.
Also shown are the line ratios for a large sample of galaxies observed by
\citet{mao10}. Neither set of observations has been corrected for beam dilution;
see Sect. \ref{sec:co32} for more discussion.}
\label{fig:lineratios32}
\end{figure}

\subsection{CO(2--1)/CO(1--0)}
\label{sec:co21}

%It is well known that in optically thin gas {\bf see Paola}, the ratio of CO(2--1) to CO(1--0) will be greater than unity
%because of the larger optical depth of the 2--1 transition.
The \cotwo\ transition traces slightly warmer
(upper level equivalent temperature $\sim$17\,K) and denser 
(critical density $\sim 10^{4}$\,\cmthree) gas than the $1-0$ transition.
In most galaxies, CO emitting gas is optically thick and the observed CO(2--1)/CO(1--0) ratios
are $\la$ 1 \citep[e.g.,][]{braine92}.
Figure \ref{fig:lineratios21} shows the distribution of the \cotwo/\coone\ line temperature ratios (\rtwo)
of our low-metallicity targets together with \rtwo\ 
ratios (corrected for beam dilution) of spiral galaxies 
from \citet{braine92}.
In Fig. \ref{fig:lineratios21}, our data are corrected to a common beam size using the exponential approach
described in Sect. \ref{sec:beam} and in Appendix \ref{sec:appendix_exponential}.
%Through CO(2--1) maps, \citet{braine92} were able to correct some of their data for the effect
%of the smaller CO(2--1) beam (37 of 51 sources); only these corrected data are shown 
%in Fig. \ref{fig:lineratios21}. 

Even after correction for beam dilution,
half of our targets have \rtwo\,$\ga 2$, higher than 
most of the normal spirals studied by \citet{braine92}.
\citet{braine92} found an extremely high ratio for another sub-Solar metallicity
starburst, NGC\,3310 with \rtwo\ $\sim 2.6$; 
however \citet{zhu09} were unable
to confirm this, finding a somewhat lower ratio (\rtwo\ $\sim 1.5$), 
after convolving the observations to a common beam size through their maps.

Although it is tempting to claim that the high \rtwo\ observed in some of these dwarf galaxies
is due to optically thin gas, the \rtwo\ values are somewhat uncertain because of the beam correction.
%First, the ratios are highly dependent on the kind of correction applied.
%The corrections are sensitive both to the source distribution
%and to the inferred exponential scale length, i.e., the source
%size; assuming a Gaussian source distribution, for example, 
%rather than an exponential,
%would almost double the \rtwo$\sim$1 values for NGC\,3353 and NGC\,7077.
First, the ratios are dependent on the kind of correction applied.
The corrections are sensitive both to the source distribution (exponential versus Gaussian)
and to the source size (exponential scale length versus Gaussian FWHM);
since we have no way of accurately determining a Gaussian FWHM, we cannot assess
how much the \rtwo\ values would change under another assumption for source
distribution.
Second, if we use the IRAM \cotwo\ observations (with a smaller beam) for Mrk\,996 and UM\,448, 
we obtain beam-corrected \rtwo$\sim$0.6 for Mrk\,996, and \rtwo$\sim$1.2 for UM\,448.
Thus the high \rtwo\ values are not altogether robust.
The only galaxy in our %(small) 
sample for which we are fairly certain that the true \rtwo\
value is $\ga$2 is NGC\,1140.
The interpretation of \rtwo\ for this galaxy will be discussed in Sect. \ref{sec:radexn1140}
where we present physical models for the molecular gas.

\begin{figure}[ht!]
\vspace{\baselineskip}
\hbox{
\centerline{
%\includegraphics[angle=0,width=0.99\linewidth]{CICOvsOH-crop.pdf}
% 27/12/2016
% redone 1/3/2017 
\includegraphics[angle=0,width=0.99\linewidth]{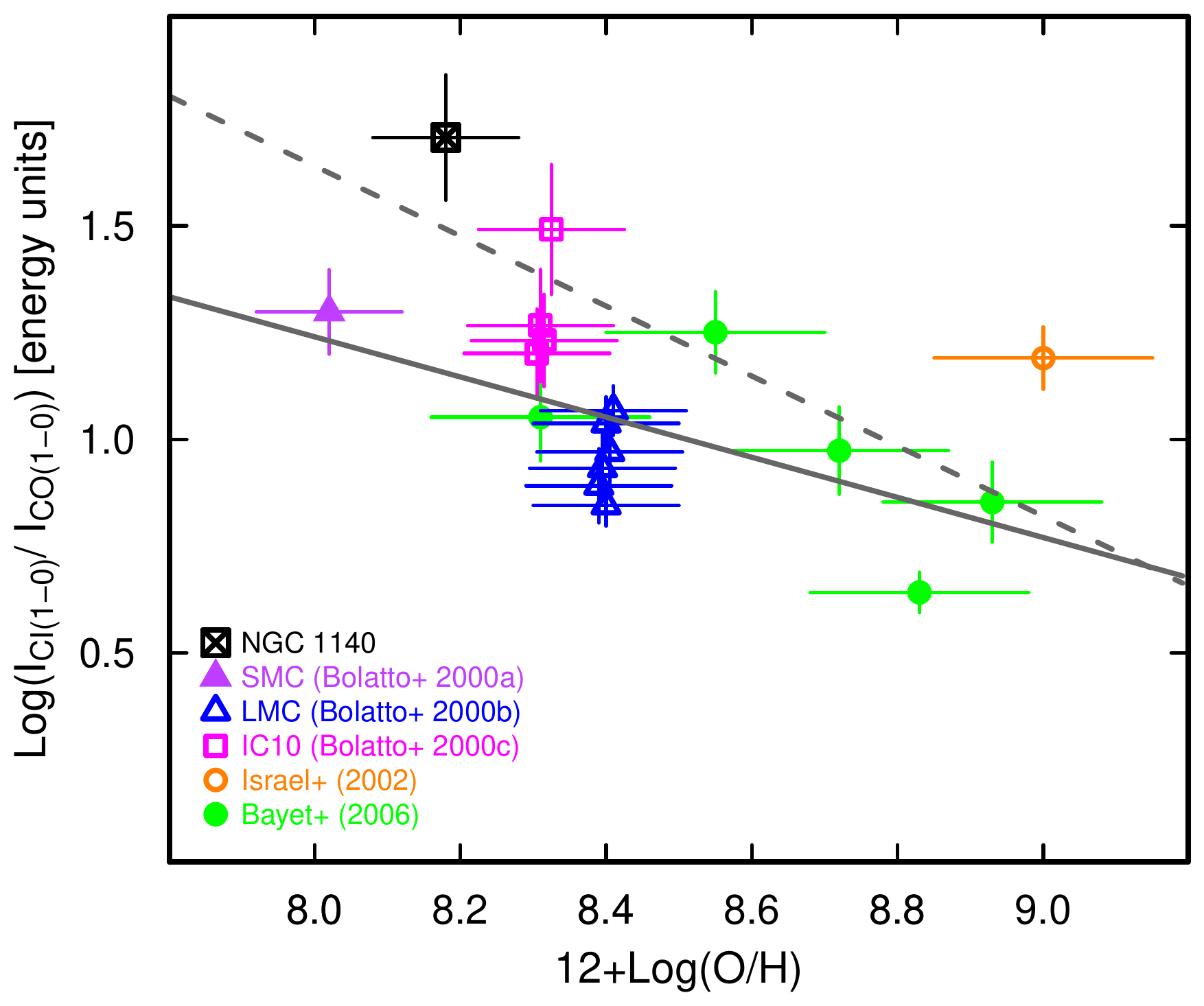}
}
}
\caption{Beam-corrected \ci/\coone\ line ratios for NGC\,1140 and
galaxies from the literature plotted against oxygen abundance \logoh.
The solid line corresponds to the data regression by \citet{bolatto00a}, and
the dashed one to the PDR model by \citet{bolatto99}.
Beam dilution has been corrected for as described in Sect. \ref{sec:beam}.
}
\label{fig:lineratiosci}
\end{figure}

\subsection{CO(3--2)/CO(1--0)}
\label{sec:co32}

% Axel asks to remove this sentence (says rubbish)
%With a higher {\bf upper level equivalent
%}
%temperature ($\sim 33$\,K) and a higher critical density
%($\sim 10^5$\,\cmthree), the \cothree\ transition traces warmer and denser gas than the lower-$J$ lines.
Figure \ref{fig:lineratios32} shows the distribution of the ratio of \cothree/\coone\ 
(\rthree\ in \tmb\ units) for the three
sources in our sample with \cothree\ detections, compared to the large sample of \citet{mao10}.
Unlike for the \cotwo/\coone\ ratio observed with the IRAM 30m, the APEX beam is $\sim$18\arcsec, 
similar to the IRAM 30-m beam of $\sim$22\arcsec\ for the 1--0 line; 
the \citet{mao10} observations were obtained with the Heinrich
Hertz Telescope with a \cothree\ beam of $\sim$22\arcsec.
Thus, we have not applied any correction for beam dilution;
the \rthree\ values shown in Fig. \ref{fig:lineratios32} should be relatively impervious
to beam corrections, but rather representative of the beam-averaged physical conditions in the molecular gas.

Of the three galaxies in our sample with \cothree\ detections (Mrk\,996 was observed but not detected, \rthree$<0.7$),
NGC\,1140 shows the largest ratio (\rthree\,$=\,2.00\,\pm\,0.45$; 
with the exponential beam correction \rthree\,$=\,1.72\,\pm\,0.39$), 
roughly equivalent to that found for NGC\,3310
(\rthree\,=\,1.9$\,\pm\,$0.52) by \citet{mao10}.
Such a ratio exceeds typical values in luminous infrared galaxies (LIRGs)
and those expected in the Orion ``hot spots''
\citep[][]{papa12}. 
\rthree\ has been historically taken as an indicator of \htwo\ density
\citep[e.g.,][]{mauersberger99}, and high \rthree\ is
possibly indicative of somewhat excited, optically thin, gas, 
although shocks and cool foreground absorbing layers are also
possible causes \citep[e.g.,][]{oka07}.
We shall explore this point further with physical radiative transfer
models in Sect. \ref{sec:radex12co}.
%although see \citet{mao10} for other conceivable explanations.

\subsection{CI(1--0)/CO(1--0)}
\label{sec:cico}

Together with CO and \cii\ (at 158\,\micron), the atomic carbon line \cione\ is an important coolant of the ISM
\citep[e.g.,][]{bolatto99,gerin00,bolatto00a,bolatto00b,bolatto00c,israel02,papa04,bayet06,israel09}.
Despite the difficulty of observation from the ground (the $^3{\rm P}_1 - ^3{\rm P}_0$ \ci\ transition lies at
609\,\micron), the potential of atomic carbon to trace molecular gas was recognized early on.
Photo-Dissociation Regions (PDRs) are thought to be larger at low metallicity,
because of the decreased attenuation of the UV field through the lower dust content
\citep[e.g.,][]{lequeux94,bolatto99}.
The brighter UV field in low \av\ environments tends to change
the structure of PDRs, enlarging the region where atomic and ionized carbon dominate,
and diminishing the region where \htwo\ is associated with CO emission
\citep[e.g.,][]{bolatto13}.
The spatial distributions of \ci\ and CO gas are similar in 
regions within the Milky Way \citep[e.g.,][]{kramer04,zhang07}, 
the Small Magellanic Cloud \citep[SMC,][]{requena16}, and
other galaxies \citep[e.g.,][]{israelbaas03,krips16}, so 
%\ci, like \cii, may be an effective tracer of CO-dark \htwo\ gas.
\ci\ seems to be tracing molecular gas. 
This is especially important in low-metallicity environments
or cosmic-ray dominated regions where CO emission is suppressed 
\citep[e.g.,][]{hopkins13,bisbas15,bisbas17}, 
or in high-redshift galaxies where \ci\ observations are facilitated \citep[e.g.,][]{walter11}. 

We have obtained a \cione\ measurement of one of our targets, NGC\,1140, as shown in
Fig. \ref{fig:lineratiosci} where we plot the \ci/\coone\ ratio (in energy units) together
with similar measurements from the literature
\citep{bolatto00a,bolatto00b,bolatto00c,israel02,bayet06}.
We have applied the correction for beam dilution as described in Sect. \ref{sec:beam} and Appendix \ref{sec:appendix_exponential}.
The regression lines are taken from the PDR model by \citet[][dashed line]{bolatto99}, and the best
fit to their data \citep{bolatto00a}.
Despite the paucity of data, the trend of enhanced \cione/\coone\ with decreasing O/H is
evident.

The expected metallicity-dependent behavior of \cione\ in PDR environments has been modeled
by various groups \citep[e.g.,][]{papa04} who confirm that the abundance of \ci\ dominates 
that of CO for large regions within a low-metallicity cloud (\av$\ga$2). 
\citet{glover16} present more recent simulations of \cione\ emission relative to \coone\ 
as a function of metallicity.
Our (beam-corrected) \cione/\coone\ velocity-integrated temperature ratio for NGC\,1140 of 0.65 (see Table \ref{tab:n1140})
%is roughly {\bf 40\% higher than their}
is more than a factor of two higher than their
predictions for the highest PDR illumination they model 
\citep[$G_0$\,=\,100 in units of the interstellar radiation radiation field by][]{draine78},
and the disagreement increases for less intense fields.
%Nevertheless, even a 50\% discrepancy shows reasonable agreement with the models.

The \ci/\coone\ ratio tends to decrease with increasing PDR density; 
the observed ratio for NGC\,1140 is higher than the predictions of 
(Solar metallicity) PDR models with volume number densities $n_0 \sim 10^4$\,\cmthree\ 
and $G_0 \sim 10^4$ \citep[][]{hollenbach91}. 
The observed \cione/\thirteencotwo\ for NGC\,1140 is a factor of $4-5$ higher than
the PDR predictions (for Solar, and twice Solar metallicities) by \citet{meijerink07}, 
for volume number densities $n_0 \ga 10^4$\,\cmthree\ 
and $G_0 \sim 10^5$.
However, they assume [\twelveco/\thirteenco]\,=\,40;
a lower isotopic ratio (see Sect. \ref{sec:radexn1140}) or a lower metallicity would reduce the disagreement. 
In Sect. \ref{sec:radexn1140}, we explore physical models of the gas in NGC\,1140 and derive
relative abundances of atomic carbon and CO.

\begin{figure*}[ht!]
\vspace{\baselineskip}
\hbox{
\centerline{
\includegraphics[angle=0,width=0.95\linewidth]{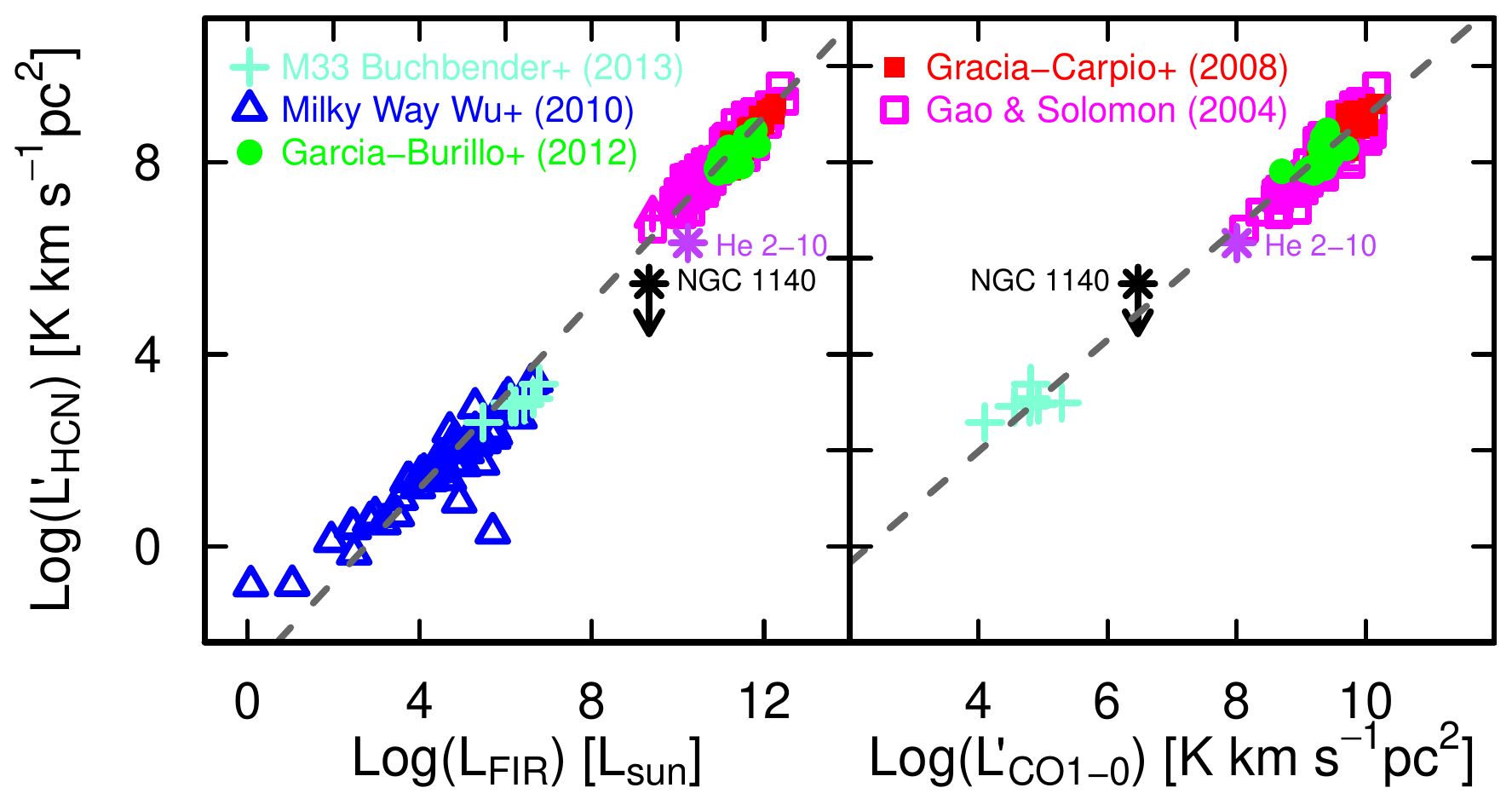}
}
}
\vspace{-\baselineskip}
\caption{HCN luminosity plotted against FIR luminosity, \lfir\ (left panel)
and CO(1--0) luminosity, \lcoone\ (right) for NGC\,1140.
and other galaxies taken from the literature (see legend
and text for more discussion).
The dashed lines show a robust unweighted regression for \lhcn\ against the two variables
plotted along the x-axis, excluding He\,2$-$10 and NGC\,1140.
}
\label{fig:hcn}
%\end{figure*}
%
%\begin{figure*}[!h]
\setcounter{figure}{6}
\vspace{\baselineskip}
\hbox{
\centerline{
\includegraphics[angle=0,width=0.95\linewidth]{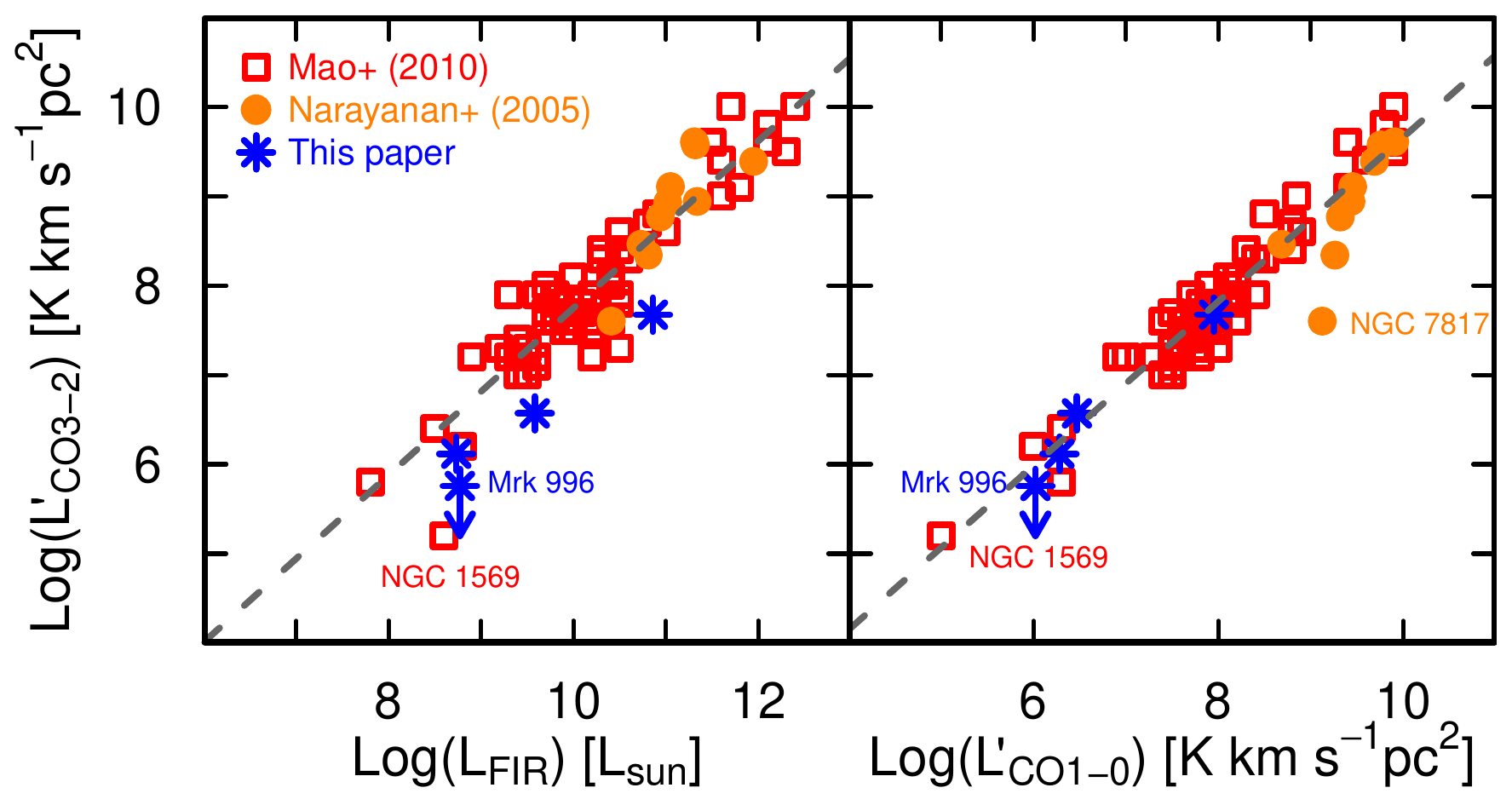}
}
}
\vspace{-\baselineskip}
\caption{CO(3-2) luminosity, \lcothree, plotted against FIR luminosity, \lfir\ (left panel)
and CO(1--0) luminosity, \lcoone\ (right) for our targets.
Also shown are values taken from \citet{mao10} and \citet{narayanan05},
see text for more discussion.
The dashed lines show a robust unweighted regression for \lcothree\ against the two variables
plotted along the x-axis, not including our data.
Some galaxies with extreme values are labeled.
The left panel suggests that CO emission traced by the \cothree\ transition is deficient
relative to \lfir\ \citep[like \coone, see][]{hunt15}, but that \cothree\ relative to \coone\
is roughly consistent, implying that the excitation conditions in these metal-poor galaxies are similar
to those in the metal-rich galaxy samples taken from the literature.
}
\label{fig:co32}
\end{figure*}

\section{Dense-gas tracers}
\label{sec:dense}

The relatively high critical densities of \cothree\ and \hcn\ (\ncrit$\sim 10^4 - 10^5$\,\cmthree)
make these transitions good tracers of dense gas, which are possibly better
correlated with SFR than the lowest-$J$ CO lines
\citep{gao04,narayanan05,narayanan08,garciaburillo12,lada12,hopkins13,greve14}.
Given the relatively high sSFRs of our targets (sSFR\,$\ga 10^{-10}$\,yr$^{-1}$, see
Paper\,I), here we examine luminosity correlations of \cothree\ and \hcn\ for these metal-poor
%dwarf 
galaxies and compare them with more massive metal-rich ones.

% to avoid ! pdfTeX error (ext4): \pdfendlink ended up in different nesting level than \pd fstartlink. 
%\clearpage

Figure \ref{fig:hcn} shows HCN luminosity plotted against FIR luminosity, \lfir\ (left panel)
and CO(1--0) luminosity, \lcoone\ (right).
The only galaxy for which we have \hcn\ observations is NGC\,1140\footnote{We do not apply beam
corrections to the upper limit, because of the similarity of the \hcn\ and \coone\ beam sizes.};
also shown in Fig. \ref{fig:hcn} are luminous (and ultraluminous) infrared galaxies from 
\citet{gao04,graciacarpio08,garciaburillo12},
together with regions within M\,33 \citep{buchbender13} and the Milky Way \citep[MW,][]{wu10};
a luminous irregular galaxy, He\,2--10, is also plotted \citep{santangelo09}.
The dashed lines show a robust regression\footnote{Here and throughout the paper we
use the ``robust'' regression algorithm, effective for minimizing the effects of outliers,
as implemented in R,
a free software environment for statistical computing and graphics,
{\it https://www.r-project.org/}.} performed on all galaxies excepting He\,2--10 and NGC\,1140; 
the slope of the \lhcn-\lfir\ correlation is $0.97\,\pm\,0.01$ and for \lhcn-\lcoone\ 
$1.16\,\pm\,0.03$ (this last without the MW points, for which there were no CO(1--0) data).
The slopes are consistent with previous work \citep[e.g.,][]{gao04,garciaburillo12},
and suggest that HCN in irregular galaxies such as He\,2--10 and NGC\,1140 is somewhat
deficient compared to \lfir\ but within normal limits compared to \lcoone.

Both He\,2--10 and NGC\,1140 contain extremely massive, compact star clusters
\citep[e.g.,][]{johnson00,degrijs04,moll07}
(Super Star Clusters, SSCs), but He\,2--10 is approximately Solar metallicity \citep{kobulnicky99}
while NGC\,1140 is $\sim$0.3\,\zsun.
Feedback from these SSCs would be expected to substantially reduce the fraction
of dense gas in these galaxies as traced by \hcn\ \citep{hopkins13}.
Indeed, with \lcoone$\sim 10^8$\,\kkmspc, the \lhcn/\lcoone\ ratio for He\,2--10 is $\sim$0.02,
and $\la$0.10 for NGC\,1140 (with \lcoone$\sim 10^6$\,\kkmspc), roughly consistent
with the model predictions by \citet{hopkins13}.
More observations of HCN in such extreme galaxies are needed to further explore
the survival of dense gas in these galaxies, and the ability of HCN to trace it 
(see also Sect. \ref{sec:radexn1140}).

Figure \ref{fig:co32} plots \lcothree\ vs. \lfir\ and \lcoone\ as in Fig. \ref{fig:hcn} for \hcn,
but here for the four galaxies in our sample which were observed in \cothree\
(Mrk\,996, NGC\,1140, NGC\,7077 UM\,448).
Like \hcn, \cothree\ is deficient relative to \lfir. % but completely consistent relative to \lcoone.
The slopes of $0.93\,\pm\,0.05$ (\lcothree\ vs. \lfir) and $0.96\,\pm\,0.04$ (\lcothree\ vs. \lcoone) 
found by fitting the combination of the \citet{mao10} and \citet{narayanan05} samples
are consistent 
with those found by \citet{mao10} and \citet{greve14}, despite different fitting techniques for
the latter;
however the \lcothree\ vs. \lcoone\ slope is significantly flatter than the one
of $\sim$1.5 found by \citet{narayanan05}
and used to calibrate the \citet{hopkins13} simulation results.
The deficit of \lcothree\ relative to \lfir\ is almost certainly due to metallicity as shown in Paper\,I;
for metal-poor galaxies, we found an offset between the CO emission relative to the SFR
in the sense that there is less CO for a given SFR value. 
This deficiency results from the metallicity dependence of the CO conversion factor to \htwo\ mass.
Interestingly, the four galaxies for which we have CO(3--2) measurements %(or upper limits) 
are not deficient in \cothree\ relative to \coone\ in the other galaxies taken from
the literature; they show similar CO luminosity ratios 
despite their low metallicity ($\sim$0.25\,\zsun).
This could be implying that excitations in these galaxies are similar to those in more
metal-rich systems, a point which we explore more fully in the next section.

% to handle fatal pdf error apparently from the goldsmith99 citation in the paragraph
%"We then attempted to correct for optical depth effects..."
%\vspace{\baselineskip}

\section{Modeling physical conditions at low metallicity }
\label{sec:models}

We now explore via modeling the physical conditions in the molecular gas
in our metal-poor targets.
Detailed physical modeling is possible only for NGC\,1140, for which we have four \twelveco\ 
and two \thirteenco\ detections (see also Sect. \ref{sec:radexn1140}). 
This is the first time that a low-metallicity galaxy outside the Local Group
can be modeled in such detail with CO;
for the remaining galaxies
we explore what can be learned from the available \twelveco\ detections.

\subsection{Local thermodynamic equilibrium}
\label{sec:lte}

For optically thin thermalized emission in Local Thermodynamic Equilibrium (LTE), antenna temperature 
is directly proportional to the column density in the upper level of the observed transition.
With knowledge of the kinetic temperature, and under the above assumptions, we can estimate the
column density of the species. 
Population (or rotation) diagrams are a useful tool for this \citep[e.g.,][]{goldsmith99}, through the analysis
of the column density per statistical weight of different energy levels as a function of their
energy above ground state.
In LTE, the statistically-weighted column densities will be a Boltzmann distribution, so that
the slope of a plot in natural logarithmic space will correspond to the inverse of the ``rotational''
temperature, equal to the kinetic temperature in the case of thermalized emission.

Despite its usefulness, there are several problems with such an approach.
First, the assumptions required are not
usually met by the physical conditions in molecular clouds;
molecular emission in the \twelveco\ lines is usually optically thick and not necessarily thermalized.
Second, to be able to correct the emission for the different beam sizes with which the various
transitions are observed requires a knowledge of the source distribution and size. 
Nevertheless, a population diagram analysis sets useful limits on the column density and temperature
and is a fruitful starting point for non-LTE considerations.

The column density of the upper energy state can be written as
\citep[e.g.,][]{goldsmith99}: 
\begin{equation}
N_u\,=\,\frac{8 \pi k \nu^2 W}{h c^3 A_{ul}} \left( \frac{\Delta \Omega_a}{\Delta \Omega_s} \right)
\left( \frac{\tau}{1 - e^{-\tau}} \right)\ ,
\label{eqn:ncol1}
\end{equation}
\noindent
where $W$ is the velocity-integrated main-beam brightness temperature, 
$A_{ul}$ the Einstein spontaneous emission coefficient, and 
$\tau$ is the optical depth of the transition.
For molecules in LTE, 
both a line's excitation temperature and a molecule's rotation temperature, which determines its relative
energy level populations, equal the kinetic temperature ($\equiv\,T$); 
%all excitation temperatures are the same ($\equiv\,T$), and
the population in each level is then given by the Boltzmann distribution:
\begin{equation}
N^{\rm thin}_u\,=\,\frac{N}{Z}\,g_u\,e^{-E_u/kT}\ ,
\label{eqn:ncol2}
\end{equation}
\noindent
where $N^{\rm thin}_u$ corresponds to optically-thin emission,
ignoring the $\tau$ term in Eqn. (\ref{eqn:ncol1});
$N$ is the total column density, $E_u$ is the upper energy level,
$g_u$ is the statistical weight, and $Z$ is the partition function.
Thus, under the assumption of LTE, and with knowledge of $T$, we can obtain the total 
molecular column density from the column density of any single transition:
\begin{equation}
\ln \frac{N^{\rm thin}_u}{g_u} + \ln C_\tau\,=\,\ln N - \ln Z - \frac{E_u}{kT} ,
\end{equation}
\label{eqn:ncol3}
\noindent
where, following \citet{goldsmith99}, we have equated the optical-depth correction factor 
$\tau/(1-e^{-\tau})$ to $C_\tau$, and $N_u$ is calculated from the first two factors in Eqn. \ref{eqn:ncol1}.
To evaluate the population diagrams numerically,
the transition frequencies, upper energy levels, Einstein spontaneous emission,
and statistical weights ($\nu$, $E_u$, $A_{ul}$, $g_u$) for the \twelveco\ and \thirteenco\ molecules 
were calculated using physical quantities from \citet{schoier05} and found to be highly consistent with 
the data from HITRAN \citep{rothman09}.

\begin{figure}[h!]
\vspace{\baselineskip}
\hbox{
\centerline{
%\includegraphics[angle=0,width=\linewidth]{NGC1140_1213CO_PopDiag_Beam_ExpFFCor-crop.pdf}
% 1/3/2017
\includegraphics[angle=0,width=\linewidth]{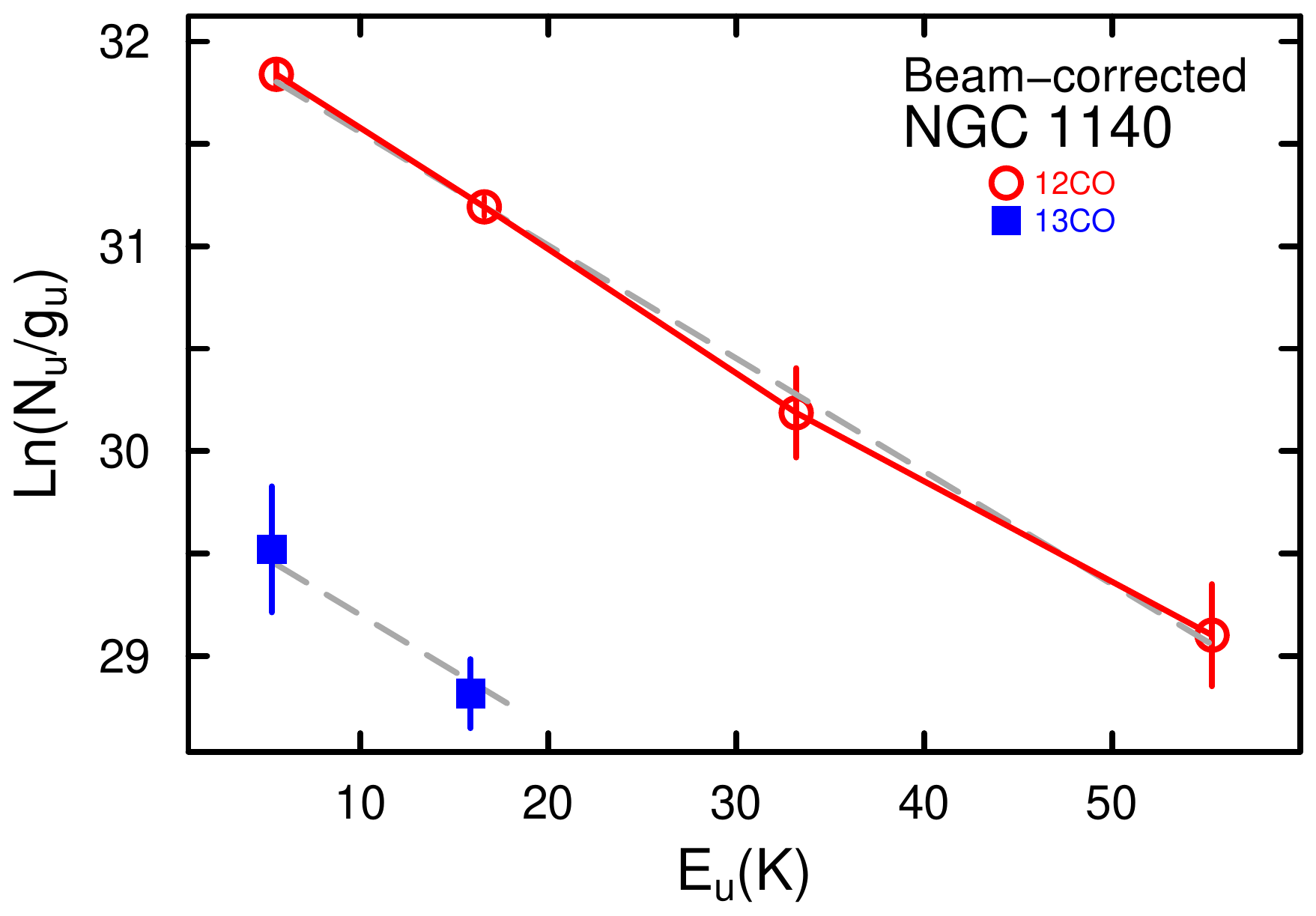}
}
}
\caption{Rotation or population level diagram for NGC\,1140.
A beam correction has been applied to the integrated intensities
(see Sect. \ref{sec:beam} for more details).
Both \twelveco\ and \thirteenco\ transitions are shown, as indicated
by open (red) circles and filled (blue) squares, respectively.
The gray dashed line for \twelveco\ corresponds to the best-fit regressions;
the temperature for \thirteenco\ is ill determined with
only two transitions, so we fixed it to the \twelveco\ value
of \tex\,=\,18.1\,K.
}
\label{fig:lte1140}
\end{figure}

\begin{figure}[h!]
\vspace{\baselineskip}
\hbox{
\centerline{
%\includegraphics[angle=0,width=\linewidth]{NGC7077UM448_12CO_PopDiag_Beam_ExpFFCor-crop.pdf}
% 1/3/2017
\includegraphics[angle=0,width=\linewidth]{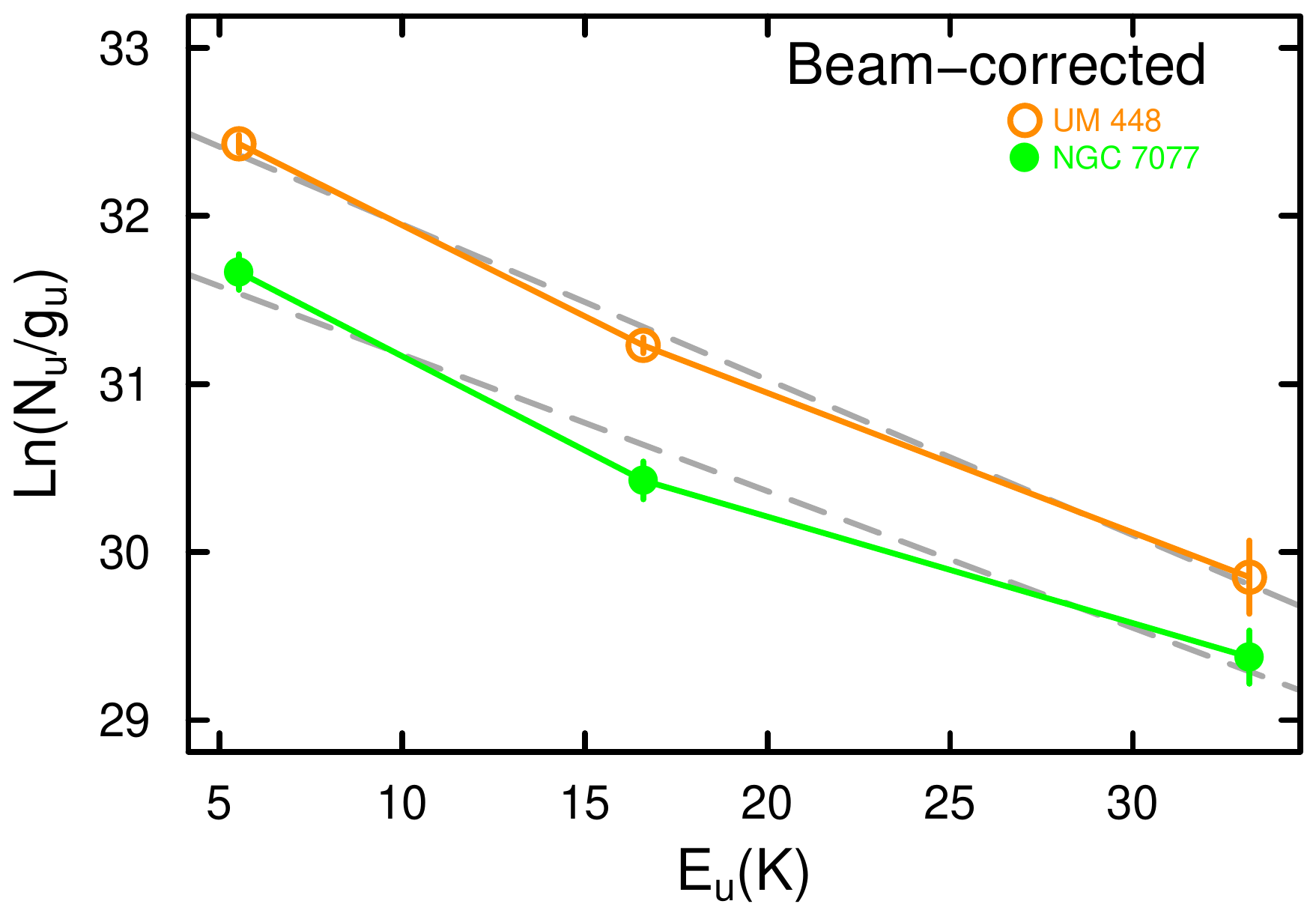}
}
}
\caption{Rotation or population level diagram for UM\,448 and NGC\,7077.
A beam correction has been applied to the integrated intensities
(see Sect. \ref{sec:beam} for more details).
The gray dashed line corresponds to the best-fit regression;
as shown by the legend,
the galaxies are differentiated by different symbols for \twelveco\ emission
(open (orange) circles and filled (green) circles, for UM\,448 and NGC\,7077, respectively).
}
\label{fig:lteothers}
\end{figure}

We first assume that the emission is optically thin, with $\tau \ll 1$ ($C_\tau\,=\,1$). 
A CO population diagram with this assumption for NGC\,1140 is shown in Fig. \ref{fig:lte1140}, 
including the two available $^{13}$CO transitions.
Here, as elsewhere, we have used the velocity-integrated \tmb\ corrected to 
a 22\arcsec\ beam size as described in Sect. \ref{sec:beam} and Appendix \ref{sec:appendix_exponential}.
By fitting the $\ln(N_u/g_u)$ vs. $E_u$ values with a linear regression
(shown as a gray dashed line),
we were able to infer the
total \twelveco\ and \thirteenco\ column densities and temperatures:
$T_{\rm 12CO}\,=\,18.1\,\pm\,1.0$\,K;
% $\log(N_{\rm 12CO})\,=\,15.8$\,\cmtwo;
% 1/3/2017
$\log(N_{\rm 12CO})\,=\,14.7$\,\cmtwo;
$T_{\rm 13CO}\,=\,15$\,K.
With only two transitions, the $^{13}$CO temperature is ill determined, so we fixed it
to $T_{\rm 12CO}$ and obtained
% 1/3/2017
$\log(N_{\rm 13CO})\,=\,13.6$\,\cmtwo.

We then attempted to correct for optical depth effects, according to \citet[][their Eqn. 27]{goldsmith99}:
the largest correction term $C_\tau$ is for \cofour\, with $C_\tau\,=\,1.01$,
and the largest optical depth $\tau\sim$0.002.
If we take these results at face value, and infer
the \twelveco/\thirteenco\ abundances
by taking the ratio of the total column densities,
we obtain a value of \twelveco/\thirteenco\,$\sim 12-13$.
To our knowledge this is one of the lowest values
ever inferred for a galaxy, but cannot be considered reliable because of the
assumption of optically thin gas in the LTE analysis.
We will explore this further in Sects. \ref{sec:radexn1140} and \ref{sec:discussion}.

The population diagrams under the optical-thin assumption for UM\,448 and NGC\,7077
(the only other galaxies for which we have \cothree\ detections) are shown in Fig. \ref{fig:lteothers}.
The observations were corrected for beam dilution as 
%before, namely by assuming
%an exponential  distribution of the source, but with an exponential folding length $r_s\,=\,0.2\,R_{\rm opt}$
%(see Sect. \ref{sec:co21}).
described in Sect. \ref{sec:beam} and Appendix \ref{sec:appendix_exponential}.
Again, from the best-fit regression, we infer $T$ and total $N_{\rm CO}$ column densities:
% $T_{\rm 12CO}\,=\,10.7\,\pm\,0.7$\,K and $\log(N_{\rm 12CO})\,=\,16.3$\,\cmtwo\ for UM\,448; and
% $T_{\rm 12CO}\,=\,12.6\,\pm\,1.9$\,K and $\log(N_{\rm 12CO})\,=\,15.6$\,\cmtwo\ for NGC\,7077.
% 1/3/2017
$T_{\rm 12CO}\,=\,10.8\,\pm\,0.8$\,K and $\log(N_{\rm 12CO})\,=\,15.0$\,\cmtwo\ for UM\,448; and
$T_{\rm 12CO}\,=\,12.3\,\pm\,2.0$\,K and $\log(N_{\rm 12CO})\,=\,14.7$\,\cmtwo\ for NGC\,7077.
As for NGC\,1140, we inferred the optical depths for the various transitions, and 
for both galaxies find that
the largest $\tau$ is for the $J=2-1$ transition with \tauco\,=\,0.001 (NGC\,7077) and \tauco\,=\,0.002 (UM\,448).

\begin{figure*}[ht!]
\vspace{\baselineskip}
\hbox{
\centerline{
%\includegraphics[angle=0,width=0.49\linewidth]{CO21_CO10_modelratiosvsTmb_obs_N1140-crop.pdf}
%\includegraphics[angle=0,width=0.49\linewidth]{CO21_CO10_modelratiosvsTB_obs_N1140_EXPNEW-crop.pdf}
%\includegraphics[angle=0,width=0.49\linewidth]{CO21_CO10_modelratiosvsTB_obs_N1140_EXPNEW_FINERRADEX-crop.pdf}
% 28/6/2017
\includegraphics[angle=0,width=0.49\linewidth]{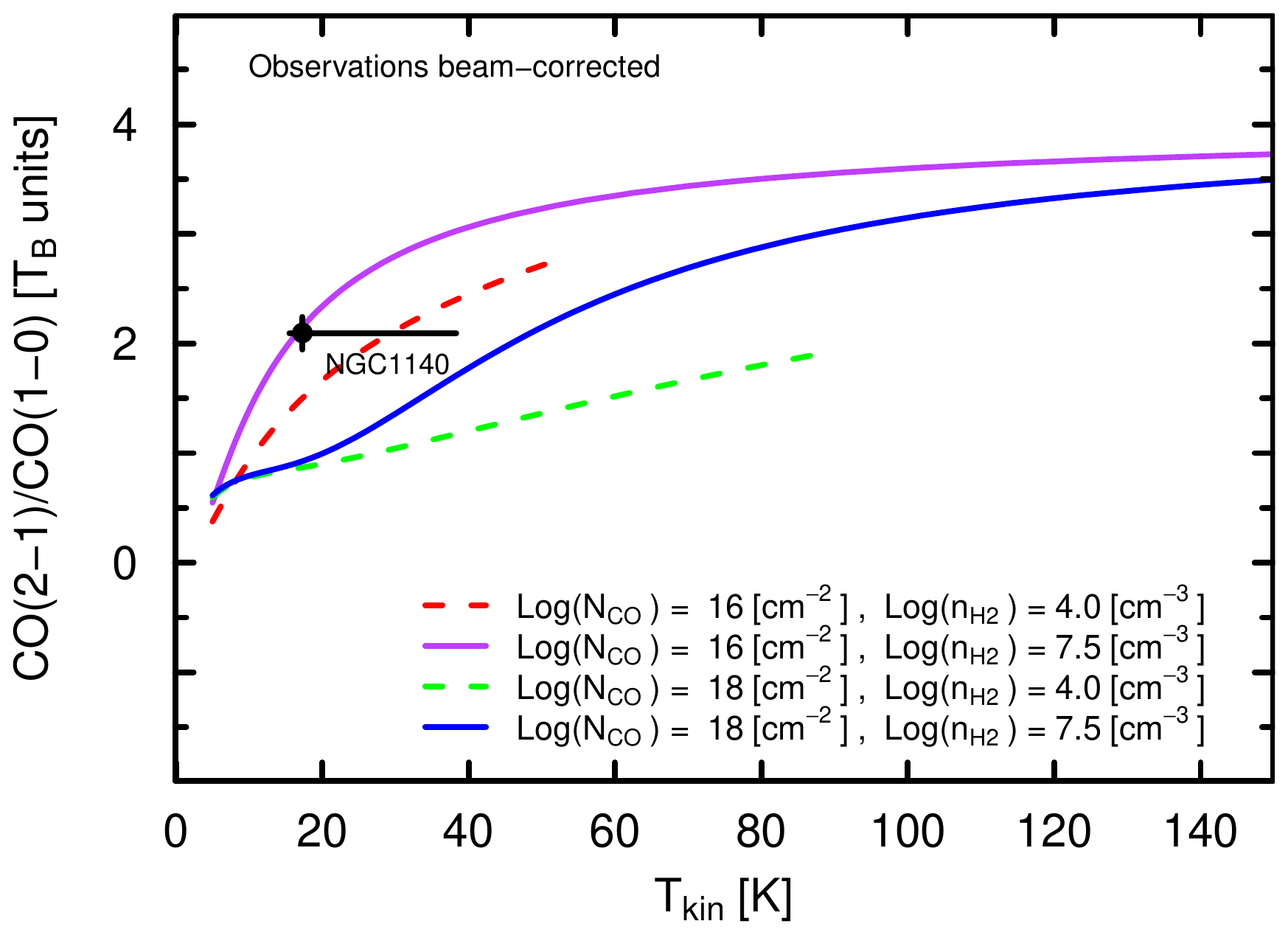}
\hspace{0.02\linewidth}
%\includegraphics[angle=0,width=0.49\linewidth]{CO32_CO10_modelratiosvsTmb_obs_N1140-crop.pdf}
%\includegraphics[angle=0,width=0.49\linewidth]{CO32_CO10_modelratiosvsTB_obs_N1140_EXPNEW-crop.pdf}
%\includegraphics[angle=0,width=0.49\linewidth]{CO32_CO10_modelratiosvsTB_obs_N1140_EXPNEW_FINERRADEX-crop.pdf}
%\includegraphics[angle=0,width=0.49\linewidth]{CO32_CO10_modelratiosvsTB_obs_N1140Others_EXPNEW_FINERRADEX-crop.pdf}
% 28/6/2017
\includegraphics[angle=0,width=0.49\linewidth]{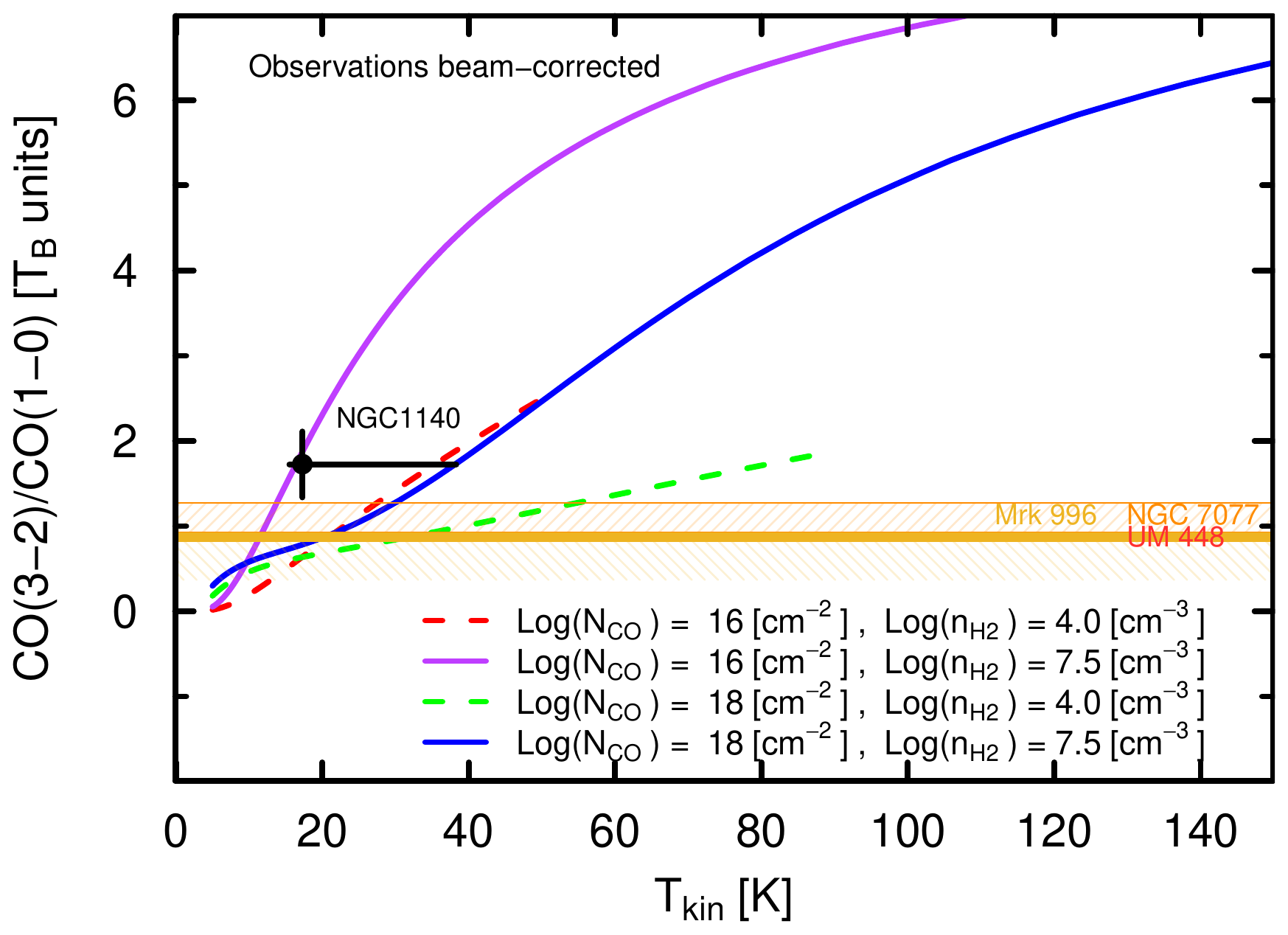}
}
}
\caption{RADEX models (with $\Delta V$\,=\,20\,\kms) of \twelveco\ line ratios plotted against \tkin:
the left panel shows \cotwo/\coone\ and the right \cothree/\coone.
%The models have been adjusted for beam dilution to be directly compared with the observed ratios of
The observations of NGC\,1140, shown as a (black) filled circle,
have been corrected for beam dilution assuming an exponential source distribution %as described in the text
as described in Sect. \ref{sec:beam}
(see also Figs. \ref{fig:lineratios21} and \ref{fig:lineratios32}).
The approximate best-fit RADEX model for NGC\,1140 
(see Sect. \ref{sec:radexn1140})
is shown in both panels as (purple) solid curves
that coincide very well with the observations of NGC\,1140. 
Also shown are three additional parameter combinations of \twelveco\ column density $N_{12CO}$
and volume density \nhtwo.
The \rthree\ parameter space occupied by the other galaxies (like Fig. \ref{fig:lineratios32}, no beam corrections) 
is shown in the right-hand panel as a hatched region and solid horizontal lines
(NGC\,7077, UM\,448, and an upper limit for Mrk\,996).
}
\label{fig:radexlineratiosn1140}
\end{figure*}

\subsection{Modeling \twelveco\ line ratios with RADEX}
\label{sec:radex12co}

To better understand the observed \twelveco\ line ratios in 
the galaxies in our sample, in particular the high ratio of NGC\,1140
compared with other, more metal-rich, galaxies
(see Figs. \ref{fig:lineratios21}, \ref{fig:lineratios32}), we have 
%corrected the RADEX models for beam dilution and 
plotted the line ratios predicted by RADEX as a function of
\tkin\ in Fig. \ref{fig:radexlineratiosn1140} 
%(observations have been corrected for beam dilution assuming an exponential source distribution).
(\rtwo\ and \rthree\ 
for NGC\,1140 have
been corrected for beam dilution as described in Sect. \ref{sec:beam} and Appendix \ref{sec:appendix_exponential}).
Two sets of column densities $N_{\rm 12CO}$ and volume densities \nhtwo\ are illustrated in
the figure; the solid (purple, upper) line passing through the locus of NGC\,1140 corresponds to the
approximate best-fit RADEX model described in Sect. \ref{sec:radexn1140}.
It is evident from Fig. \ref{fig:radexlineratiosn1140} that the combination of moderately low column 
density (and thus low \tauco) and high \nhtwo\ is a necessary ingredient for the high line ratios 
observed in NGC\,1140.
Line ratios would be high also in the case of high $N_{\rm 12CO}$ and high \nhtwo\ (see lower blue solid line), 
but only at high temperatures;
this is the case in NGC\,1068 %for example 
where gas is warm and dense in the circumnuclear disk
around the active nucleus \citep{viti14}.
We discuss below in Sect. \ref{sec:carbon} why we consider high temperatures
in NGC\,1140 to be unlikely.
In the case of more moderate \nhtwo\ (here illustrated as $10^{4}$\,\cmthree), there is no \tkin\ %column density (or \tkin)
that can sufficiently excite the upper $J$ level.

The remaining galaxies for which we have \cothree\ observations
(Mrk\,996, NGC\,7077, UM\,448) are also shown in the right panel of Fig. \ref{fig:radexlineratiosn1140}.
The \rthree\ values of these are less extreme, and can be well modeled with \htwo\ volume
densities as low as $10^4$\,\cmthree.
In these galaxies, 
the gas does not necessarily have to be cool and optically thin (i.e., low $N_{\rm 12CO}$)
in order to achieve the observed \rthree; 
it can be warmer and have a higher column density than in NGC\,1140.

\subsection{RADEX models for NGC\,1140}
\label{sec:radexn1140}

To determine the excitation conditions of the molecular gas in NGC\,1140, we
have fitted the six \twelveco\ and \thirteenco\ detected transitions with
the radiative transfer code RADEX developed by \citet{vandertak07},
using the uniform sphere approximation to calculate the escape probabilities.
A one-zone model was adopted, namely
the kinetic temperature and \htwo\ volume density were required to be the same for
all molecules, implying that all transitions sample the same gas.
We calculated RADEX predictions for %several linewidths including 
\dv\,=10 and 20\,\kms, covering
the following range of parameters
sampled in %$\sim$0.3\,dex 
logarithmic steps\footnote{Temperature was sampled linearly.}: 
\begin{itemize}
\item
\htwo\ volume densities \nhtwo\,=\,$10^2 - 10^8$\,\cmthree; %15 density points
\item
kinetic temperature \tkin\,=\,$5 - 200$\,K; % 20 temperature points
\item
CO column densities \nco\,=\,$10^{12} - 10^{20}$\,\cmtwo. % 19 values
\end{itemize}
All parameters were calculated assuming a cosmic microwave background 
radiation temperature of $T$\,=\,2.73\,K.
The \thirteenco\ abundance was included in the calculation through a second RADEX grid
for the isotope; the column density \nthirteenco\ was adjusted by sampling
a range of abundances (from \ntwelveco/\nthirteenco\,=\,1 to 100),
and refining the fineness of the grid once the initial best-fit parameters were established.
Ultimately, $\ga$400\,000 total models in the combined \twelveco\ $+$ \thirteenco\ grid
were sampled.

The observed emission is generally only a small fraction of integrated line intensity from RADEX 
(and Large-Velocity Gradient, LVG) models.
%The fraction of these two quantities, $F\,=\,(\Delta \varv_{\rm obs} T_{\rm obs})/(\Delta \varv_{\rm model} T_{\rm model})$,
The fraction of these two quantities, $F\,=\,(\Delta \varv_{\rm obs} T_{\rm obs})/(\Delta V_{\rm model} T_{\rm model})$,
is the filling factor.
We incorporated $F$ as a free parameter in the models, and fit
the grid of RADEX models via a $\chi^2$ minimization technique \citep[e.g.,][]{nikolic07}: 
\begin{equation}
\displaystyle
\chi^2\,=\,\sum_{i=1}^6 \left( \frac{F\ I_i^{\rm mod} - I_i^{\rm obs}}{\Delta I_i^{\rm obs}} \right)^2
\label{eqn:chisqintensities}
\end{equation}
\noindent
where $I_i^{\rm obs}$ is the velocity-integrated \tmb\ of the $i^{th}$ transition, 
$I_i^{\rm mod}$ the $i^{th}$ model prediction,
and $\Delta I_i^{\rm obs}$ are the 1$\sigma$ uncertainties given in Table \ref{tab:n1140}.

We also experimented with a minimization technique not including the filling
factor $F$, but rather based on line ratios:
\begin{equation}
\displaystyle
\chi^2\,=\,\sum_{j=1}^5 \left( \frac{R_j^{\rm mod} - R_j^{\rm obs}}{\Delta R_j^{\rm obs}} \right)^2 
\label{eqn:chisqratios}
\end{equation}
\noindent
where $R_j^{\rm obs}$ is the $j^{th}$ velocity-integrated temperature ratio,
$R_j^{\rm mod}$ is the corresponding model line ratio,
and $\Delta R_j^{\rm obs}$ are the 1$\sigma$ uncertainties.
These fits do not explicitly require equality of the filling factor $F$ and thus
may be more accommodating. 
We experimented with several sets of line ratios, in order to avoid
potentially biasing the fits to any of the individual ratios.

\begin{figure*}[ht!]
\vspace{\baselineskip}
\hbox{
\centerline{
%\includegraphics[angle=0,height=0.43\linewidth]{Chisquare_NEWffnewlineratios_denstempNGC1140_dv=10_color-crop.pdf}
% 2/3/2017
% \includegraphics[angle=0,height=0.43\linewidth]{Chisquare_newlineratiosNEWBeamCor_denstempNGC1140_dv=10_color-crop.pdf}
% 5/3/2017
\includegraphics[angle=0,height=0.43\linewidth]{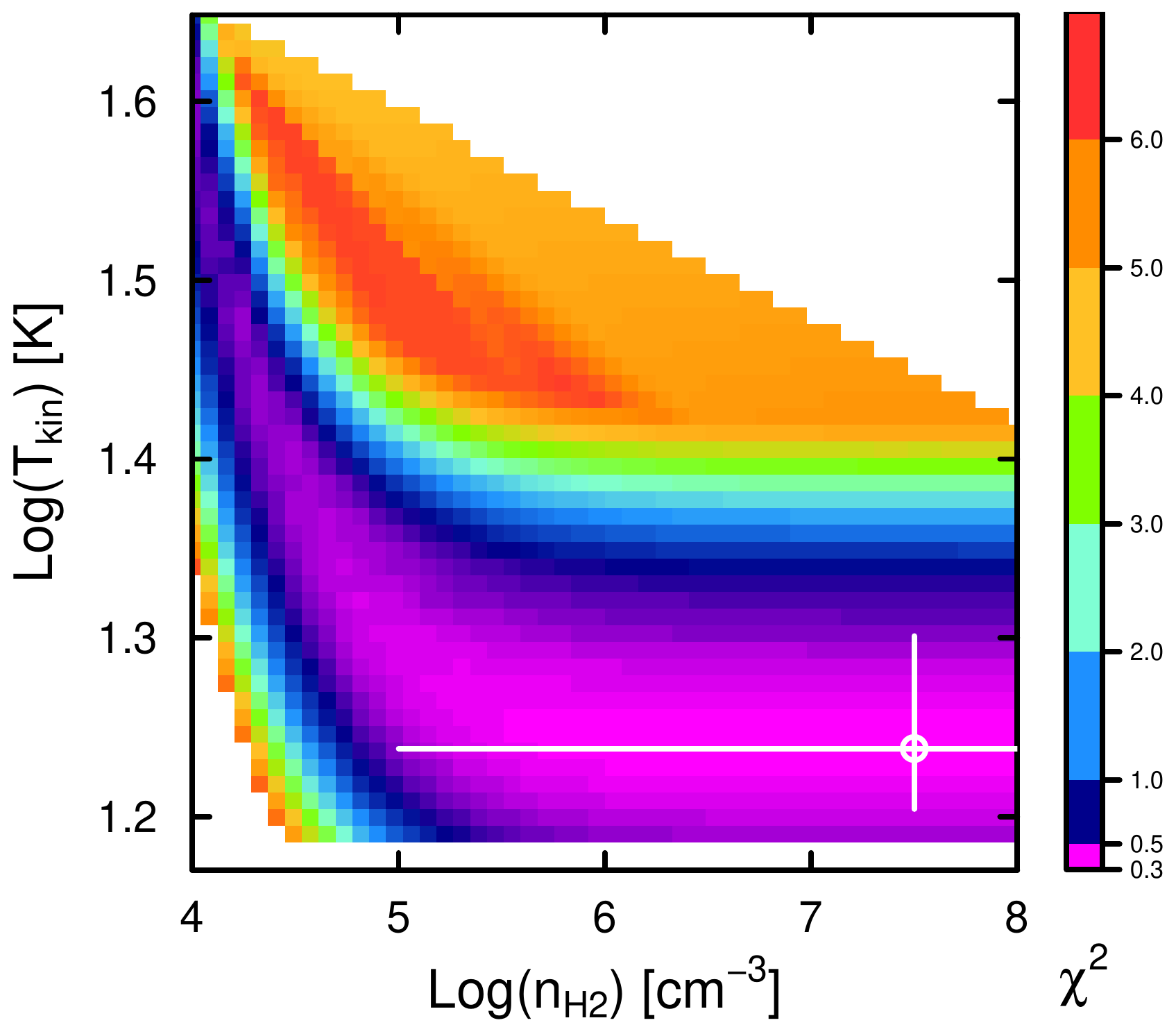}
% test
%\includegraphics[angle=0,width=0.49\linewidth]{/home/hunt/statistics/moleculesBCDs_PaperII/filename_crop.png}
%\includegraphics[angle=0,width=0.98\linewidth]{Chisquare_newlineratios_denstempNGC1140_dv=10_blue-crop.pdf}
\hspace{0.02\linewidth}
%\includegraphics[angle=0,height=0.43\linewidth]{Chisquare_NEWffnewlineratios_Ncol13SizeNGC1140_dv=10_colorNEW-crop.pdf}
% 2/3/2017
%\includegraphics[angle=0,height=0.43\linewidth]{Chisquare_newlineratiosNEWBeamCor_NCOTau_NGC1140_dv=10_color-crop.pdf}
% 5/3/2017
\includegraphics[angle=0,height=0.43\linewidth]{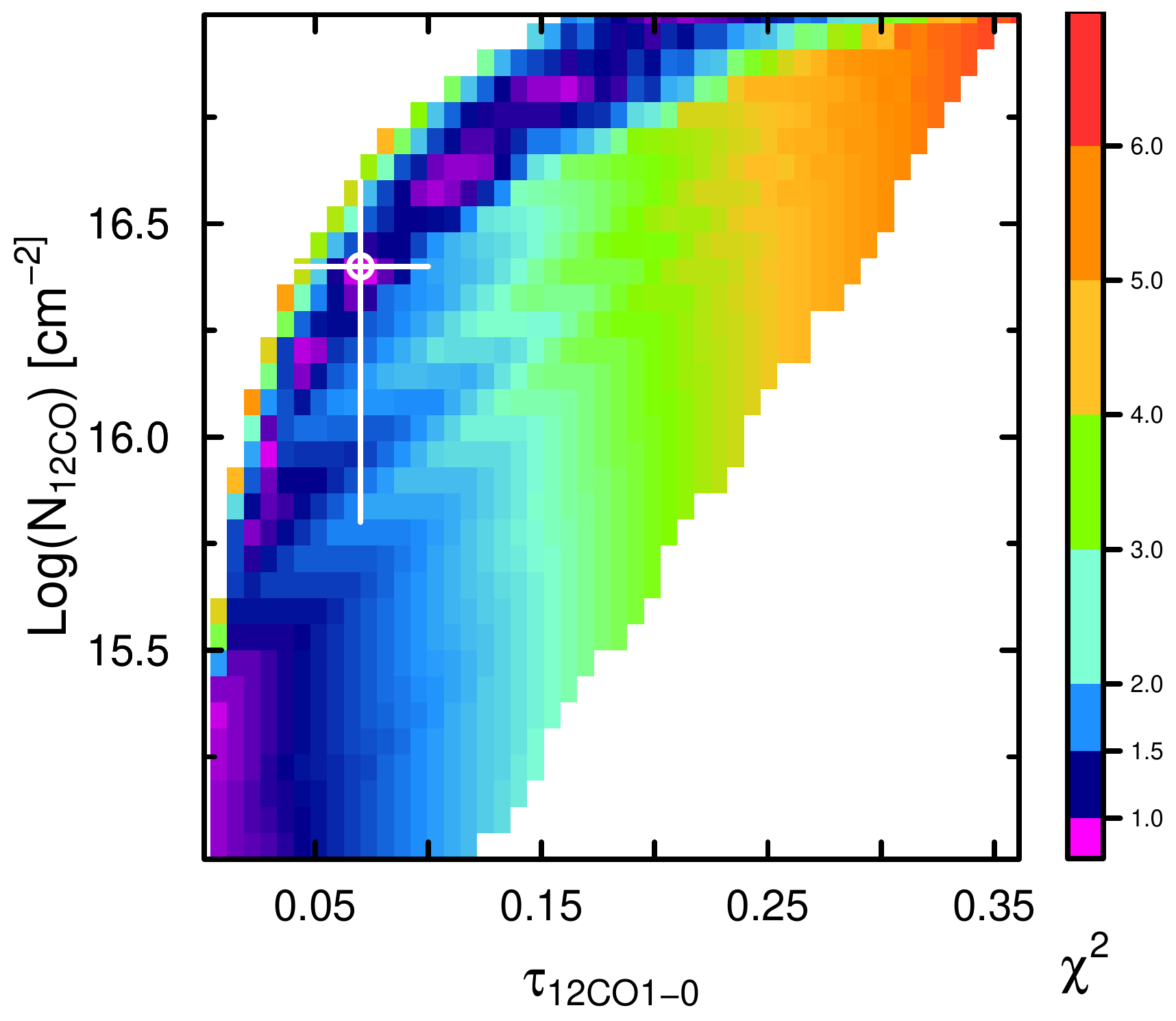}
}
}
\caption{RADEX fit $\chi^2$ surface (for the ``Set 1" line ratios) 
%with an exponential source distribution)
with the exponential beam correction
as a function of \tkin\ and \nhtwo\ in the left panel,
and as a function of \ntwelveco\ and $\tau$(\coone) in the right. 
%considering a range of source exponential radii (2\arcsec\,$\leq r_s \leq 4$\arcsec) in left panel;
%as a function of $N_{\rm 13CO}$ and $r_s$ in the right (these exponential folding radii are
%equivalent to $\theta_s\,=\,2\,r_s\,\ln(2)$ as shown in Appendix \ref{sec:appendix_exponential}).
The RADEX models shown here have $\Delta V$\,=\,20\,\kms.
The associated $\chi^2$ values are shown as a side-bar color table.
The $\chi^2$ color table in the right panel is different from the left because
the interpolation used for the graphic display suffers from the coarser resolution
in column density.
The best-fit values are shown as an open (white) circle;
the error bars shown consider a slightly more limited range in $\chi^2$
respect to  Table \ref{tab:radexn1140}. 
}
\label{fig:chisquaren1140}
\end{figure*}

\begin{figure}[ht!]
\vspace{\baselineskip}
\hbox{
\centerline{
\includegraphics[angle=0,height=0.87\linewidth]{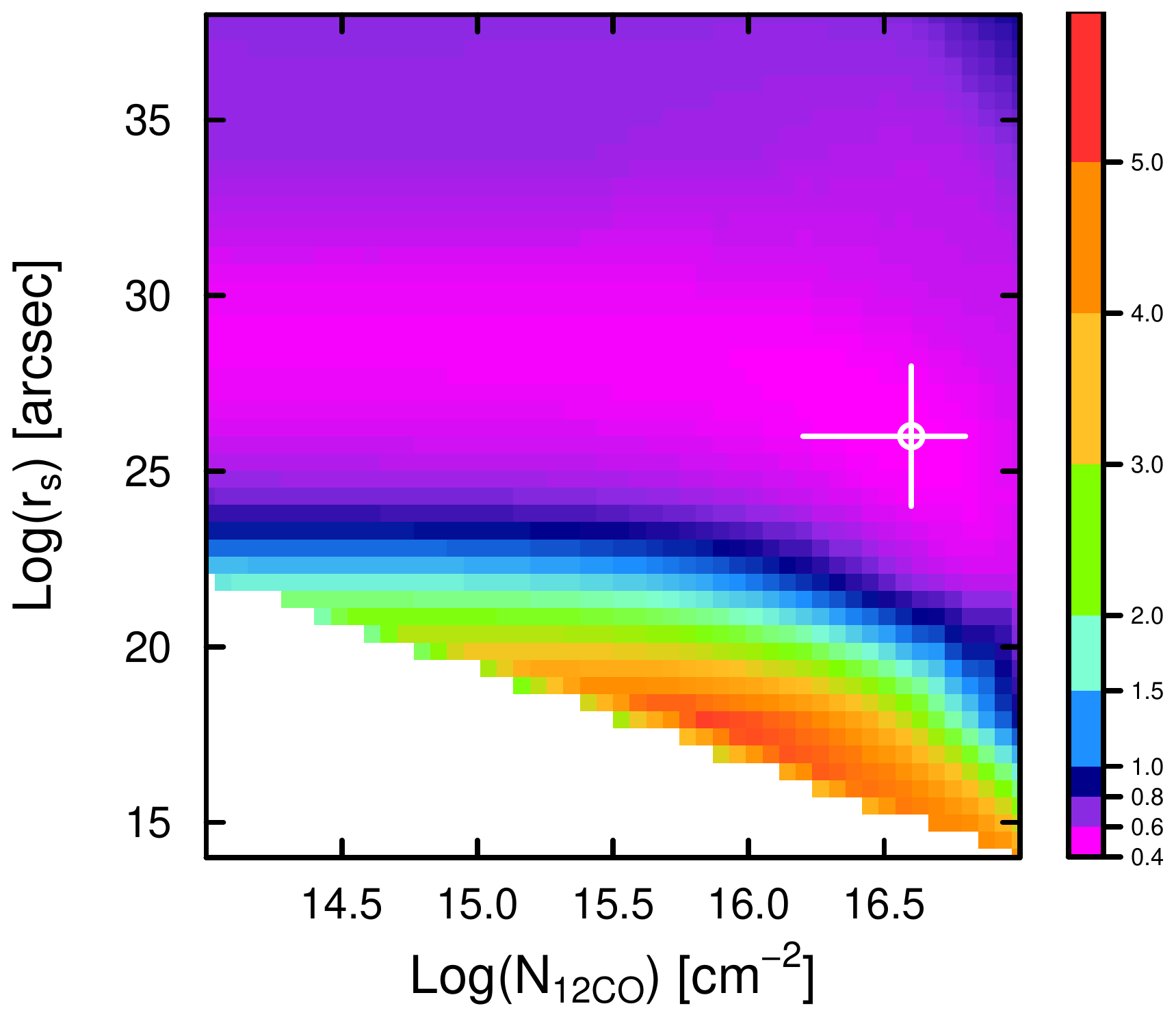}
}
}
\caption{RADEX fit $\chi^2$ surface (for the ``Set 1" line ratios, 
assuming the Gaussian-derived beam correction factors)
as a function of $N_{\rm 12CO}$ and Gaussian FWHM $\theta_s$ in the left panel,
The RADEX models shown here have $\Delta V$\,=\,20\,\kms;
the associated $\chi^2$ values are shown as a side-bar color table.
The best-fit values are shown as an open (white) circle;
the error bars shown consider a slightly more limited range in $\chi^2$
respect to  Table \ref{tab:radexn1140}. 
}
\label{fig:chisquarencolsizen1140}
\end{figure}

As described in Sect. \ref{sec:beam}, we explored two different methods for
correcting the velocity-integrated \tmb\ (and uncertainties) to a 22\arcsec\ beam size: 
(1)~analyzing the cool-dust distribution and assuming that the molecular gas is distributed like the dust
(see Appendix \ref{sec:appendix_exponential});
and
(2)~assuming the molecular gas is distributed in a Gaussian, and letting the Gaussian's FWHM
($\theta_s$) be a free parameter in the fit
(see Appendix \ref{sec:appendix_gaussian}).
We used both methods to calculate the observed $I_i^{\rm obs}$ in Eqns. (\ref{eqn:chisqintensities}, \ref{eqn:chisqratios}).
In the first case based on the fitted exponential function of the dust emission, 
 $I_i^{\rm obs}$ were calculated independently of the fitting procedure.
Thus for the six intensities (Eqn. \ref{eqn:chisqintensities}), 
we fit five parameters 
(\nhtwo, \tkin, \nco(\twelveco), [\twelveco]/[\thirteenco], $F$), 
and for the five independent line ratios (Eqn. \ref{eqn:chisqratios}), we fit four parameters 
(same as before, but not including $F$). 
For the second case of Gaussian beam corrections, $I_i^{\rm obs}$ were calculated within the fit procedure;
%Thus both fits are {\bf only slightly over-determined, and the} reduced $\chi^2_\nu$ is equivalent to $\chi^2$.
we explored $\theta_s$ ranging from 2\arcsec\ to 40\arcsec, and narrowed the step size in successive iterations.
Thus for the Gaussian beam corrections, the fitting procedure comprised an additional parameter, $\theta_s$,
and the minimum $\chi^2$ value includes this dependence. 
All the $\chi^2$ values quoted here are not reduced, but rather the total summed $\chi^2$.

% -----------------------------------------------------------------
% --- Table x: RADEX fit for NGC1140  
%
\begin{center}
\begin{table*}
      \caption[]{Best-fit RADEX parameters for NGC\,1140$^{\rm a}$} 
\label{tab:radexn1140}
\resizebox{\linewidth}{!}{
%\addtolength{\tabcolsep}{7pt}
{%\small
%\tiny
\begin{tabular}{llcccccccccc}
\hline
\multicolumn{1}{c}{\dv} &
\multicolumn{1}{c}{$\chi^2$} &
\multicolumn{6}{c}{$\tau$} &
\multicolumn{1}{c}{\tkin} &
\multicolumn{1}{c}{Log(\nhtwo)} &
\multicolumn{1}{c}{Log($N_{12CO}$)} &
\multicolumn{1}{c}{\twelveco/\thirteenco} \\
%\multicolumn{1}{c}{$\theta_s$} \\
\multicolumn{1}{c}{(\kms)} &&
\multicolumn{1}{c}{\coone} &
\multicolumn{1}{c}{\cotwo} &
\multicolumn{1}{c}{\cothree} &
\multicolumn{1}{c}{\cofour} &
\multicolumn{1}{c}{\thirteencoone} &
\multicolumn{1}{c}{\thirteencotwo} &
\multicolumn{1}{c}{(K)} &
\multicolumn{1}{c}{(\cmthree)} &
\multicolumn{1}{c}{(\cmtwo)} \\
%\multicolumn{1}{c}{(arcsec)} \\
\multicolumn{1}{c}{(1)} & 
\multicolumn{1}{c}{(2)} & 
\multicolumn{1}{c}{(3)} & 
\multicolumn{1}{c}{(4)} & 
\multicolumn{1}{c}{(5)} & 
\multicolumn{1}{c}{(6)} & 
\multicolumn{1}{c}{(7)} & 
\multicolumn{1}{c}{(8)} & 
\multicolumn{1}{c}{(9)} & 
\multicolumn{1}{c}{(10)} & 
\multicolumn{1}{c}{(11)} & 
\multicolumn{1}{c}{(12)} \\
%\multicolumn{1}{c}{(13)} \\
\hline
\\
% with R rbest <- 3 > rbest*2*log(2*rbest) [1] 10.75056
% 
% 4/3/2017
\multicolumn{12}{c}{Fit to intensities with exponential beam correction} \\
%10 &  1.58 &    0.03 &    0.07 &    0.07 &    0.05 &   0.002 &   0.007 &  18.2 &   6.0 &  15.7 & 10 \\
% $-$ & $ 1.58 -  2.29$ & $ 0.03 -  0.03$ & $ 0.07 -  0.07$ & $ 0.07 -  0.08$ & $ 0.05 -  0.05$ & $0.002 - 0.003$ & $0.006 - 0.007$ &  $ 18.2 -  18.2$ & $  5.2 -   8.0$ & $ 15.7 -  15.7$ & $8 - 14$  \\
% 5/3/2017 finer RADEX grid
10 &  0.94 &    0.02 &    0.06 &    0.06 &    0.04 &   0.002 &   0.006 &  17.3 &   7.4 &  15.6 &  8 \\
 $-$ & $ 0.94 -  1.93$ & $ 0.02 -  0.04$ & $ 0.05 -  0.10$ & $ 0.06 -  0.10$ & $ 0.04 -  0.06$ & $0.002 - 0.004$ & $0.005 - 0.010$ &  $ 16.1 -  18.7$ & $  4.9 -   8.0$ & $ 15.6 -  15.8$ & $ 8 - 10$  \\
%20 &  2.15 &    0.01 &    0.04 &    0.04 &    0.02 &   0.001 &   0.003 &  18.2 &   7.2 &  15.7 & 10 \\
% $-$ & $ 2.15 -  2.91$ & $ 0.01 -  0.01$ & $ 0.04 -  0.04$ & $ 0.04 -  0.04$ & $ 0.02 -  0.02$ & $0.001 - 0.001$ & $0.003 - 0.004$ &  $ 18.2 -  18.2$ & $  4.8 -   8.0$ & $ 15.7 -  15.7$ & $8 - 14$  \\
% 5/3/2017 finer RADEX grid
20 &  0.93 &    0.02 &    0.05 &    0.05 &    0.03 &   0.002 &   0.005 &  16.7 &   7.5 &  15.8 &  8 \\
 $-$ & $ 0.93 -  1.93$ & $ 0.01 -  0.05$ & $ 0.03 -  0.13$ & $ 0.03 -  0.13$ & $ 0.02 -  0.08$ & $0.001 - 0.005$ & $0.003 - 0.012$ &  $ 16.1 -  19.4$ & $  4.9 -   8.0$ & $ 15.6 -  16.2$ & $ 8 - 10$  \\
\\
\multicolumn{12}{c}{Fit to intensities with Gaussian beam correction$^{\mathrm b}$ } \\
10 &  0.95 &    0.02 &    0.06 &    0.06 &    0.04 &   0.002 &   0.005 &  18.0 &   7.1 &  15.6 & 10 \\
 $-$ & $ 0.95 -  1.95$ & $ 0.01 -  0.06$ & $ 0.05 -  0.15$ & $ 0.06 -  0.16$ & $ 0.04 -  0.10$ & $0.001 - 0.005$ & $0.004 - 0.014$ &  $ 16.1 -  26.3$ & $  4.3 -   8.0$ & $ 15.6 -  16.0$ & $8 - 10$  \\
20 &  0.85 &    0.01 &    0.03 &    0.03 &    0.02 &   0.001 &   0.003 &  17.3 &   5.9 &  15.6 & 10 \\
 $-$ & $ 0.85 -  1.85$ & $ 0.01 -  0.05$ & $ 0.03 -  0.13$ & $ 0.03 -  0.13$ & $ 0.02 -  0.08$ & $0.001 - 0.005$ & $0.002 - 0.012$ &  $ 16.1 -  24.4$ & $  4.5 -   8.0$ & $ 15.6 -  16.2$ & $8 - 10$  \\
\\
%\multicolumn{12}{c}{Fit to ``assorted'' line ratios } \\
\multicolumn{12}{c}{Fit to ``Set 1'' line ratios$^{\mathrm c}$ with exponential beam correction } \\
%10 &  0.77 &    0.06 &    0.14 &    0.16 &    0.09 &   0.005 &   0.014 &  18.2 &   5.2 &  16.0 & 10 &  4.2$^{\rm b}$ \\
% $-$ & $ 0.77 -  1.69$ & $ 0.01 -  0.20$ & $ 0.01 -  0.60$ & $ 0.01 -  0.83$ & $ 0.01 -  0.57$ & $0.000 - 0.018$ & $0.001 - 0.057$ &  $ 15.1 -  38.0$ & $  4.0 -   8.0$ & $ 15.0 -  16.7$ & $10 - 10$ & $ 4.2 -  4.2^{\rm b}$ \\
% 1/3/2017 coarse RADEX grid
%10 &  0.43 &    0.11 &    0.28 &    0.30 &    0.19 &   0.010 &   0.026 &  18.2 &   6.0 &  16.3 & 10 \\
% $-$ & $ 0.43 -  1.39$ & $ 0.01 -  0.20$ & $ 0.02 -  0.60$ & $ 0.02 -  0.83$ & $ 0.01 -  0.57$ & $0.001 - 0.018$ & $0.002 - 0.057$ &  $ 15.1 -  38.0$ & $  4.0 -   8.0$ & $ 15.0 -  16.7$ & $8 - 12$  \\
% 5/3/2017 finer RADEX grid
10 &  0.39 &    0.09 &    0.23 &    0.24 &    0.14 &   0.009 &   0.022 &  17.3 &   7.9 &  16.2 & 10 \\
 $-$ & $ 0.39 -  1.39$ & $ 0.00 -  0.24$ & $ 0.00 -  0.72$ & $ 0.00 -  1.01$ & $ 0.00 -  0.75$ & $0.000 - 0.022$ & $0.000 - 0.068$ &  $ 15.5 -  44.5$ & $  4.0 -   8.0$ & $ 14.0 -  16.8$ & $ 8 - 10$  \\
% 4/3/2017 coarse RADEX grid
%20 &  0.49 &    0.14 &    0.34 &    0.37 &    0.23 &   0.012 &   0.032 &  18.2 &   7.6 &  16.7 & 10 \\
% $-$ & $ 0.49 -  1.48$ & $ 0.00 -  0.20$ & $ 0.01 -  0.78$ & $ 0.01 -  1.38$ & $ 0.00 -  1.20$ & $0.000 - 0.018$ & $0.001 - 0.078$ &  $ 15.1 -  45.7$ & $  4.0 -   8.0$ & $ 15.0 -  17.3$ & $ 8 - 14$  \\
% 5/3/2017 finer RADEX grid
20 &  0.38 &    0.07 &    0.18 &    0.19 &    0.11 &   0.007 &   0.017 &  17.3 &   7.5 &  16.4 & 10 \\
 $-$ & $ 0.38 -  1.37$ & $ 0.00 -  0.23$ & $ 0.00 -  0.63$ & $ 0.00 -  0.85$ & $ 0.00 -  0.58$ & $0.000 - 0.021$ & $0.000 - 0.060$ &  $ 15.5 -  41.3$ & $  4.0 -   8.0$ & $ 14.0 -  17.0$ & $ 8 - 14$  \\
\\
\multicolumn{12}{c}{Fit to ``Set 1'' line ratios$^{\mathrm c}$ with Gaussian beam correction$^{\mathrm d}$ } \\
10 &  0.44 &    0.09 &    0.23 &    0.24 &    0.14 &   0.009 &   0.022 &  17.3 &   7.1 &  16.2 & 10 \\
 $-$ & $ 0.44 -  1.44$ & $ 0.00 -  0.33$ & $ 0.00 -  0.85$ & $ 0.00 -  1.03$ & $ 0.00 -  0.75$ & $0.000 - 0.030$ & $0.000 - 0.080$ &  $ 15.5 -  44.5$ & $  4.0 -   8.0$ & $ 14.0 -  16.8$ & $8 - 10$  \\
20 &  0.43 &    0.12 &    0.29 &    0.30 &    0.18 &   0.011 &   0.028 &  17.3 &   6.9 &  16.6 & 10 \\
 $-$ & $ 0.43 -  1.43$ & $ 0.00 -  0.29$ & $ 0.00 -  0.74$ & $ 0.00 -  0.85$ & $ 0.00 -  0.59$ & $0.000 - 0.027$ & $0.000 - 0.070$ &  $ 15.5 -  44.5$ & $  4.0 -   8.0$ & $ 14.0 -  17.0$ & $8 - 14$  \\
\\
%\multicolumn{12}{c}{Fit to ``different'' line ratios } \\
\multicolumn{12}{c}{Fit to ``other'' sets of line ratios$^{\mathrm c}$ with exponential beam correction } \\
% 3/3/2017
%10 &  $1.07 - 1.20$ &    0.05 &    0.14 &    0.15 &    0.09 &   0.005 &   0.013 &  18.2 &   5.6 &  16.0 & 10 \\
% $-$ & $ 1.07 -  2.20$ & $ 0.01 -  0.11$ & $ 0.01 -  0.29$ & $ 0.01 -  0.31$ & $ 0.01 -  0.19$ & $0.000 - 0.010$ & $0.001 - 0.027$ &  $ 18.2 -  18.2$ & $  5.2 -   8.0$ & $ 15.0 -  16.3$ & $8 - 14$  \\
% 5/3/2017 finer RADEX grid
10 &  0.85 &    0.06 &    0.15 &    0.15 &    0.09 &   0.005 &   0.014 &  17.3 &   7.8 &  16.0 & 10 \\
 $-$ & $ 0.85 -  1.84$ & $ 0.00 -  0.27$ & $ 0.00 -  0.91$ & $ 0.00 -  1.40$ & $ 0.00 -  1.20$ & $0.000 - 0.016$ & $0.000 - 0.056$ &  $ 16.1 -  44.5$ & $  4.0 -   8.0$ & $ 14.0 -  17.0$ & $ 8 - 20$  \\
% 4/3/2017
%20 &  $1.07 - 1.20$ &    0.05 &    0.14 &    0.15 &    0.09 &   0.005 &   0.013 &  18.2 &   5.6 &  16.3 & 10 \\
% $-$ & $ 1.07 -  2.21$ & $ 0.00 -  0.14$ & $ 0.01 -  0.35$ & $ 0.01 -  0.38$ & $ 0.01 -  0.23$ & $0.000 - 0.013$ & $0.001 - 0.033$ &  $ 18.2 -  18.2$ & $  5.2 -   8.0$ & $ 15.0 -  16.7$ & $ 8 - 14$  \\
% 5/3/2017 finer RADEX grid
20 &  0.83 &    0.05 &    0.12 &    0.12 &    0.07 &   0.004 &   0.011 &  17.3 &   8.0 &  16.2 & 10 \\
 $-$ & $ 0.83 -  1.83$ & $ 0.00 -  0.24$ & $ 0.00 -  0.64$ & $ 0.00 -  0.78$ & $ 0.00 -  0.56$ & $0.000 - 0.014$ & $0.000 - 0.038$ &  $ 16.1 -  27.3$ & $  4.3 -   8.0$ & $ 14.0 -  17.0$ & $ 8 - 20$  \\
\\
\hline
%\\
\multicolumn{1}{c}{\dv} &
\multicolumn{1}{c}{$\chi^2$} &
\multicolumn{5}{c}{$\tau$} & &
\multicolumn{1}{c}{\tkin} &
\multicolumn{1}{c}{Log(\nhtwo)} &
\multicolumn{1}{c}{Log($N_{12CO}$)} &
\multicolumn{1}{c}{\twelveco/C} \\
%\multicolumn{1}{c}{$\theta_s$} \\
\multicolumn{1}{c}{(\kms)} & &
\multicolumn{1}{c}{\coone} &
\multicolumn{1}{c}{\cotwo} &
\multicolumn{1}{c}{\cothree} &
\multicolumn{1}{c}{\cofour} &
\multicolumn{1}{c}{\ci(1--0)} & &
\multicolumn{1}{c}{(K)} &
\multicolumn{1}{c}{(\cmthree)} &
\multicolumn{1}{c}{(\cmtwo)} \\
%\multicolumn{1}{c}{(arcsec)} \\
\hline
\\
\multicolumn{12}{c}{Fit to intensities with exponential beam correction} \\
% need finer RADEX grid
%10 &  1.24 &    0.03 &    0.07 &    0.07 &    0.05 &    0.06 & &  18.2 &   6.0 &  15.7 &  0.02 \\
% $-$ & $ 1.24 -  1.73$ & $ 0.03 -  0.03$ & $ 0.07 -  0.07$ & $ 0.07 -  0.08$ & $ 0.05 -  0.05$ & $0.061 - 0.061$ & &  $ 18.2 -  18.2$ & $  5.2 -   8.0$ & $ 15.7 -  15.7$ & $ 0.02 -  0.02$ \\
% 9/3/2017 finer RADEX grid now with chi2min+1
% ff =  0.10 to  0.16
%10 &  0.55 &    0.02 &    0.06 &    0.06 &    0.04 &    0.05 & &  17.3 &   7.1 &  15.6 &  0.02 \\
% $-$ & $ 0.55 -  1.54$ & $ 0.02 -  0.04$ & $ 0.06 -  0.10$ & $ 0.06 -  0.10$ & $ 0.03 -  0.06$ & $0.047 - 0.077$ & &  $ 16.1 -  18.7$ & $  5.1 -   8.0$ & $ 15.6 -  15.8$ & $ 0.02 -  0.02$ \\
% 9/3/2017 finer RADEX grid now with chi2min+1, and REAL (i.e., difference between NCI and NCO12, rather than approximate values used in @abundance)
10 &  0.55 &    0.02 &    0.06 &    0.06 &    0.04 &    0.05 & &  17.3 &   7.1 &  15.6 &  0.03 \\
 $-$ & $ 0.55 -  1.54$ & $ 0.02 -  0.04$ & $ 0.06 -  0.10$ & $ 0.06 -  0.10$ & $ 0.03 -  0.06$ & $0.047 - 0.077$ & &  $ 16.1 -  18.7$ & $  5.1 -   8.0$ & $ 15.6 -  15.8$ & $ 0.03 -  0.03$ \\
%20 &  1.89 &    0.01 &    0.03 &    0.04 &    0.02 &    0.03 & &  18.2 &   6.8 &  15.7 &  0.02 \\
% $-$ & $ 1.89 -  2.04$ & $ 0.01 -  0.01$ & $ 0.03 -  0.04$ & $ 0.04 -  0.04$ & $ 0.02 -  0.02$ & $0.031 - 0.031$ & &  $ 18.2 -  18.2$ & $  5.6 -   8.0$ & $ 15.7 -  15.7$ & $ 0.02 -  0.02$ \\
% 8/3/2017 finer RADEX grid now with chi2min+1
% ff =  0.04 to  0.16 confirmed
%20 &  0.66 &    0.02 &    0.05 &    0.05 &    0.03 &    0.04 & &  17.3 &   6.2 &  15.8 &  0.02 \\
% $-$ & $ 0.66 -  1.65$ & $ 0.01 -  0.05$ & $ 0.03 -  0.13$ & $ 0.03 -  0.13$ & $ 0.02 -  0.08$ & $0.023 - 0.097$ & &  $ 16.1 -  18.7$ & $  4.9 -   8.0$ & $ 15.6 -  16.2$ & $ 0.02 -  0.02$ \\
% 9/3/2017 finer RADEX grid now with chi2min+1, and REAL (i.e., difference between NCI and NCO12, rather than approximate values used in @abundance)
20 &  0.66 &    0.02 &    0.05 &    0.05 &    0.03 &    0.04 & &  17.3 &   6.2 &  15.8 &  0.03 \\
 $-$ & $ 0.66 -  1.65$ & $ 0.01 -  0.05$ & $ 0.03 -  0.13$ & $ 0.03 -  0.13$ & $ 0.02 -  0.08$ & $0.023 - 0.097$ & &  $ 16.1 -  18.7$ & $  4.9 -   8.0$ & $ 15.6 -  16.2$ & $ 0.03 -  0.03$ \\
\\
\multicolumn{12}{c}{Fit to intensities with Gaussian beam correction$^{\mathrm e}$ } \\
10 &  0.75 &    0.04 &    0.10 &    0.10 &    0.06 &    0.08 & &  16.7 &   7.4 &  15.8 &  0.03 \\
 $-$ & $ 0.75 -  1.75$ & $ 0.00 -  0.04$ & $ 0.07 -  0.10$ & $ 0.09 -  0.11$ & $ 0.06 -  0.07$ & $0.015 - 0.077$ & &  $ 16.1 -  41.3$ & $  4.0 -   8.0$ & $ 15.8 -  15.8$ & $ 0.03 -  0.10$ \\
20 &  0.66 &    0.02 &    0.05 &    0.05 &    0.03 &    0.04 & &  16.7 &   7.5 &  15.8 &  0.03 \\
 $-$ & $ 0.66 -  1.66$ & $ 0.02 -  0.02$ & $ 0.04 -  0.05$ & $ 0.05 -  0.05$ & $ 0.03 -  0.03$ & $0.037 - 0.039$ & &  $ 16.1 -  18.7$ & $  4.9 -   8.0$ & $ 15.8 -  15.8$ & $ 0.03 -  0.03$ \\
\\
\multicolumn{12}{c}{Fit to \ci/\twelveco\ ratios with exponential beam correction} \\
%10 &  0.04 &    0.40 &    1.18 &    1.45 &    1.04 &    0.62 & &  21.9 &   4.8 &  17.0 &  0.05 & 17 \\
%$-$ & $ 0.04 -  0.54$ & $ 0.00 -  0.60$ & $ 0.00 -  2.27$ & $ 0.00 -  4.05$ & $ 0.00 -  4.79$ & $0.000 - 0.620$ & &  $ 18.2 -  79.5$ & $  4.0 -   8.0$ & $ 14.0 -  17.7$ & $ 0.02 -  0.20$ & $15 - 30$ \\
% 18/11/2016
%10 &  0.20 &    0.20 &    0.60 &    0.74 &    0.52 &    0.31 & &  21.9 &   4.8 &  16.7 &  0.05  \\
% $-$ & $ 0.20 -  0.70$ & $ 0.00 -  1.09$ & $ 0.00 -  2.83$ & $ 0.00 -  4.69$ & $ 0.00 -  5.17$ & $0.000 - 1.530$ & &  $ 15.1 -  66.1$ & $  4.0 -   8.0$ & $ 14.0 -  17.7$ & $ 0.02 -  0.20$ \\
% 1/3/2017
%10 &  0.36 &    0.40 &    1.18 &    1.45 &    1.04 &    0.62 & &  21.9 &   4.8 &  17.0 &  0.04 \\
% $-$ & $ 0.36 -  0.84$ & $ 0.07 -  1.09$ & $ 0.56 -  2.83$ & $ 0.69 -  4.69$ & $ 0.50 -  5.17$ & $0.275 - 1.530$ & &  $ 18.2 -  79.5$ & $  4.0 -   8.0$ & $ 16.7 -  17.7$ & $ 0.03 -  0.30$ \\
% 9/3/2017 finer RADEX grid now with chi2min+1
%10 &  0.03 &    0.19 &    0.67 &    0.93 &    0.70 &    0.30 & &  26.3 &   4.5 &  16.8 &  0.05 \\
% $-$ & $ 0.03 -  1.03$ & $ 0.00 -  0.94$ & $ 0.00 -  2.86$ & $ 0.00 -  4.32$ & $ 0.00 -  4.25$ & $0.000 - 1.927$ & &  $ 16.1 -  44.5$ & $  4.0 -   8.0$ & $ 14.0 -  17.6$ & $ 0.02 -  0.15$ \\
% 9/3/2017 finer RADEX grid now with chi2min+1, and REAL (i.e., difference between NCI and NCO12, rather than approximate values used in @abundance)
10 &  0.03 &    0.19 &    0.67 &    0.93 &    0.70 &    0.30 & &  26.3 &   4.5 &  16.8 &  0.06 \\
 $-$ & $ 0.03 -  1.03$ & $ 0.00 -  0.94$ & $ 0.00 -  2.86$ & $ 0.00 -  4.32$ & $ 0.00 -  4.25$ & $0.000 - 1.927$ & &  $ 16.1 -  44.5$ & $  4.0 -   8.0$ & $ 14.0 -  17.6$ & $ 0.03 -  0.16$ \\
% 7/3/2017
%20 &  0.24 &    0.20 &    0.60 &    0.74 &    0.52 &    0.31 & &  21.9 &   4.8 &  17.0 &  0.04 \\
% $-$ & $ 0.24 -  0.72$ & $ 0.00 -  0.53$ & $ 0.00 -  1.94$ & $ 0.00 -  2.88$ & $ 0.00 -  2.71$ & $0.000 - 0.770$ & &  $ 18.2 -  45.7$ & $  4.0 -   8.0$ & $ 14.0 -  17.7$ & $ 0.02 -  0.10$ \\
% 9/3/2017 finer RADEX grid now with chi2min+1
% note that 0.07 is also consistent (within RADEX model grid resolution) with CI X
%20 &  0.03 &    0.10 &    0.68 &    1.14 &    0.94 &    0.20 & &  38.3 &   4.1 &  17.2 &  0.10 \\
% $-$ & $ 0.03 -  1.03$ & $ 0.00 -  0.97$ & $ 0.00 -  2.89$ & $ 0.00 -  3.88$ & $ 0.00 -  3.64$ & $0.000 - 1.546$ & &  $ 16.1 -  44.5$ & $  4.0 -   8.0$ & $ 14.0 -  17.8$ & $ 0.02 -  0.15$ \\
% 9/3/2017 finer RADEX grid now with chi2min+1, and REAL (i.e., difference between NCI and NCO12, rather than approximate values used in @abundance)
20 &  0.03 &    0.10 &    0.68 &    1.14 &    0.94 &    0.20 & &  38.3 &   4.1 &  17.2 &  0.10 \\
 $-$ & $ 0.03 -  1.03$ & $ 0.00 -  0.97$ & $ 0.00 -  2.89$ & $ 0.00 -  3.88$ & $ 0.00 -  3.64$ & $0.000 - 1.546$ & &  $ 16.1 -  44.5$ & $  4.0 -   8.0$ & $ 14.0 -  17.8$ & $ 0.03 -  0.16$ \\
\\
\multicolumn{12}{c}{Fit to \ci/\twelveco\ ratios with Gaussian beam correction$^{\mathrm f}$ } \\
10 &  0.01 &    0.34 &    1.09 &    1.43 &    1.08 &    0.48 & &  24.4 &   4.6 &  17.0 &  0.05 \\
 $-$ & $ 0.01 -  1.01$ & $ 0.00 -  1.19$ & $ 0.00 -  3.88$ & $ 0.00 -  6.06$ & $ 0.00 -  6.40$ & $0.000 - 1.927$ & &  $ 15.5 -  44.5$ & $  4.0 -   8.0$ & $ 14.0 -  17.8$ & $ 0.03 -  0.25$ \\
20 &  0.01 &    0.24 &    1.10 &    1.70 &    1.46 &    0.34 & &  32.9 &   4.2 &  17.4 &  0.10 \\
 $-$ & $ 0.01 -  1.01$ & $ 0.00 -  1.22$ & $ 0.00 -  3.93$ & $ 0.00 -  6.09$ & $ 0.00 -  6.58$ & $0.000 - 1.537$ & &  $ 15.5 -  44.5$ & $  4.0 -   8.0$ & $ 14.0 -  18.2$ & $ 0.03 -  0.25$ \\
\\
\hline
\end{tabular}
}
}
\vspace{0.5\baselineskip}
\begin{description}
\item
[$^{\mathrm{a}}$] 
The parameter ranges given in the second line for all sets of fits correspond to 
the full range found over the range in $\chi^2$ from $\chi^2_{\rm min}$ to
$\chi^2_{\rm min}+1$ %($\chi^2_{\rm min}$ to $\chi^2_{\rm min}+0.5$ for \ci), 
i.e., marginalizing over the full range of posteriors.
\item
[$^{\mathrm{b}}$]
The intensity model gives a best-fit Gaussian FWHM 
$\theta_s$\,=\,31\arcsec\ (\dv\,=\,10\,\kms), and
$\theta_s$\,=\,29\arcsec\ (\dv\,=\,20\,\kms).
%This fit was obtained using the ``(1--0)" line ratios,
%and the exponential distribution best-fit folding radius of 3\arcsec\ has been converted to
%a FWHM measurement given for $\theta_s$ as described in Appendix \ref{sec:appendix};
\item
[$^{\mathrm{c}}$] 
``Set 1" was taken with the ratios relative to the adjacent lower-$J$ line:
%%(hereafter ``assorted'') 
%(hereafter ``Set 1'':
[\cotwo/\coone, \cothree/\cotwo, \cofour/\cothree, \thirteencoone/\coone, \thirteencotwo/\thirteencoone].
The ``other'' sets of line ratios with various $J$ transitions in the numerators/denominators give 
virtually identical results so we report here only a representative example with
[\cotwo/\coone, \cothree/\coone, \cofour/\coone, \thirteencotwo/\thirteencoone, \cotwo/\thirteencotwo]. 
%Two used assorted ratios with the higher-$J$ lines relative to either \cotwo\ or \coone\ 
%(hereafter ``Sets 2, 3'');
%and the last took 
%the \twelveco\ lines relative to \coone, and \thirteencotwo\ line relative to \thirteencoone;
%%(hereafter ``(1--0)'') 
%(hereafter ``Set 4''). 
%The exponential folding radii $r_s$ have been converted to
%the FWHM measurement given for $\theta_s$ as described in Appendix \ref{sec:appendix};
\item
[$^{\mathrm{d}}$]
The Set 1 line ratio models give a best-fit Gaussian FWHM $\theta_s$\,=\,27\arcsec\ (\dv\,=\,10\,\kms), and
$\theta_s$\,=\,26\arcsec\ (\dv\,=\,20\,\kms). 
\item
[$^{\mathrm{e}}$]
The \cione\ line ratio models give a best-fit Gaussian FWHM $\theta_s$\,=\,26\arcsec\ (\dv\,=\,10\,\kms), and
$\theta_s$\,=\,27\arcsec\ (\dv\,=\,20\,\kms).
\item
[$^{\mathrm{f}}$]
The \cione\ line ratio models give a best-fit Gaussian FWHM $\theta_s$\,=\,23\arcsec\ (\dv\,=\,10\,\kms), and
$\theta_s$\,=\,24\arcsec\ (\dv\,=\,20\,\kms).
%\item
%[$^{\mathrm{d}}$]
%This fit was obtained using \cione\ to \twelveco\ line ratios. 
\end{description}
\end{table*}
\end{center}
% -----------------------------------------------------------------

Table \ref{tab:radexn1140} gives the best-fit parameters obtained for NGC\,1140.
%the nominal best fits correspond to line-ratio Sets 1 and 2. 
%As reported in Table \ref{tab:radexn1140}, 
The different fitting methods 
and even the different beam correction approaches
give very similar results, 
namely cool temperatures, high volume densities, 
low CO opacities, and low \twelveco/\thirteenco\ abundance ratios.
The main difference among the fits is the CO column density, \ntwelveco, which
for a given $\Delta V$ varies by a factor of $\sim$4. 
Because the opacity in the RADEX models $\tau\propto N_{\rm co}/\Delta V$, and $\tau$ governs the line
ratios and thus must remain relatively constant, 
for higher $\Delta V$, \ntwelveco\ is consequently higher %commensurately higher, 
within the resolution of our RADEX grid.
For higher \ntwelveco,
the filling factor $F$ is also commensurately lower, which we will discuss in
Sect. \ref{sec:ff}.
In any case, the low values of $\chi^2$ imply that a one-zone spherical cloud with uniform physical
properties gives a good fit to the molecular gas emission in NGC\,1140.
The extremely low \twelveco/\thirteenco\ abundance ratio of $\sim 8 - 20$
is obtained for NGC\,1140 even with a non-LTE analysis; 
we discuss this unexpected result further in Sect. \ref{sec:fractionation}.

Figure \ref{fig:chisquaren1140} shows the $\chi^2$ surface for the Set 1 line ratios\footnote{``Set 1" was 
taken with the ratios relative to the adjacent lower-$J$ line, specifically:
[\cotwo/\coone, \cothree/\cotwo, \cofour/\cothree, \thirteencoone/\coone, \thirteencotwo/\thirteencoone].
See Table \ref{tab:radexn1140} for more details.}
with the exponential beam correction
as a function of the fitted
parameters (\tkin\ vs. \nhtwo\ in the left panel, 
%and $\theta_s$ vs. $N_{\rm 13CO}$ in the right).
%and $r_s$ vs. $N_{\rm 13CO}$ in the right).
and \ntwelveco\ vs. $\tau$[\coone] in the right).
In both panels, the error bars correspond to $\chi^2 \la 1$ at the best-fit value,
rather than the full posterior range given in Table \ref{tab:radexn1140}.
%{\bf Even though the $\chi^2$ surface is for the Set 1 line ratios,
%the minimum for Set 2 is visible in the right panel at $\tau\sim$0.06 and \ntwelveco$\sim 10^{16}$\,\cmtwo.
%}
It can also be seen from Fig. \ref{fig:chisquaren1140} that the parameters are fairly
well constrained by the model; specifically, %the line ratios require that
the volume density \nhtwo\ cannot be much lower than %the lowest of the best-fit values ($2\times10^5$\,\cmthree)},
$\sim 10^5$\,\cmthree,
and because of the well-known degeneracy between \tkin\ and \nhtwo,
the temperature also cannot be much greater than $\sim$20\,K. %the best-fit value of $\sim$18\,K.

Figure \ref{fig:chisquarencolsizen1140} gives the $\chi^2$ surface for the Set 1 line ratios
using the Gaussian beam correction plotted against the \twelveco\ column density
$N_{\rm 12CO}$ and the Gaussian FWHM $\theta_s$.
The fit is somewhat sensitive to the assumed $\theta_s$;
for $\theta_s \la$15\arcsec\ and for $\theta_s \ga$34\arcsec, $\chi^2>$ 10, and thus is not plotted.
The best-fit Gaussian sizes range from 26\arcsec\ to 31\arcsec,
implying that under the assumption of a Gaussian distribution,
the macroscopic size of the emission is $\ga$2\,kpc.
This would imply that the CO emission is more extended than the dust
(FWHM($r_s$) $\sim$850\,pc, see Appendix \ref{sec:appendix_exponential}),
which seems unlikely.
%Instead, the molecular gas is probably less concentrated than a Gaussian,
%possibly more extended in an exponential like the dust.
In any case, even the largest CO beam is sampling less than half of the galaxy ISM;
as shown in Fig. \ref{fig:beam_ngc1140}, 
50\% of the dust emission lies within an effective radius (diameter) of $\sim$20\arcsec\ (40\arcsec, 3.8\,kpc).

It should be emphasized that small $\chi^2$ values are not easy to achieve, even
with several trials with many different formulations of beam corrections, in addition
to the ones described in Sect. \ref{sec:beam} and in the Appendices.
The small $\chi^2$ values of the fits in Table \ref{tab:radexn1140} are telling us that
for reasonable assumptions for beam corrections, the single-zone RADEX
models can well accommodate the data.
Moreover,
the physical conditions inferred from the different fitting techniques are mutually consistent,
independently of the kind of beam correction.
Because of this similarity, in what follows, we will use the
exponential beam corrections described in Appendix \ref{sec:appendix_exponential};
the virtue of these is that they can be measured or estimated for all the galaxies
observed, under the assumption that on kpc scales the molecular gas is following the dust.

\subsubsection{Filling factor}
\label{sec:ff}

The mean \twelveco\ line width of the various components in NGC\,1140 (see Table \ref{tab:n1140}) is $\sim21$\,\kms;
we thus prefer the RADEX fits with $\Delta V$\,=\,20\,\kms.
For $\Delta V$\,=\,20\,\kms,
fitting the intensities as in Eqn. (\ref{eqn:chisqintensities}) gives a best-fit filling factor $F$ of $0.10\,\pm\,0.06$.
This best-fit $F$ is rather large, but related to the somewhat low \ntwelveco\ value of $10^{15.8}$\,\cmtwo; %dex(15.8)\,\cmtwo;
as discussed below, the fits of the line ratios as in Eqn. (\ref{eqn:chisqratios}) give 
higher \ntwelveco\ thus consequently lower, possibly more realistic, values of $F$.

For the line-ratio fits,
the best-fit %(Set 1) 
RADEX model for NGC\,1140 predicts a velocity-integrated brightness temperature 
% 5/3/2017
% see readme_5
$T_B($\coone$)$\,=\,21.0\,\kkms; from Table \ref{tab:n1140}, %\ref{tab:corrections}, 
we have a beam-corrected velocity-integrated \tmb\ of 0.56\,\kkms.
Dividing the two values gives a nominal filling factor of $\sim$0.03.
%and considering the line width $\Delta \varv$ of 10\,\kms\ used for the RADEX models,
%we would obtain $F\sim$0.016.
Similar filling factors are derived for the other transitions including \thirteenco, 
as a consequence of the assumption of a single-zone model.
If we compare with the model from the ``other'' $\Delta V$\,=\,20\,\kms\ line-ratio fits
(see Table \ref{sec:radexn1140}), with lower \ntwelveco\ and $\tau$
(see Table \ref{tab:radexn1140}), we obtain a slightly higher value, $F\sim$0.04. 
Using the main-beam peak temperatures rather than velocity-integrated values (see Table \ref{tab:n1140}), and 
adjusting to the common 22\arcsec\ beam size, % by applying the factor given in Table \ref{tab:corrections}.
%For \coone, in 5.2\,\kms\ wide channels, 
%\tmb(peak)$\sim$10\,mK; the beam correction factor for the \coone\ beam is roughly unity. 
%% dv=20 --> RADEX TB = 21.00 K.km/s/20 km/s = 1.05 K; observed TB = 10mK
%Comparing this with the best-fit RADEX model, divided by $\Delta V$, 
%we find a filling factor of $\sim$0.01.
%For \cofour, in 10.2\,\kms\ channels,
%\tmb(peak)$\sim$27\,mK, but for this transition the correction factor to a 22\arcsec\ beam is 0.68;
%thus correcting to a common beam size as used in the models, \tmb(peak)$\sim$18\,mK.
%The best-fit RADEX model predicts 16.37\,\kkms\ for 10\,\kms\ line widths, so
%The best-fit RADEX model predicts 27.0\,\kkms\ for 10\,\kms\ line widths, so
%The best-fit RADEX model predicts \cofour\ to be 12.3\,\kkms\ for 10\,\kms\ line widths (\tb$\sim$1.2\,K), so
gives filling factors of $\sim$0.01.
%and accounting for the velocity filling factor
%($\Delta\varv$\,=\,0.52, the ratio of the channel width of the observations and the RADEX $\Delta\varv$),

These filling factors of a few percent for the molecular gas in NGC\,1140
are similar to those found from high-resolution ALMA maps
of 30 Doradus in the Large Magellanic Cloud \citep[LMC,][]{indebetouw13}.
They are also similar to those obtained by a similar $\chi^2$ analysis
of non-LTE models of single-dish line ratios in the SMC and the LMC \citep{nikolic07};
these last are as high as $\sim$0.14 (for what they consider to be good fits with $\chi^2 \la 2$), consistent with $F\,=\,0.1$ 
for NGC\,1140 obtained from the intensity fitting.

\begin{figure}[h!]
\vspace{\baselineskip}
\hbox{
\centerline{
%\includegraphics[angle=0,width=\linewidth]{NGC1140_COCoolingCurve_NewRatios-crop.pdf}
%\includegraphics[angle=0,width=\linewidth]{NGC1140_COCoolingCurveJykms_NewRatios-crop.pdf}
%\includegraphics[angle=0,width=\linewidth]{NGC1140_COCoolingCurveJykms_nu2_NewRatios_FFEXP-crop.pdf}
% 8/3/2017
\includegraphics[angle=0,width=\linewidth]{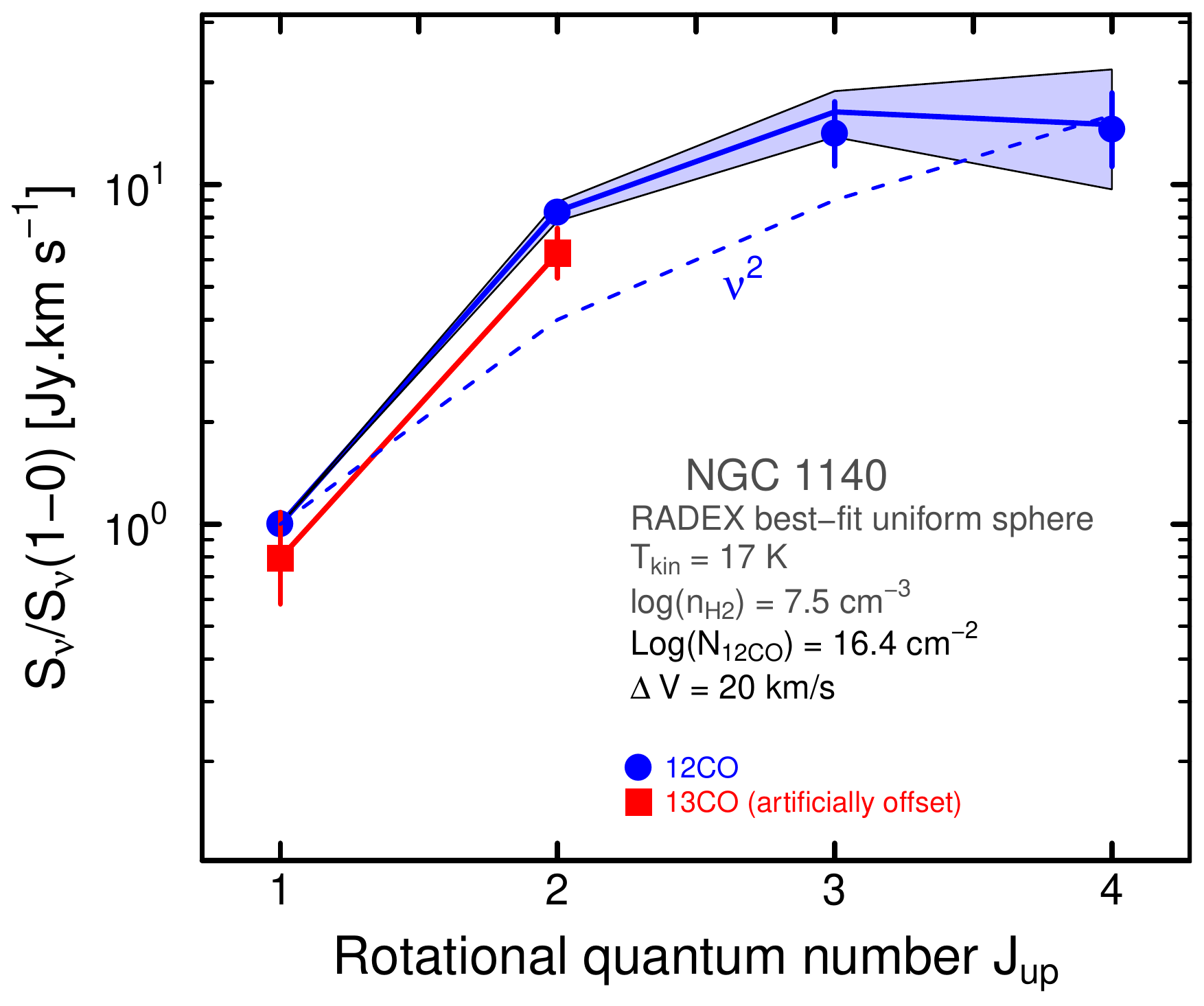}
}
}
\caption{RADEX best-fit model compared to observations for the CO excitation ladder of NGC\,1140;
the observations have been corrected for beam size as described in Sect. \ref{sec:beam}, 
and normalized to the CO(1--0) flux.
The observations are shown 
as filled blue circles (\twelveco) and red squares (\thirteenco), and
%as the line segments, and 
the range of the allowed RADEX models by the light-blue region (for \twelveco);
\thirteenco\ points have arbitrarily offset relative to \twelveco\ for clarity.
%The filled (blue) circles correspond to \twelveco, and the filled (red) squares
%to \thirteenco\ 
The dashed curve shows the $\nu^2$ behavior (here normalized to $J\,=\,1$) expected
for optically-thick emission.
}
\label{fig:cocooling}
\end{figure}

\begin{figure*}[ht!]
% \begin{figure*}[H]
%\vspace{\baselineskip}
%\begin{center}
%\begin{minipage}{\textwidth}
\hbox{
\centerline{
\includegraphics[angle=0,width=0.45\linewidth]{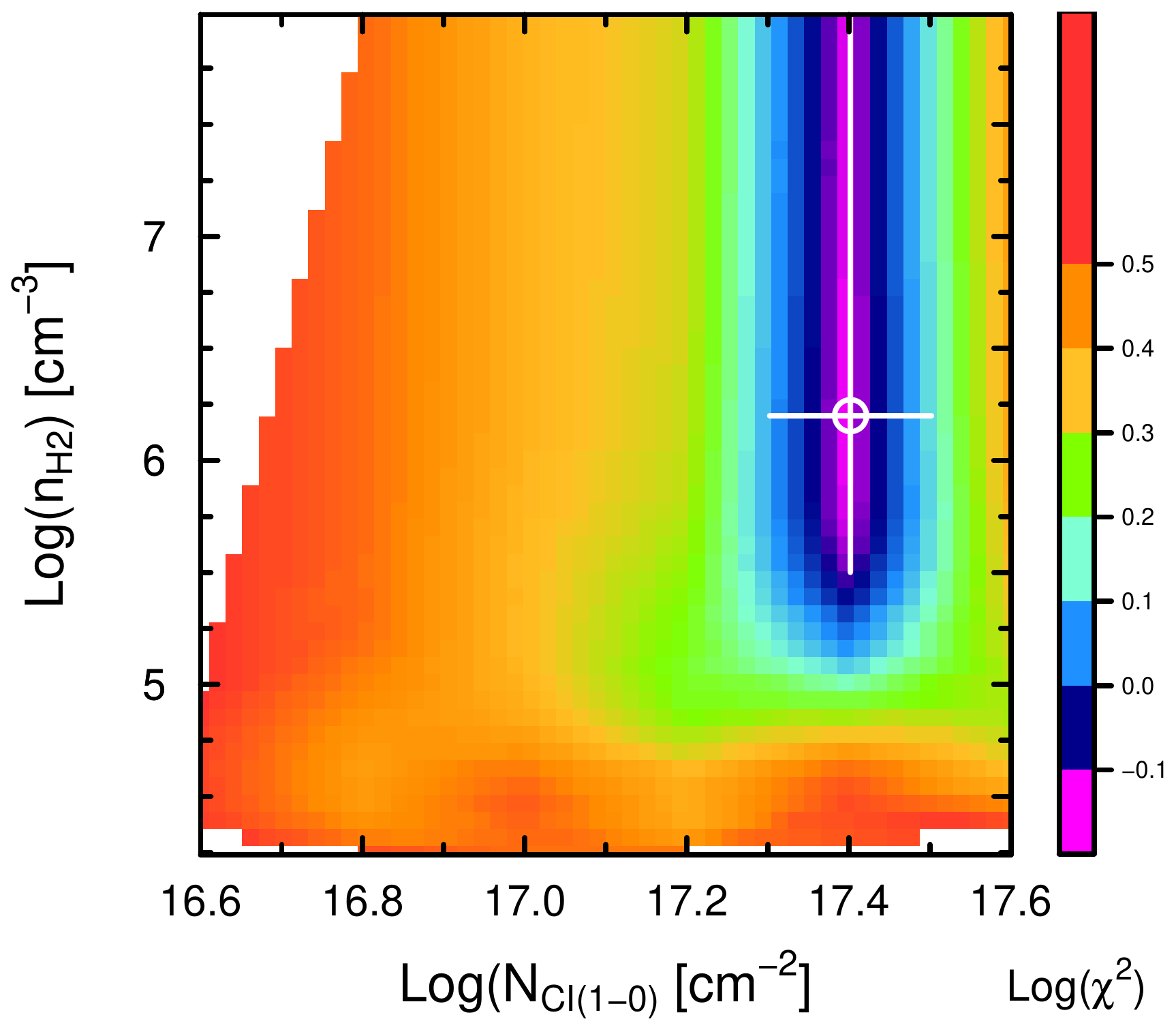}
\hspace{0.02\linewidth}
\includegraphics[angle=0,width=0.45\linewidth]{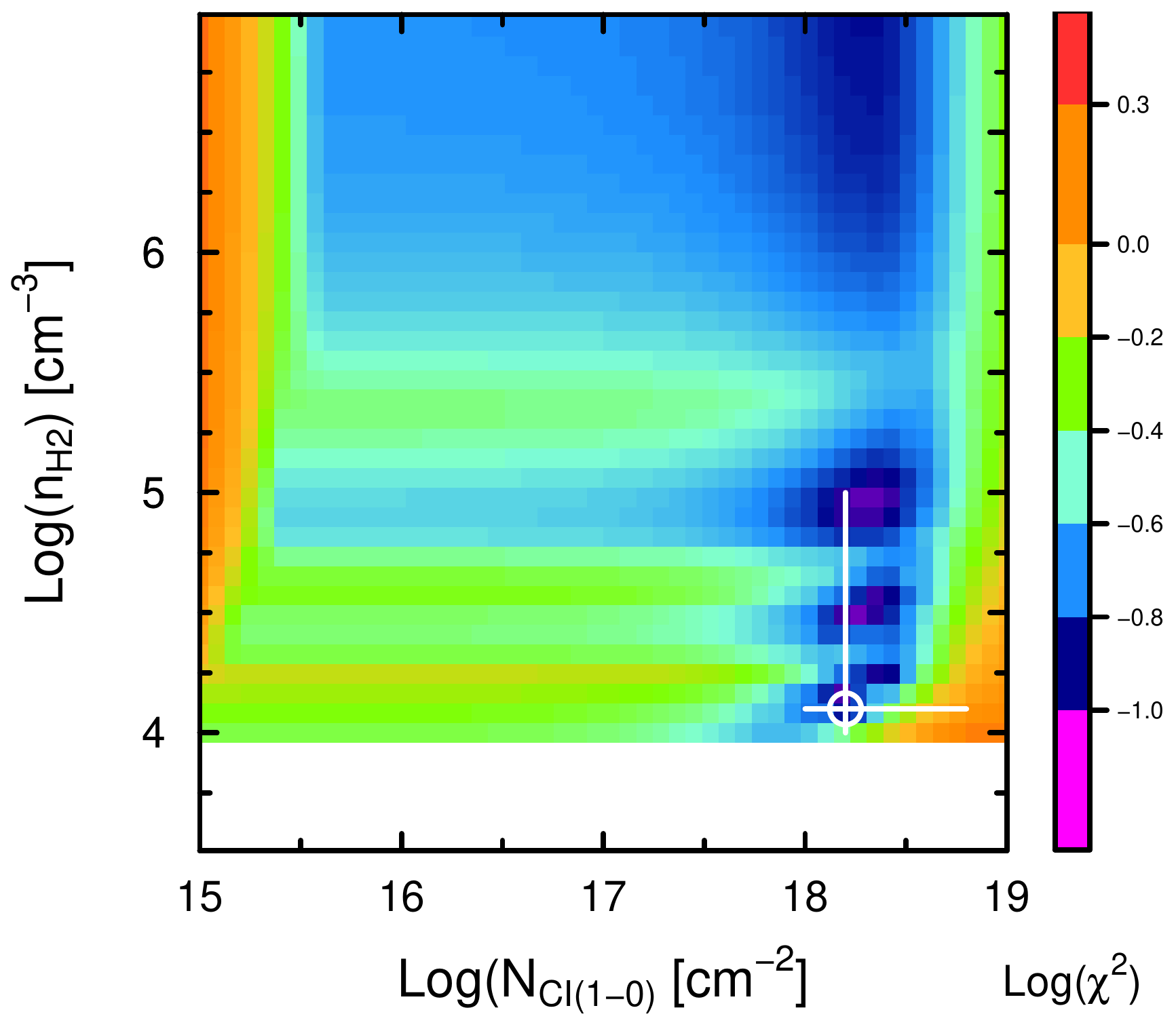}
}
}
%\vspace{-2\baselineskip}
%\captionof{figure}{...}\label{visina8}
\caption{RADEX fit of the $\chi^2$ (logarithmic) surface of the \ci/\twelveco\ fits 
as a function of \nhtwo\ and $N_{\rm CI(1-0)}$;
the left panel shows the intensity fits (Eqn. \ref{eqn:chisqintensities}) 
and the right the ratio ones (Eqn. \ref{eqn:chisqratios}). 
The RADEX models shown here have $\Delta V$\,=\,20\,\kms.
The associated Log($\chi^2$) values are shown as a side-bar color table.
The best-fit values are shown as an open (white) circle;
the error bars shown consider a slightly more limited range in $\chi^2$
respect to  Table \ref{tab:radexn1140}. 
%The magenta-colored region with the smallest $\chi^2$ values (0.04-0.08)
%lies beneath the size of the open circle showing the best fit.
}
\label{fig:chisquarecin1140}
\end{figure*}
%\end{center}
%\end{minipage}

\subsubsection{CO excitation ladder}
\label{sec:cooling}

The modeled CO excitation ladders for NGC\,1140 are shown in Fig. \ref{fig:cocooling}.
As in Fig. \ref{fig:radexlineratiosn1140}, 
the observations have been corrected for beam size as described in Sect. \ref{sec:beam}
and Appendix \ref{sec:appendix_exponential}.
The line segments show the best-fit RADEX model for both \twelveco\ and \thirteenco\ transitions
adopting the best fit for the Set 1 line ratios with $\Delta V$\,=\,20\,\kms;
the blue area corresponds to the range of ratios allowed by the models
(for $\chi^2$\,=\,$\chi^2({\rm min})+1$).
%The similarity of the \thirteenco\ line ratios to the \twelveco\
%suggests that the physical conditions of the gas emitting the rarer isotope 
%are similar to those for \twelveco.
% from Santi
The similar \twelveco\ and \thirteenco\ slopes for NGC\,1140 (the data points
follow parallel lines) suggest that even the \twelveco\ emission may be optically
thin, as also implied by the best-fit results in Table \ref{tab:radexn1140}.
%and the X[\twelveco/\thirteenco] abundance ratio may be unexpectedly low.
The CO gas is in an excited state, more similar to the center of M\,82
\citep[][]{weiss05} than to the Milky Way \citep[e.g.,][]{carilli13}.
This is not a surprising result given the population of SSCs in the
nuclear region of NGC\,1140.

The low optical depths of even the \twelveco\ lines steepen the 
%cooling curve 
excitation ladder
at low $J$, steeper than $\nu^2$ as would be expected for 
thermalized, optically thick lines. %in LTE.
The shape is an indirect consequence of the normalization
to $J$=1--0, due to the increase of \tauco\ with increasing $J$ up to the inflection in the excitation.
Because \coone\ has the smallest $\tau$ of the \twelveco\ transitions,
with a different normalization, it would drop below the $\nu^2$ trend.
The results of the RADEX fit (see Table \ref{tab:radexn1140}) are telling us that 
the $J=4$ level is sparsely populated because of the decreased \tauco\
relative to the lower-$J$ transitions; this produces the flattening in the
excitation ladder.
%cooling curve.

\subsubsection{Atomic carbon}
\label{sec:carbon}

We have performed 
additional RADEX fits to the \twelveco\ and \cione\ emission,
both with intensities as in Eqn. (\ref{eqn:chisqintensities}) and with
ratios (Eqn. \ref{eqn:chisqratios}).
Fitted parameters include, as before,
(\nhtwo, \tkin, \nco(\twelveco), $F$), but now [\twelveco]/[\twelvec], rather than [\twelveco]/[\thirteenco].
%using the same methodology as described above (here there are four independent ratios,
%rather than five).
Here, again, the implicit assumption is that atomic carbon and CO occupy the same volume, 
and are uniformly intermixed within an ISM of a single \htwo\ kinetic temperature and volume density;
such an assumption is consistent with the findings of \citet{requena16} for
\hii\ regions in the SMC and of \citet{okada15} for the LMC.
The results of the fits are given in Table \ref{tab:radexn1140} and shown graphically
in Fig. \ref{fig:chisquarecin1140} where the $\chi^2$ surfaces (in logarithmic terms)
are shown as a function of \nhtwo\ and $N_{\rm CI(1-0)}$ column density.
As for the previous CO analysis, we prefer the $\Delta V$\,=\,20\,\kms\ fits, 
roughly the mean component line width observed in NGC\,1140.

%As expected because the data are dominated by CO transitions,
The intensity fit of the \twelveco\ and \ci\ emission gives parameters
very similar to the CO fit, although the inferred \nhtwo\ for the \cione\ fit is lower.
On the other hand, the fit to the \cione\ ratios (with \twelveco) gives higher \tkin,
lower \nhtwo, and 25 times higher \nco, higher than any other fit we performed.
Folding in the best-fit [\twelveco]/[\twelvec] abundance ratio, which varies by a factor of
$\sim$3 between the two fits, shows that the best-fit column density of atomic carbon $N_C$ 
is also different for both fits:
%$\sim$dex(17.3)\,\cmtwo\ in the intensity case and
%$\sim$dex(18.2)\,\cmtwo\ for the ratios.
$\sim 10^{17.3}$\,\cmtwo\ in the intensity case and
$\sim 10^{18.2}$\,\cmtwo\ for the ratios.
Consequently, the filling factors for the two fits also differ.
The best-fit $F$ for the \ci\ intensity fits is 0.10 as for (the intensity fitted) CO, but for the \ci\ ratio
fits, it is much lower, $F\la$0.01.
This is because the higher \nco\ requires a lower filling factor to accommodate
the RADEX predictions with the observations.
We found a similar behavior in Sect. \ref{sec:radexn1140} for the CO ratio fits relative to the intensity ones.
The reasons for the disagreement are unclear,
although
the low values of $\chi^2$ for all the \ci\ fits make it unlikely that the model
is failing to capture the physical conditions in the clouds.
In any case, these values of $F$ are between 2 and 20 times lower than
those found for \ci\ emission in the Magellanic Clouds by \citet{pineda17};
however their observations probe physical scales more than 50 times smaller,
so a comparison is not straightforward.

Independently of the disagreement, our RADEX fits for \ci\ all find 
that the atomic carbon column density must be larger than \nco,
implying a [\twelveco]/[\twelvec] abundance of $<$1.
For the intensity fit, with lower \nco, this requires a very low [\twelveco]/[\twelvec]
of $\sim$0.03, while for the ratio fit with higher \nco, [\twelveco]/[\twelvec]$\sim$0.1.
This would indicate that 
{\it atomic carbon is at least 10 times more abundant than \twelveco}
in NGC\,1140. 
Such a result would be consistent with the theoretical predictions of \citet{bialy15} for
interstellar ion-molecule gas-phase chemistry in galaxies;
their models predict that [\twelveco]/[\twelvec] abundance ratios decrease
with decreasing metallicity and with higher ionization parameters.
On the other hand, observational studies of luminous (metal-rich) galaxies found
that atomic carbon is {\it less} abundant than CO, [\twelveco]/[\twelvec] $\sim 4-8$
\citep[e.g.,][]{israel09,israel15}. 
However, the observed \cione/\cofour\ ratio (in flux units) for NGC\,1140 ($\sim 0.7$) is more than 4$\sigma$
higher than the mean of the galaxies observed by \citet{israel15}, so the relatively high
abundance of atomic carbon is perhaps not altogether surprising.
Although our \cione\ detection is marginal ($\sim 3\sigma$, see Table \ref{tab:n1140}),
to reduce the inferred [\twelveco]/[\twelvec] ratio to unity or less would require a reduction in
the \cione\ intensity by more than a factor of 10; this seems unlikely given the trends shown
in Fig. \ref{fig:lineratiosci}.

%Interestingly, 
Independent one-zone model fitting of the
\twelveco, \thirteenco, and \cione\ emission of NGC\,1140
gives similar physical conditions. 
Specifically, by fitting a vast grid of LVG models 
\citep{weiss07}
to the NGC\,1140 data,
and considering dust continuum, and molecular and atomic abundances relative to \htwo, 
we obtain
high densities (\nhtwo\,$\ga 10^5$\,\cmthree),
cool temperatures (\tkin\,$\sim$20\,K),
low [\twelveco]/[\thirteenco] abundance ratios ($\sim 13$),
low [\twelveco]/[\twelvec] abundance ratios ($\sim 0.13$),
and moderate column densities (\nco\,$\sim 5\times10^{16}$\,\cmtwo),
similar to the RADEX results.
However, with these models, a ``warm" solution
with \tkin\,$\sim$52\,K is also possible. 
This solution would result in higher abundance ratios
([\twelveco]/[\thirteenco] $\sim 36$, [\twelveco]/[\twelvec] $\sim 0.3$),
and higher \nco\ ($\sim 7\times10^{17}$\,\cmtwo).
Nevertheless, in this case there would be an unrealistically low gas-to-dust mass
ratio (GDR), $\sim$62;
this is lower than even the Milky Way \citep[GDR\,$\sim$ 100,][]{draine07sings},
which is unlikely considering the low metallicity of NGC\,1140.
With \zzsun\,$\sim$0.3 for NGC\,1140, and given the linear trend of GDR with metallicity in this abundance range \citep[e.g.,][]{remy14},
we would expect a GDR of $\sim$300,
not far from the GDR of $\sim$380, inferred from the ``cool'' solution
obtained with the \citet{weiss07} LVG models. 
%of Weiss et al.
Thus, the ``cool" solution with \tkin\,$\sim$\,20\,K seems more likely for NGC\,1140. 

The ``warm'' %Weiss et al. (2017) 
solution with the \citet{weiss07} LVG models for \twelveco, \thirteenco, and \cione\
is not far from the ``warm'' solution (with \tkin\,$\sim 38$\,K) 
that emerged from the RADEX fitting of \cione\ and \twelveco\ line ratios
(see Table \ref{tab:radexn1140}).
However, our RADEX models for \thirteenco\ and \twelveco\ do not allow the higher \tkin\ found with
the \citet{weiss07} models; we obtain $\chi^2 > 10$ for \tkin\,$> 46$\,K, 
and a mean $\chi^2$ of $\sim 6-7$ for $40 \leq$\,\tkin\,$\leq 46$\,K,
compared to $\chi^2$ values of $< 1$ for our best fits.
Based on the much lower $\chi^2$ values for the CO fits with cool \tkin,
we do not consider further the ``warm" RADEX fits. 
Observations of higher-$J$ CO lines and \citwo\ would help settle this potential ambiguity. 

Combined with the [\twelveco]/[\thirteenco] abundance ratio of $\sim 10-12$ found above
(see also Sect. \ref{sec:lte}), this would imply a [\twelvec]/[\thirteenco] 
abundance ratio of $\ga 100-300$,
within the range of values found by \citet{israel15} for galaxies with low CO/\cione] ratios.
However, this may be a specious comparison because of the unusually low 
[\twelveco]/[\thirteenco] abundance ratio which, in some sense, compensates for the extremely low  [\twelveco]/[\twelvec]
found here for NGC\,1140.
As already mentioned,
we discuss further the low [\twelveco]/[\thirteenco] abundance ratio found for NGC\,1140 in
Sect. \ref{sec:fractionation}.

\begin{figure}[ht!]
\vspace{\baselineskip}
\hbox{
\centerline{
\includegraphics[angle=0,width=\linewidth]{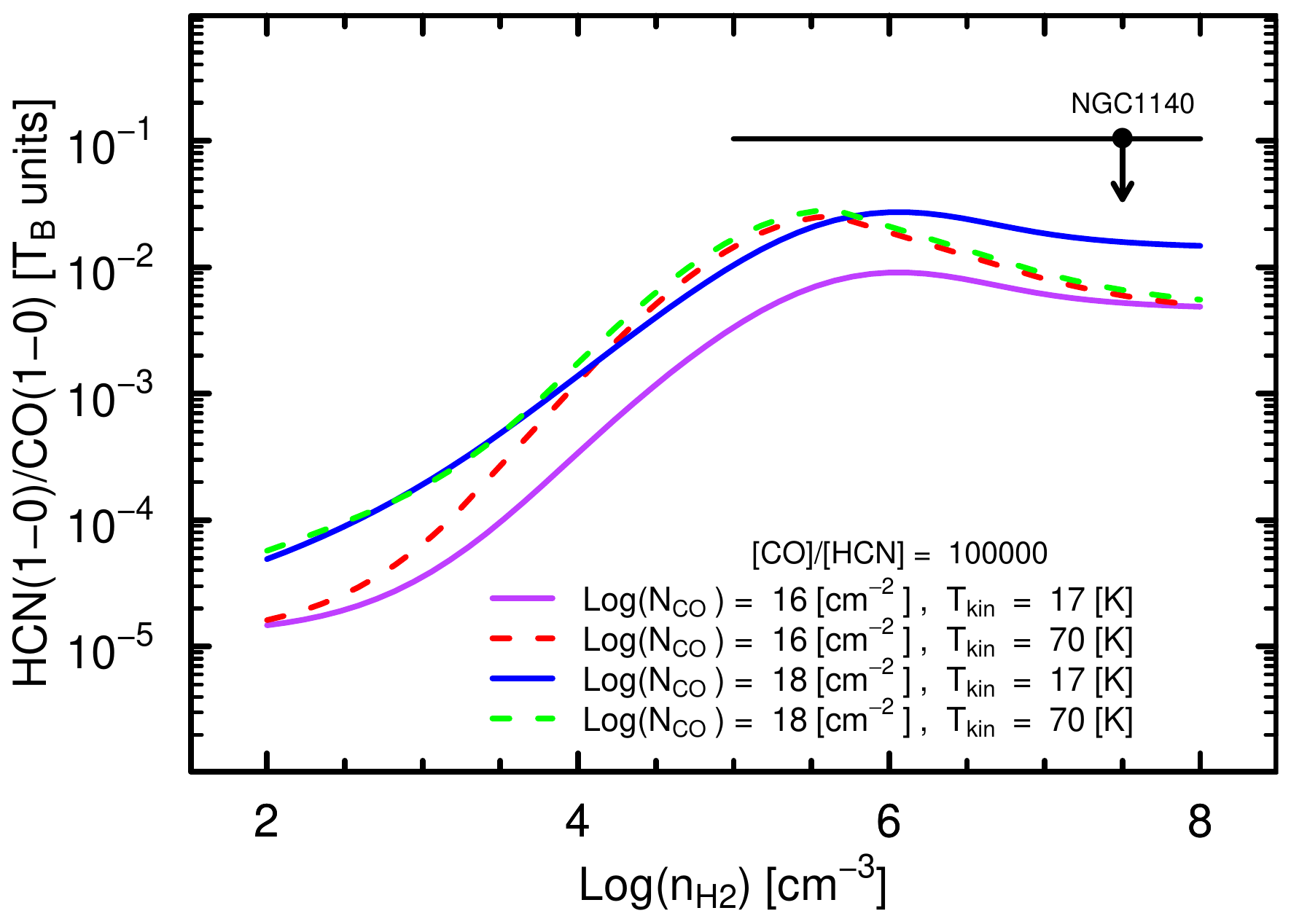}
}
}
\caption{RADEX models of \hcn/\coone\ line ratios plotted against \nhtwo, and
assuming [CO]/[HCN]$\,=\,10^5$.
Here the $\Delta V$\,=\,20\,\kms\ models are illustrated. 
The upper limit \hcn/\coone\ ratio for NGC\,1140, shown as a (black) filled circle,
has not been corrected for beam dilution because of the similar beam sizes for the
two transitions. 
The approximate best-fit RADEX model for NGC\,1140 is shown in both panels as (purple) solid curves
that are consistent with the non-detection of \hcn. 
Also shown are three additional parameter combinations of \twelveco\ column density $N_{12CO}$
and volume density \nhtwo.
}
\label{fig:radexhcn}
\end{figure}

\subsubsection{Modeling HCN with RADEX}
\label{sec:hcn}

We did not detect \hcn\ in NGC\,1140, even though the RADEX best fit suggests that
there is dense gas in this galaxy.
Figure \ref{fig:radexhcn} shows the predictions of RADEX models for the \hcn/\coone\
line ratio plotted against \htwo\ volume density, \nhtwo\ and
assuming [CO]/[HCN]$\,=\,10^5$.
The implicit assumption here, almost certainly incorrect, is that the filling factor of the
emitting gas is the same in both transitions.
Fig. \ref{fig:radexhcn} suggests that the abundance ratio [CO]/[HCN] must be high $\ga 10^5$.
%this is the abundance ratio used for the line ratios shown in Fig. \ref{fig:radexhcn}. 
Relative to the best-fit
RADEX model for NGC\,1140,  [CO]/[HCN]$\,=\,10^5$ corresponds to an HCN column density of $\sim 2\times10^{11}$\,\cmtwo,
an order of magnitude lower than in diffuse clouds in the Galaxy \citep{listz01}.

The upper limit (UL) for the \hcn/\coone\ line temperature ratio of NGC\,1140 is $\sim0.1$, higher
than the detected \hcn/\coone\ ratio of $\sim$0.05 in the prominent star-formation region N\,113 of the LMC \citep[][]{wang09}.
Thus, even after 18\,hr with the IRAM 30m, our observations are not able to
constrain the HCN content of NGC\,1140.
More observations of dense-gas tracers are needed at low metallicity, especially
outside the Local Group where conditions can be even more extreme than in the
massive star-forming regions in the Magellanic Clouds.

\section{Discussion}
\label{sec:discussion}

Unlike some previous work covering a similar range of CO $J$ values \citep[e.g.,][]{meier01,israel05,nikolic07}, 
our one-zone non-LTE models give very good fits to the 
%(four to five) independent line ratios
CO and atomic carbon emission in NGC\,1140.
%this is true both with the inclusion of the \thirteenco\ transitions and with atomic
%carbon \ci.
Part of this success 
is almost certainly due to the lack of high-$J$ transitions in our data set;
we are fitting only the cool, dense clouds in this galaxy, while there may be
a substantial contribution of warmer, more diffuse gas.
It may also be a consequence of the extreme conditions in this galaxy.
None of the six dwarf galaxies observed by \citet{cormier14} in \coone\ and \cotwo\ 
have \rtwo$>$1, independently of the beam size of the observations or the upper limits.
The N\,113 region in the LMC has \rtwo$\la$1 after beam deconvolution
\citep[\rtwo$\sim$1.2 before, see][]{wang09}, and
\citet{minamidani08} found maximum values of \rthree$\ga$1
in only a handful of the 33 LMC regions observed.
These ratios are much higher in NGC\,1140 (\rtwo\,=\,2.1, \rthree\,=\,1.7, see Figs. 
\ref{fig:lineratios21}, \ref{fig:lineratios32}, and Table \ref{tab:corrections}),
independently of the beam corrections applied.
Because of these high line ratios, and the implied more extreme physical conditions dominating the ISM of this galaxy,
one-zone models may be more effective. 

The CO excitation in NGC\,1140 is starburst-like, similar to M\,82, but as mentioned
before, this is not particularly surprising given the presence of at least 6 SSCs within
its central region \citep{degrijs04,moll07}.
However, unlike M\,82, where the gas is warm (\tkin$\sim 60-200$\,K) and not particularly dense
(\nhtwo$\sim 10^3$\,\cmthree) \citep{weiss05,muhle07}, in NGC\,1140 the gas is dense and cool.
Given the relatively high excitation, it is not straightforward to understand
why the CO has not been photo-dissociated through the intense radiation field,
although as discussed below, the dense, small-filling factor clouds may be
one part of the explanation.
High atomic gas columns may also contribute to self-shielding as discussed below
and suggested by \citet{wong09} for the LMC.

Finding evidence for cool, dense gas, \nhtwo$\ga 10^6$\,\cmthree\ in NGC\,1140 %, and also NGC\,7077, UM\,448 (and Mrk\,996),
was unexpected.
However, at least one other low-metallicity galaxy outside the Local Group shows similar properties.
In Haro\,11, PDR
modeling \citep{cormier12,cormier14} also suggests that the molecular gas is dense, with 
hydrogen densities $n_H$\,$\sim 10^{5-6}$\,\cmthree. %very similar to the best-fit value of 
%\nhtwo\,$\sim 10^{5.2}$\,\cmthree\ for NGC\,1140 (see Table \ref{tab:radexn1140}).
%\nhtwo$\sim 1.2 \times 10^5$\,\cmthree, very similar to the best-fit value of 
%$1.6 \times 10^5$\,\cmthree\ for NGC\,1140 (see Table \ref{tab:radexn1140}).
The \citet{cormier14} PDR model for Haro\,11 implies that CO emission is produced at intermediate
depths within the cloud ($3 < A_V < 6$\,mag), and for larger \av, the gas temperature is $\sim$30\,K,
somewhat warmer than the (non-LTE) gas temperature \tkin\ of $\sim$17\,K we find for NGC\,1140.

Another unexpected result is the low optical depth of the \twelveco\ transitions.
Our RADEX fits (see Table \ref{tab:radexn1140}, $\Delta V$\,=\,20\,\kms)
suggest that the \twelveco\
and \thirteenco\ transitions have a maximum \tauco\ $\sim 0.2$ for \cothree.
The best fit of \twelveco\ with \ci\ gives higher optical depths, with \tauco\ $\sim 1$ for \cothree\
but lower for the other transitions.
Apparently even these low optical depths allow sufficient self-shielding for detectable \twelveco\
and \thirteenco\ emission.
This could be possible because the clouds are dense and occupy only a small fraction
of the macroscopic source size (see Sect. \ref{sec:ff}).

It could also be that high 
%\hi\ 
column densities are contributing to the shielding of the
molecular gas.
\citet{hunter94} find a peak 
\hi\ column \nhi\ 
of $2.4\times10^{21}$\,\cmtwo\ (averaged over a 30\arcsec$\times$37\arcsec\ beam),
%equivalent to $\sim$38\,\msunpc.
%of $2.4\times10^{21}$\,\cmtwo,
$\sim$10 times higher than the \Nhtwo\ of $2\times10^{20}$\,\cmtwo\ estimated from \nco$\sim 10^{16}$\,\cmtwo,
assuming a roughly Solar abundance ratio [\htwo]/[CO]$\sim 5\times10^{-5}$ \citep[e.g.,][]{sakamoto99}.
Even the disk-averaged \nhi\ in NGC\,1140 of $\sim 10^{21}$\,\cmtwo\ 
\citep{fumagalli10} is 5 times higher than 
the \htwo\ column that we would infer from \nco.
As proposed by \citet{wong09}, high \hi\ 
columns may even be necessary (but not sufficient) for the detection of CO in low-metallicity
galaxies. 
However, the LVG abundance fits using the models of \citet{weiss07} indicate that
the [CO]/[\htwo] abundance ratio in NGC\,1140 is extremely low:
%$\sim$ 10 times lower than the Milky Way for the ``warm'' solution, and
$\sim$ 100 times lower than Solar for the ``cool'' solution ([CO]/[\htwo]$\sim 8\times 10^{-7}$).
This would imply that the \htwo\ columns could be comparable to, if not exceed, the \hi,
thus dominating the column density contribution necessary for self-shielding. 
%Thus, although the beam-averaged CO optical depths and column densities are low, within the individual
%dense cloud complexes self shielding could allow CO emission to emerge, despite the effects of
%photo-dissociation.
%The emission we are observing could arise from the dense, but small, somewhat eroded molecular clouds,
%possibly similar to the dense cores of cometary globules observed around \hii\ regions
%in the Galaxy \citep[e.g.,][]{lefloch94}.
%This hypothesis seems consistent with the high molecular gas surface densities observed
%with resolved observations of other metal-poor %dwarf 
%starbursts, including II\,Zw\,40 \citep{kepley16} 
%and the LMC \citep{indebetouw13}.
%It could also imply that the molecular clouds in NGC\,1140 have undergone significant
%reshaping from turbulence, as suggested by the three-dimensional hydrodynamical models by
%\citet{tremblin12}.

Such a low [CO]/[\htwo] abundance and cold \tkin\ in NGC\,1140 could be the result of a high level of
photo-dissociation of CO, leaving only the densest, best shielded cores.
This is a similar situation to ammonia (\nhthree) in the LMC where it is only
detected in N\,159\,W \citep{ott10}.
In this region, the \nhthree\ is cold with \tkin\ $\sim$16\,K, and has very low abundance;
\nhthree\ abundance is between 1.5 and 5 orders of magnitude lower than
observed in Galactic star-forming regions, but similar to that %the low abundance
found in the late-stage starburst M\,82 \citep{weiss01}.
\nhthree\ is particularly sensitive to UV radiation
\citep{sato83}, 
but formaldehyde (\htwoco) is less vulnerable to photo-dissociation than \nhthree,
and in N\,159\,W is warmer, \tkin$\sim$30-35\,K \citep{tang17}.
The implication is that \htwoco\ is sampling regions that are more exposed to radiation
of young massive stars, while \nhthree\ resides in the most shielded knots of dense gas
\citep{tang17}.
The stronger radiation field in NGC\,1140
from the concentration of SSCs would annihilate \nhthree,
and possibly also \htwoco, but CO could survive,
thus playing the role that \nhthree\ plays in the LMC, 
tracing the surviving cool pockets of dense gas.
%Detections: Only in N159W. The first and (to my
%knowledge so far only) detection of NH3 in the
%LMC. And even here the column density is only
%6 x 10^12 cm^-2, yielding a fractional abundance
%of only ~4 x 10^-10. In dark clouds, the fractional
%abundance is of order 10^-7 (e.g., Benson & Myers 1983, ApJ 270, 589), and in hot cores
%10^-5 (e.g. Mauersberger et al. 1987, A&A 173,
%352). While (only) in the late stage starburst M82,
%it is similarly low (Weiss et al. 2001,
%ApJ 554, L143). NH3 is particularly sensitive
%to UV radiation (e.g., Sato M. & Lee I.C. 1983,
%J. Chem. Phys. 78, 4515).
%
%Low is not only the abundance but also the NH3-
%derived T_kin: ~16 K in N159W
%A stronger UV radiation field would annihilate
%the NH3 (and possibly also H2CO, where T_kin is
%30 - 35 K with T_dust = 30-40 K (Tang et al. 2017,
%A&A 600, A16) in N159W, see their Table 3) and
%CO would then play the role, which NH3 plays
%in the LMC.

Such a scenario may occur in any specific
low-metallicity star-forming region. That it is
seen over a $\sim$2\,kpc area of a galaxy with six
SSCs requires that the evaporation of dense
molecular gas occurs roughly simultaneously.
Thus, 
%So would guess, that 
galaxies like NGC 1140
are probably rare. 
Consistently with observations,
the starburst must also be either
%And that the starburst is either
quite young (all activity is occurring at almost the same
time) or %that this is 
the last one of several
possibly very short starburst episodes
\citep[e.g.,][]{degrijs04,moll07}.
With 6 SSCs, all of them clearly surpassing
30\,Doradus in the LMC, this last starburst episode must then
be at least (if not the only one) the strongest
one.
More extensive observations of molecular tracers are necessary to test this scenario,
and constrain models 
of CO/\htwo\ abundance ratios in NGC\,1140, and, more generally, in metal-poor starbursts.

Our RADEX fits with \twelveco\ and \ci\ also give an unusually high atomic carbon
abundance (see Sect. \ref{sec:carbon}), %\twelveco/\ci\ 
C/CO $\ga$ 10.
This result is premised on the assumption that \ci\ and CO occupy the same
volume in a cloud characterized by a single kinetic temperature
and volume density \citep[e.g.,][]{okada15,requena16}.
Such a high \ci\ abundance relative to CO is more extreme than found in the
SMC with the LVG analysis of \citet{requena16}; there the clouds tend to
be less dense (\nhtwo\,$\sim 10^4$\,\cmthree), and warmer (\tkin\,$\sim 30-50$\,K),
although of similar CO column density \citep[\nco\,$\sim 2-4\times10^{16}$\,\cmtwo, see also][]{nikolic07}.
Spatially-resolved observations of \ci\ and CO in NGC\,1140 would help understand
whether the high \ci\ abundance we find is dictated by a different spatial distribution
relative to CO, unlike what is observed in the Magellanic Clouds.

\begin{figure}[h!]
\vspace{\baselineskip}
\hbox{
\centerline{
%\includegraphics[angle=0,width=\linewidth]{12CO13COvsN12CO-crop.pdf}
% 10/3/2017
\includegraphics[angle=0,width=\linewidth]{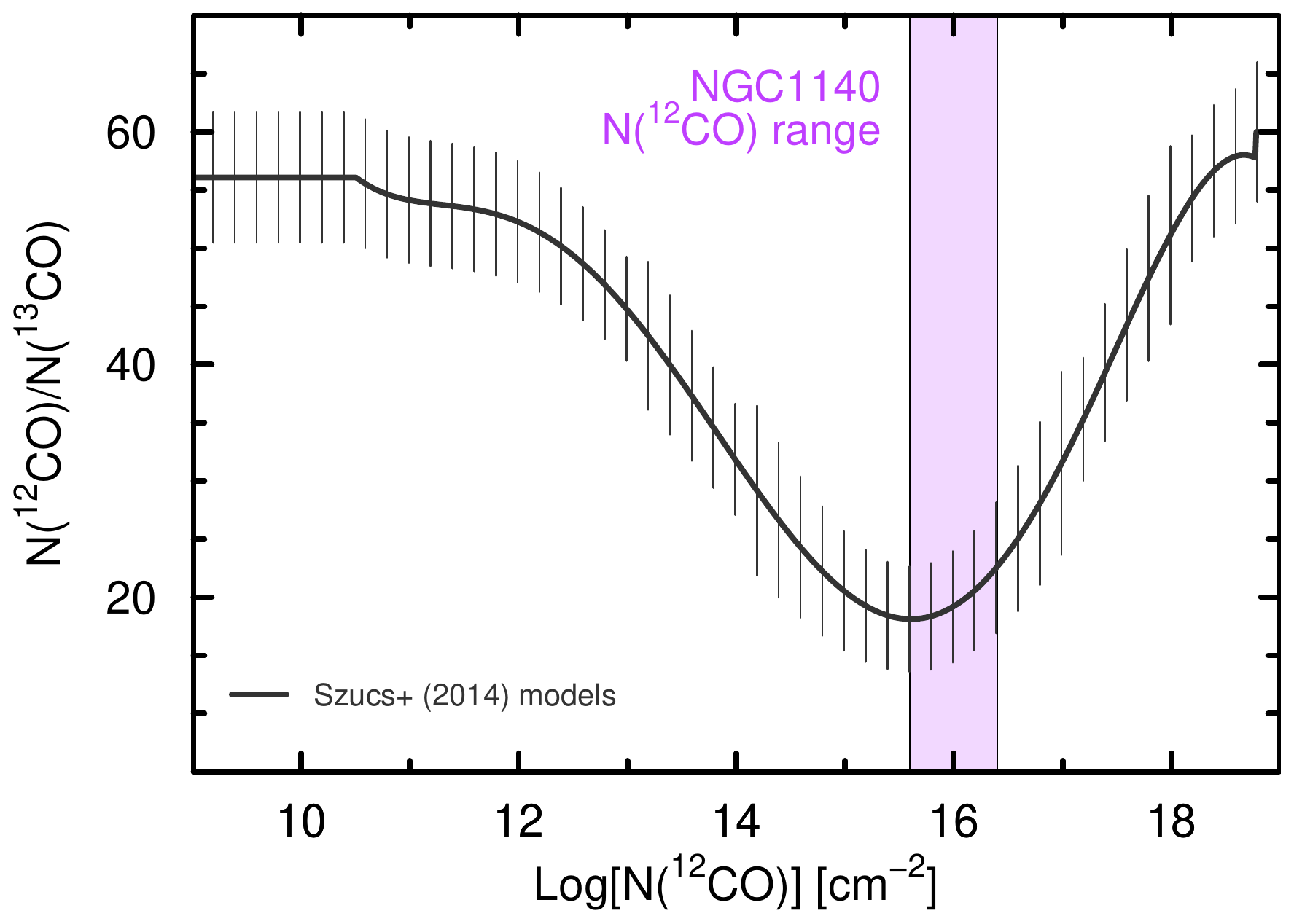}
}
}
\caption{\twelveco/\thirteenco\ abundance ratio plotted against
$N$(\twelveco).
The shaded rectangle shows the ranges of values from RADEX fits %and LTE fits
($\Delta V$\,=\,20\,\kms)
of NGC\,1140.
The solid line corresponds to the models of the \twelveco/\thirteenco\
abundance ratio vs. the \twelveco\ column density by
\citet{szucs14}.
The error bars for these models (vertical black lines) are (very roughly) estimated from their Fig. 8.
\label{fig:12co13co}
}
\end{figure}

\subsection{ \thirteenco\ charge-exchange reactions and fractionation}
\label{sec:fractionation}

Another surprising result from our RADEX analysis of NGC\,1140 is the
unusually low [\twelveco]/[\thirteenco] abundance ratio, 
$\sim 8-20$, lower even than the value
of $\sim$24 found in the Galactic Center \citep{langer90}. 
Isotopic ratios are generally governed by two competing processes:
photo-selective dissociation that would tend to decrease the relative \thirteenco\
abundance, and low-temperature carbon isotope exchange reactions that would tend
to increase it.

Low [\twelveco]/[\thirteenco] abundance ratios are typically found in older 
intermediate-mass stellar populations ($\ga$500\,Myr), in which $^{13}$C is produced in 
nuclear reactions in the cores of asymptotic giant branch (AGB) stars, and ``dredged up'' through
convective mixing where it is ejected into the ISM in stellar winds
\citep[e.g.,][]{milam05,milam09}.
However, the SSCs in NGC\,1140 are quite young, with ages ranging from $\sim$ 5 to 12\,Myr,
and the overall starburst is younger than $\sim$55\,Myr
\citep{degrijs04,moll07}.
These young ages make it unlikely that the isotopic ratio in NGC\,1140 has been lowered
by stellar reprocessing \citep[e.g.,][]{boothroyd99,pavlenko03,henkel14}.

Instead, the extremely low \twelveco/\thirteenco\ abundance ratio for NGC\,1140
could be explained by charge-exchange reactions, i.e., fractionation
\citep[e.g.,][]{watson76}.
Isotopic selective photo-dissociation of CO is effective only in diffuse gas (\nhtwo $\la$ 100\,\cmthree);
in denser regions with higher column density but moderate extinction (1\,mag\,$\la$\,\av\,$\la$ 3\,mag),
fractionation reactions become important:
$$
^{13}{\rm C}^+ + {\rm ^{12}CO}\,\leftrightharpoons\,{\rm ^{12}C}^+ \,+\, {\rm ^{13}CO} \,+\, \Delta E\, {\rm (\,=\,35\,K)}.
$$
At intermediate \nco\ (and $T \la 35$\,K), the rightmost (exothermic) reaction dominates, 
enhancing \thirteenco\ and leading to a reduced [\twelveco]/[\thirteenco] abundance ratio. 
Thus, if molecular gas in low-metallicity galaxies is dense and cool, but with moderate \nco\ as in NGC\,1140,
it is possible that fractionation drives low \twelveco/\thirteenco\ ratios.
%{\bf xxx see Paola's comment}

The low value of [\twelveco]/[\thirteenco] we obtain for NGC\,1140 is
roughly consistent with (although slightly higher than) the trends of CO column density 
and \thirteenco\ fractionation found by \citet{rollig13} and \citet{szucs14}.
Fig. \ref{fig:12co13co} shows the trend of \twelveco/\thirteenco\ abundance
ratio and \twelveco\ column density given by \citet{szucs14};
the range of $N$(\twelveco) for NGC\,1140 as estimated by the best-fit
RADEX (and LTE) models is shown as a shaded rectangle.
\citet{szucs14} show that there is a ``sweet spot" in \nco\ for 
maximizing the fractionation process, as long as the temperature
barrier of $\sim 35$\,K is not exceeded.
Interestingly, the best-fit \tkin\ for NGC\,1140 is $\sim 16-18$\,K
(see Fig. \ref{fig:chisquaren1140} and Table \ref{tab:radexn1140}), consistent with the low temperatures
necessary for optimizing \thirteenco\ fractionation.
In NGC\,1140, the low X(\twelveco)/X(\thirteenco) abundance ratio could be %is apparently
fostered by the moderately low \nco\ and the low temperature, which in turn are probably related to the
high \nhtwo\ volume density.
We are pursuing an observational program to establish whether such a phenomenon is common in metal-poor
starbursts.

\subsection{The \htwo\ mass for NGC\,1140 revealed with atomic carbon}
\label{sec:n1140h2mass}

In Paper\,I, we derived the metallicity dependence for the CO-to-\htwo\ conversion factor,
\aco\,$\propto (Z/Z_\odot)^{-2}$, which, at the metallicity of NGC\,1140,
is roughly equivalent to the exponential variation with abundance found by \citet{wolfire10}.
Here we compare the molecular mass for NGC\,1140 using \aco\ with 
%and compare it with masses estimated in two different ways:
%(1)\,what would be inferred based on our RADEX results for CO,
%and (2)\,
the molecular mass estimated from \ci\ luminosity according to \citet{glover16}. %\footnote{In all cases, helium is included in the derived gas masses.}.
After correcting to a 22\arcsec\ beam size as in Table \ref{tab:n1140} (see also Table \ref{tab:corrections}), $L^\prime({\rm CO})$\,=\,$2.9 \times 10^6$\,\kkmspc.
Using \aco\,=\, 3.2\,\msun\,(\kkms\,pc$^{2})^{-1}$ \citep[not including helium,][]{saintonge11}, and scaling
with $(Z/Z_\odot)^{-2}$ (see Paper I) for NGC\,1140 \citep[\zzsun\,=\,0.31; \logoh\,=\,8.18, and assuming \logoh$_\odot$\,=\,8.69 from][]{asplund09},
we find a total molecular gas mass of $\sim 9.3\times10^7$\,\msun.
%dex(6.585)$\times$1.03\,\kkmspc, and
%included the aperture correction given in Table \ref{tab:corrections} of 1.03 for a 22\arcsec\ beam.

\citet{glover16} use their hydrodynamical simulations of \cione\ emission to estimate \htwo\ masses, and 
provide a \ci\ luminosity-to-\htwo\ conversion factor.
At the highest $G_0$ modeled by \citet{glover16},
they obtain a mean value $X_{CI}\,=\,8.8 \times 10^{20}$\,\cmtwo\,(\kkms)$^{-1}$ for $Z$\,$\sim$\,\zsun\ and an approximately
linear increase %variation 
with decreasing metallicity.
For NGC\,1140 the \cione\ luminosity %$L^\prime$(\cione)\,=\,$8.8 \times 10^5$\,\kkmspc\
$L^\prime$(\cione)\,=\,$1.6\times 10^6$\,\kkmspc\ %dex(6.138$\times$1.96) 
(from Table \ref{tab:n1140}, including the aperture correction for a 22\arcsec\ beam).
Converting the \citet{glover16} $X_{CI}$ to equivalent \aco\ units, we would find
$\alpha_{CI}$\,=\,53.2\,\msun\,(\kkmspc)$^{-1}$ for 0.31 \zzsun, and a resulting
molecular mass \mhtwo\ of $\sim 8.7 \times 10^7$\,\msun.
This value is within 10\% of what we infer with the quadratic metallicity scaling of \aco,
despite the larger observed \cione/\coone\ ratio (see Sect. \ref{sec:cico}) relative to their models\footnote{In these
calculations, unlike Paper I, we have not included the factor of 1.64 to correct to total flux in NGC\,1140;
these values are corrected to 22\arcsec\ as in Tables \ref{tab:n1140} and \ref{tab:corrections}}.
These values of molecular gas mass for NGC\,1140 are within 15\% of those estimated 
using the models of \citet{weiss07}.

\section{Summary and conclusions}
\label{sec:conclusions}

We have presented \twelveco\ observations of ten galaxies, with detections for eight.
With metallicities ranging from 
\logoh\,$\sim$7.7 to 8.4 (0.1\,\zsun\ to 0.5\,\zsun), at and below the abundance of the SMC,
this is the largest sample of metal-poor galaxies with CO detections so far obtained outside the Local Group.
For one of the galaxies, NGC\,1140, we report additional \thirteenco, \cione, and \hcn\ measurements. 
Our main conclusions are the following:
\begin{itemize}
\item
After correcting for differences in beam sizes,
NGC\,1140 shows high velocity-integrated temperature ratios of
\cotwo/\coone\ (\rtwo\,=\,2.1\,$\pm$\,0.06) 
and \cothree/\coone\ 
(\rthree\,=\,2.0\,$\pm$\,0.45, or
%\rthree\,=\,1.7\,$\pm$\,0.21 with the correction to 22\arcsec\ beams), while 
\rthree\,=\,1.7\,$\pm$\,0.39 with the correction to 22\arcsec\ beams), while 
the other galaxies have less extreme ratios
(see Figs. \ref{fig:lineratios21}, \ref{fig:lineratios32}).
\item
A comparison of dense-gas tracers such as \cothree\ 
(of the four observed metal-poor galaxies) and \hcn\ (upper limit for NGC\,1140)
with data in the literature shows a deficit in \cothree\ relative to SFR
(see Fig. \ref{fig:co32}), similar to the \coone\ deficit found in Paper I.
However, gas excitation at low metallicity measured by comparing \cothree\ to \coone\ luminosities
seems similar to that of more metal-rich systems.
\hcn\ was observed but not detected in NGC\,1140.
The lower limit of \hcn\ for NGC\,1140 may be due to a combination
of stellar feedback \citep[e.g.,][]{hopkins13} and 
an extremely high [CO]/[HCN] abundance ratio ($\ga 10^5$)
probably because of the low metallicity (see Fig. \ref{fig:radexhcn}).
\item
Fitting LTE models to the \cothree, \cotwo, and \coone\ emission in NGC\,1140, NGC\,7077, 
and UM\,448 suggests that CO column densities are moderate and temperatures are low.
These fits also suggest low optical depths, but the lack of higher-$J$ lines for
NGC\,7077 and UM\,448 prevents confirming this as a general result with radiative-transfer models.
\item
Fitting physical models (RADEX) of the \twelveco\ and \thirteenco\ emission in NGC\,1140
suggests that the molecular gas is 
cool (\tkin\,$\la$\,20\,K), 
dense (\nhtwo\,$\ga 10^6$\,\cmthree),
with moderate CO column density
(\nco\,$\sim \, 10^{16}$\,\cmtwo)
and low filling factor ($F\,\sim 0.01-0.1$).
\twelveco\ optical depths are fairly low
(\tauco\,$\la$0.2), and we speculate that the CO survives
photo-dissociation because of the high \hi\ 
column density in this galaxy.
The molecular excitation in NGC\,1140 is starburst-like, similar to M\,82
(see Fig. \ref{fig:cocooling}).
\item
The [\twelveco]/[\thirteenco] abundance ratio in NGC\,1140 inferred
from the fit is very low,
$\sim 8-20$, lower even than the value of $\sim 24$ found in the Galactic center
\citep{langer90}.
Because the starburst in NGC\,1140 is quite young, it is difficult to
interpret this low ratio as due to stellar reprocessing by older
intermediate-mass populations \citep[e.g.,][]{milam05,milam09}.
Instead, we attribute it to enhanced CO fractionation,
from the combination of moderate column densities
and cool gas (see Fig. \ref{fig:12co13co}), as predicted by the chemical models
of \citet{rollig13} and \citet{szucs14}.
\item
Fitting RADEX models of the \twelveco\ and \cione\ emission in NGC\,1140
gives physical conditions similar to the \twelveco$+$\thirteenco\ fits,
although there is a possibility for warmer (\tkin\,$\sim$\,38\,K),
less dense (\nhtwo\,$\sim\,10^4$\,\cmtwo) gas at higher column densities.
Independent fitting results for NGC\,1140 using the models of \citet{weiss07}
%suggests that such a ``warm'' solution results in an unrealistically
%low gas-to-dust mass ratio ($\sim$62 vs. $\sim$100 for Solar);
%thus we have considered the cooler solution (\tkin\,$\la$\,20\,K) as more reliable.
give similar results.
%Independently of the temperature, 
Both our RADEX fits and the LVG fits 
%Weiss et al. (2017) 
suggest that atomic carbon is at least 10 times
more abundant than \twelveco\ in NGC\,1140.
\end{itemize}

For the first time outside the Local Group, 
the six CO transitions measured for NGC\,1140, together with \cione, have
enabled an analysis of physical conditions in the
molecular gas of a low-metallicity galaxy with at least 6 SSCs.
%as well as a determination of the 
The unusual 
[\twelveco]/[\thirteenco] and [\twelveco]/[\twelvec] abundance ratios,
and the cool, dense gas at moderate column densities,
may be the consequence of the SSCs in this galaxy and their feedback effect 
on the ISM.
Future work will attempt to put this speculation on a more quantitative footing,
both with models of ISM chemistry and with more observations of the molecular ISM in 
low-metallicity starbursts.

%%%%%%%%%%%%%%%%%%%%%%%% acknowledgments
\begin{acknowledgements}
We acknowledge the anonymous referee for her/his useful comments and suggestions.
LKH is grateful to Carlo Giovanardi for interesting discussions and mathematical insights for the beam
corrections.
SGB acknowledges economic support from grants ESP2015-68964-P and AYA2016-76682-C3-2-P.
We warmly thank the IRAM staff, both in Granada
and at Pico Veleta, for their capable management of the logistics and the telescope/receiver operations.
We are also indebted to the APEX service observing team for their dedication
in the challenging high-frequency observations obtained for NGC\,1140.
We gratefully acknowledge the International Space Science Institute (Bern) for hospitality during the 
conception of this paper.
Heavy use was made of the NASA/IPAC Extragalactic Database (NED).
\end{acknowledgements}
%%%%%%%%%%%%%%%%%%%%%%%%%%%%%%%%%%%%%

\appendix
\section{Exponential beam size correction}
\label{sec:appendix_exponential}

Following \citet{liu17}, we have corrected the velocity-integrated \tmb\ values
to a common beam size according to the distribution of cool dust, thus
explicitly assuming that the molecular gas and the dust have the same distribution. 
In order to apply such a correction also to the two galaxies (Mrk\,0996, NGC\,7077)
for which this was not possible, we have adopted an analytical approach;
such an approach also tends to average out photometric uncertainties and possibly
improves the reliability of the correction.

\begin{figure*}[h!]
\vspace{\baselineskip}
\hbox{
\centerline{
\includegraphics[angle=0,width=0.45\linewidth]{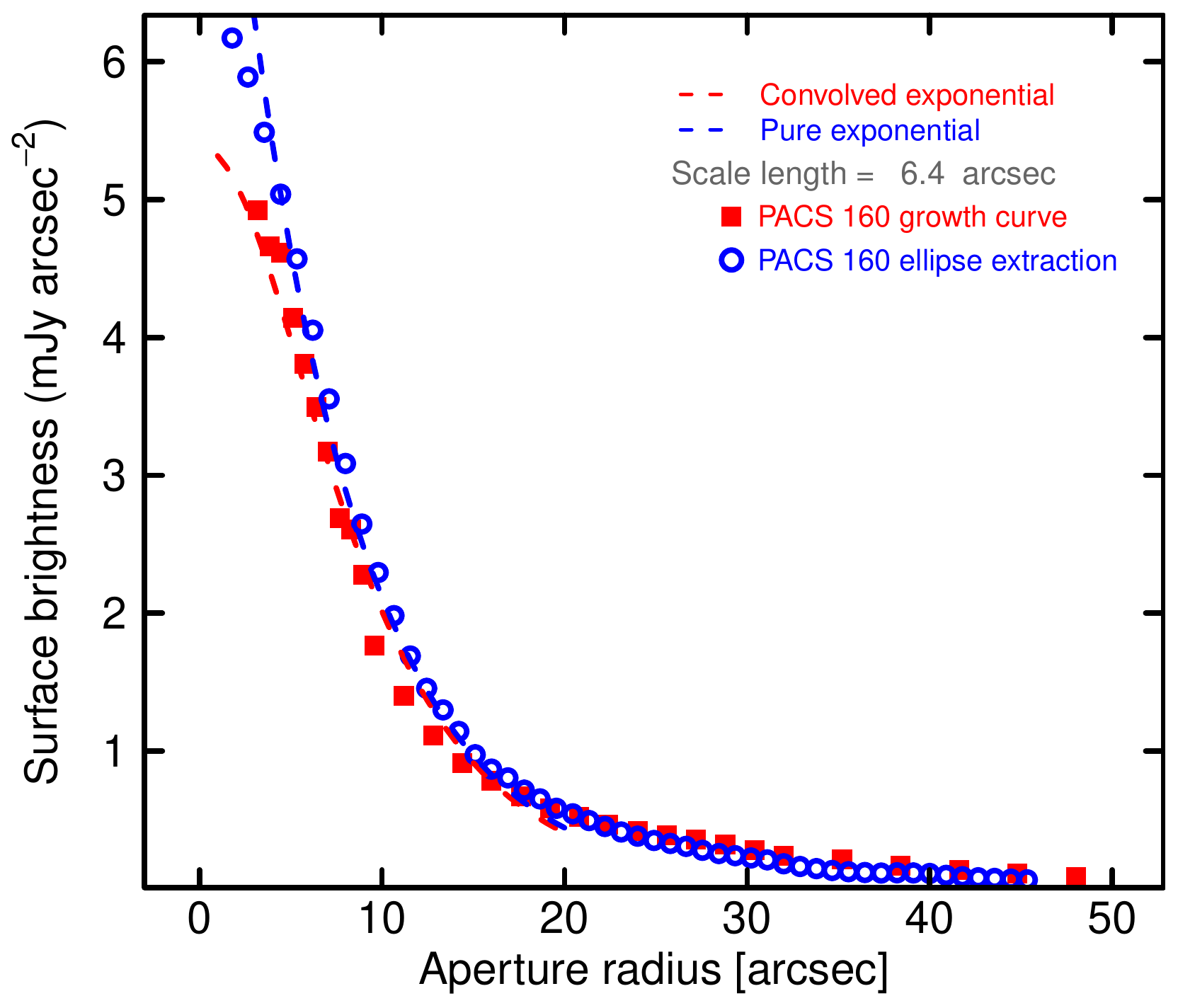}
\includegraphics[angle=0,width=0.45\linewidth]{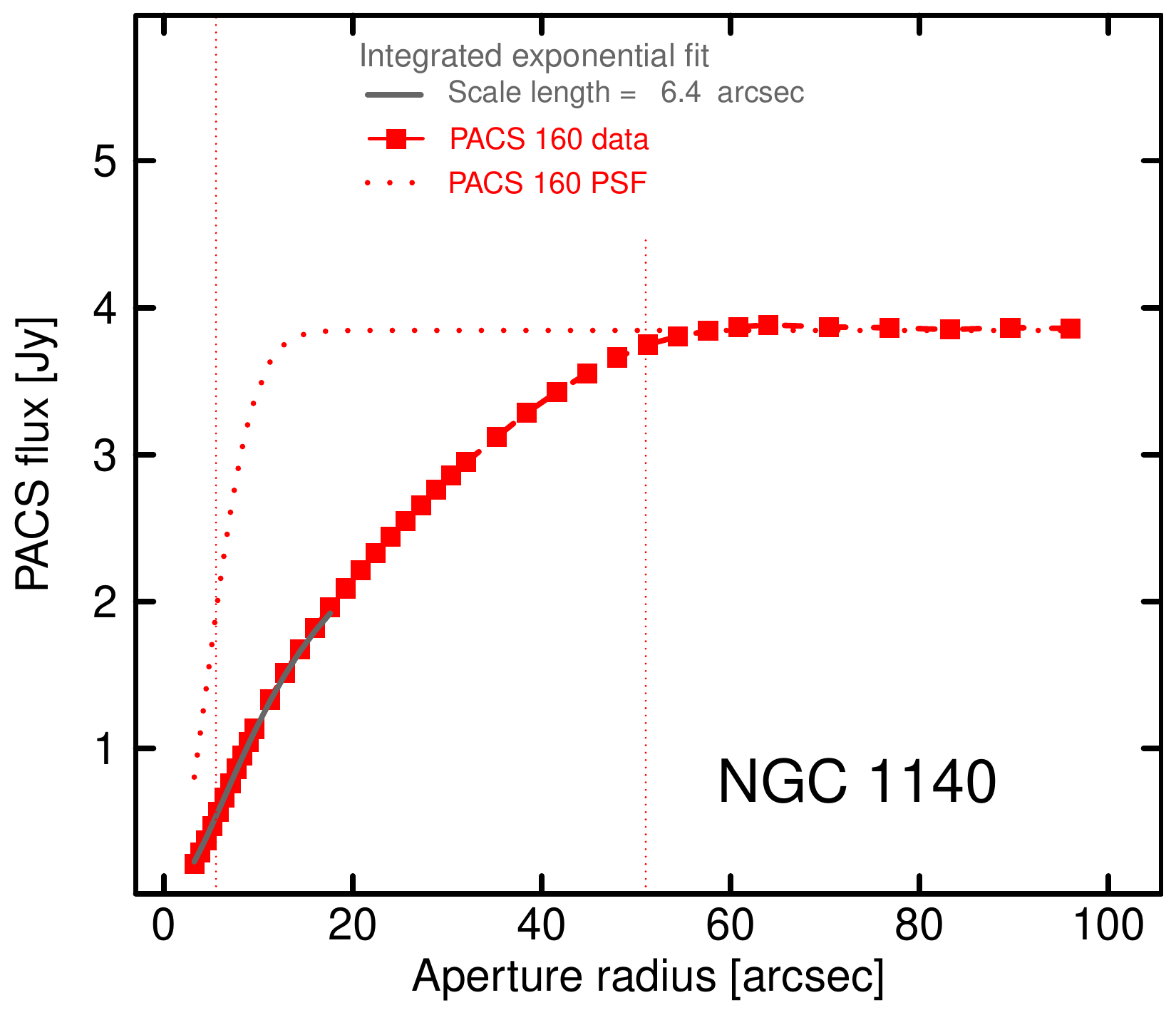}
}
}
\caption{{\it Left panel:} PACS 160\,\micron\ radial surface brightness profile of NGC\,1140;
open (blue) circles give the profile obtained from azimuthally averaging the image, 
and filled (red) squares the analogous profile obtained from the first derivative of the photometric
growth curve.
The lower dashed curve show an exponential distribution of scale length 6\farcs4 convolved with the PACS
160\,\micron\ beam (assuming it is Gaussian), and the upper one the same exponential but without convolution.
{\it Right panel:} PACS 160\,\micron\ growth curve centered on the brightness peak.
The left-most dotted vertical line gives the PACS beam radius ($\sim$5\farcs5) and the right-most one
the optical radius of NGC\,1140 taken from NED (51\arcsec).
The solid grey curve corresponds to the best-fit exponential integral as described in the text.
}
\label{fig:beam_ngc1140}
\end{figure*}

\begin{figure*}[h!]
\vspace{\baselineskip}
\hbox{
\centerline{
\includegraphics[angle=0,width=0.45\linewidth]{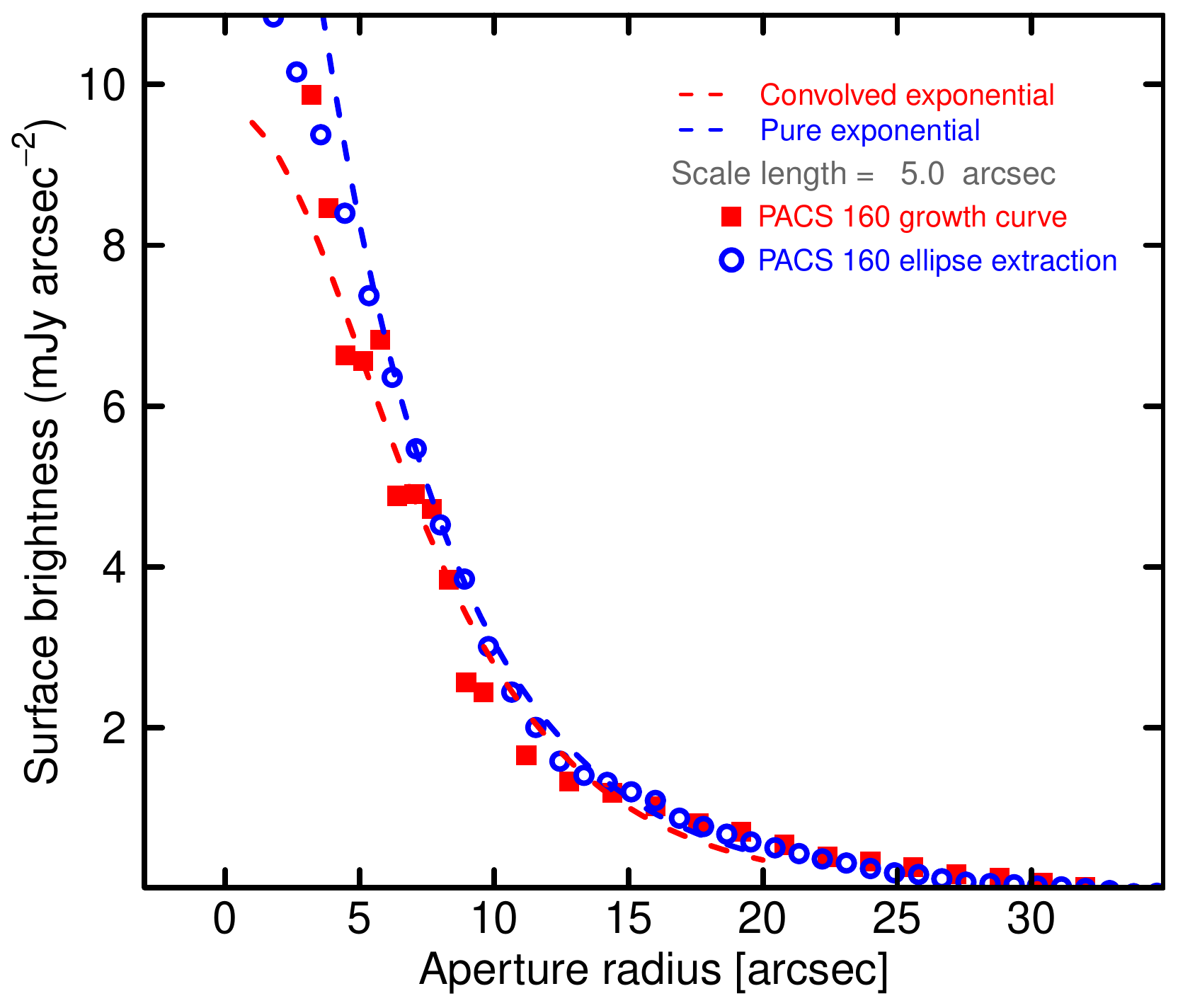}
\includegraphics[angle=0,width=0.45\linewidth]{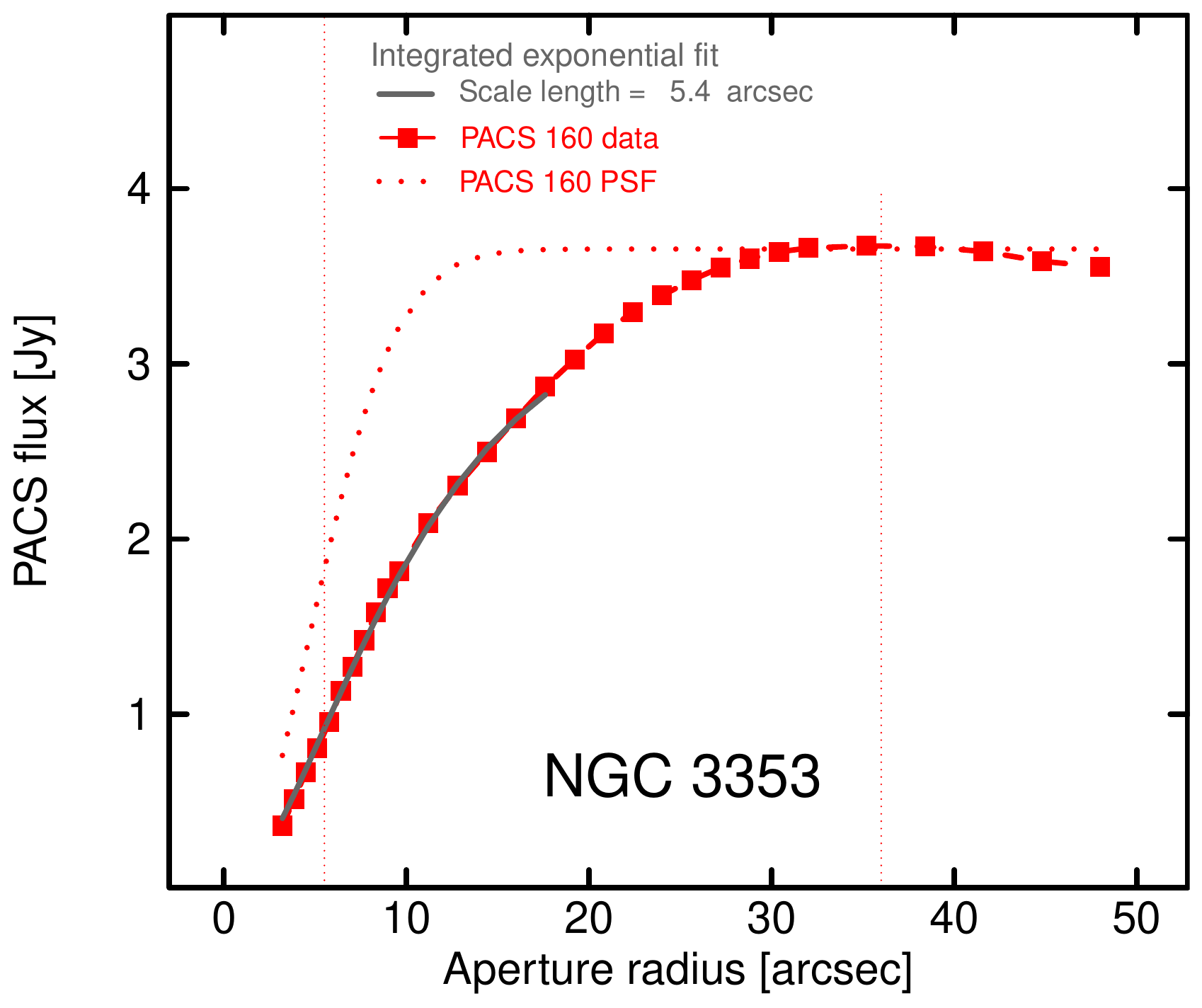}
}
}
\caption{{\it Left panel:} PACS 160\,\micron\ radial surface brightness profile of NGC\,3353 (Haro\,3);
open (blue) circles give the profile obtained from averaging the image over fixed-position-angle ellipses,
and filled (red) squares the analogous profile obtained from the first derivative of the photometric
growth curve.
As in Fig. \ref{fig:beam_ngc1140},
the dashed curves show the exponential convolved with the PACS 160\,\micron\ beam (lower),
and the upper the exponential without convolution; the best-fit scale length is 5\farcs0. 
{\it Right panel:} PACS 160\,\micron\ growth curve centered on the brightness peak.
The left-most dotted vertical line gives the PACS beam radius ($\sim$5\farcs5) and the right-most one
the optical radius of NGC\,3353 taken from NED (36\arcsec).
The solid grey curve corresponds to the best-fit exponential integral as described in the text;
in the case of NGC\,3353, the growth-curve best fit scalelength is slightly different from
the radial profile one and we have adopted the growth-curve value for beam corrections.
}
\label{fig:beam_ngc3353}
\end{figure*}

\begin{figure*}[h!]
\vspace{\baselineskip}
\hbox{
\centerline{
\includegraphics[angle=0,width=0.45\linewidth]{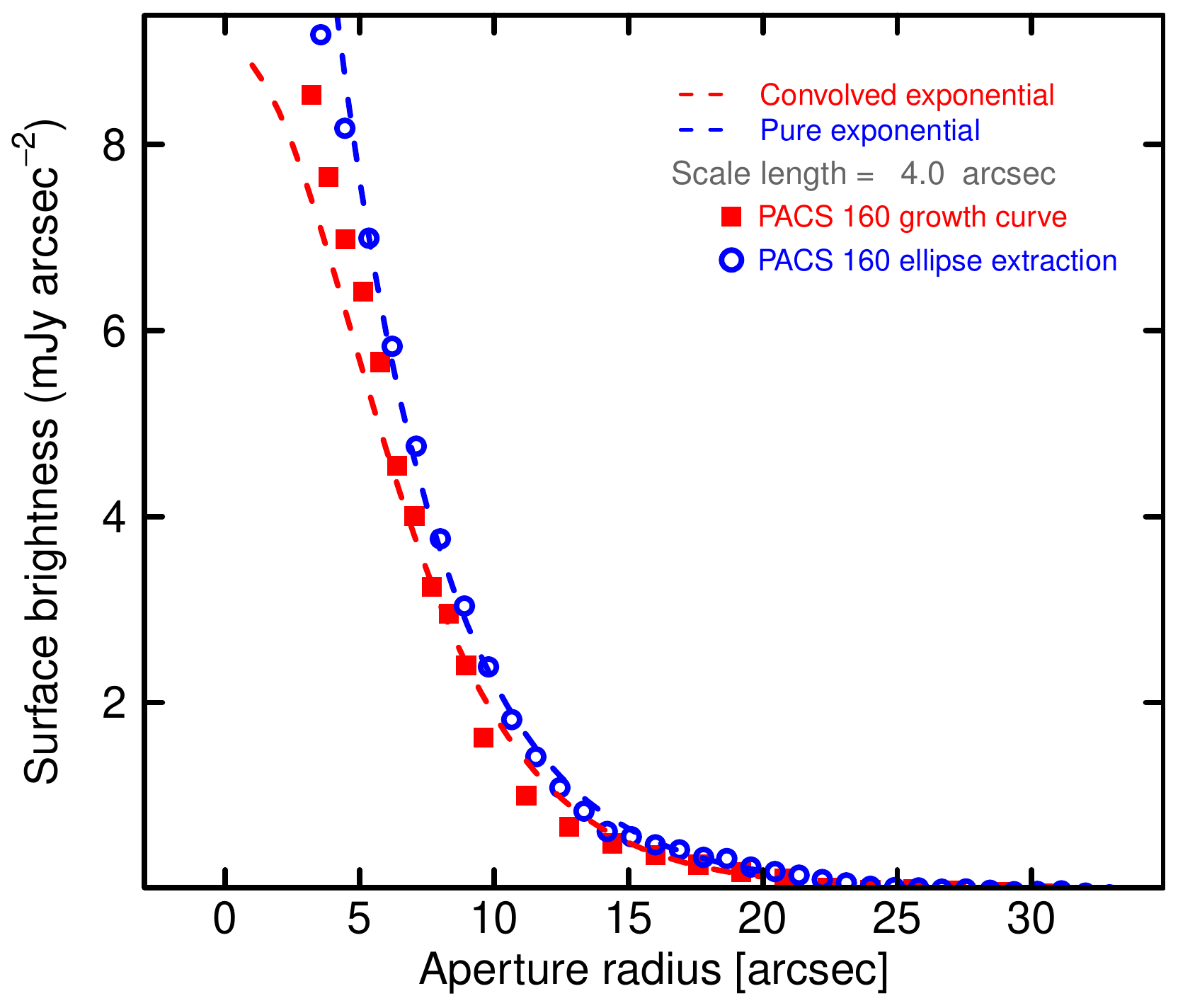}
\includegraphics[angle=0,width=0.45\linewidth]{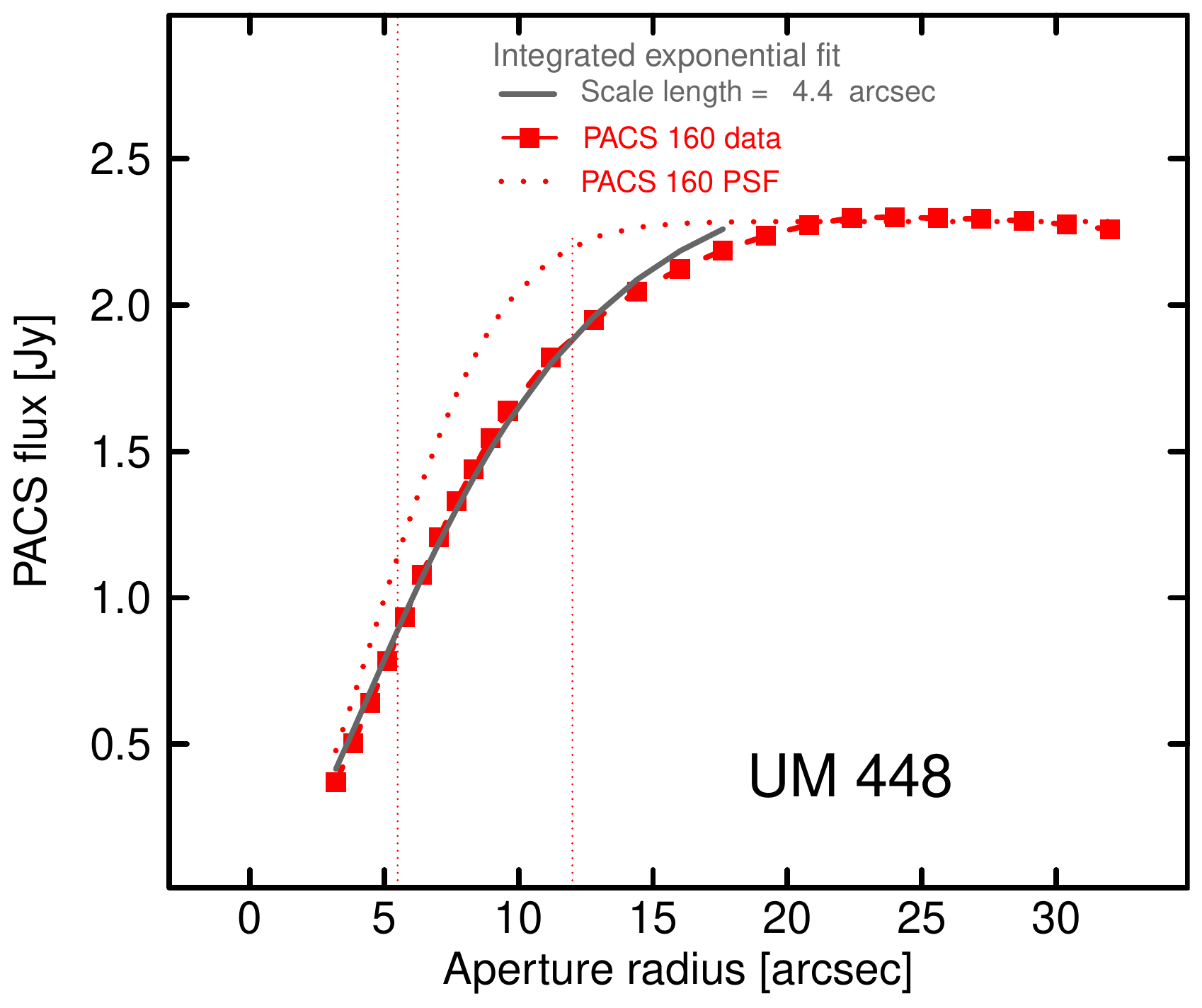}
}
}
\caption{{\it Left panel:} PACS 160\,\micron\ radial surface brightness profile of UM\,448;
open (blue) circles give the profile obtained from averaging the image over fixed-position-angle ellipses,
and filled (red) squares the analogous profile obtained from the first derivative of the photometric
growth curve.
As in Fig. \ref{fig:beam_ngc1140},
the dashed curves show the exponential convolved with the PACS 160\,\micron\ beam (lower),
and the upper the exponential without convolution; the best-fit scale length is 4\farcs0. 
{\it Right panel:} PACS 160\,\micron\ growth curve centered on the brightness peak.
The left-most dotted vertical line gives the PACS beam radius ($\sim$5\farcs5) and the right-most one
the optical radius of UM\,448 taken from NED (12\arcsec).
The solid grey curve corresponds to the best-fit exponential integral as described in the text;
like NGC\,3353, the growth-curve best fit scalelength for UM\,448 is slightly different from
the radial profile one and we have adopted the growth-curve value for beam corrections.
}
\label{fig:beam_um448}
\end{figure*}

\begin{figure*}[h!]
\vspace{\baselineskip}
\hbox{
\centerline{
\includegraphics[angle=0,width=0.45\linewidth]{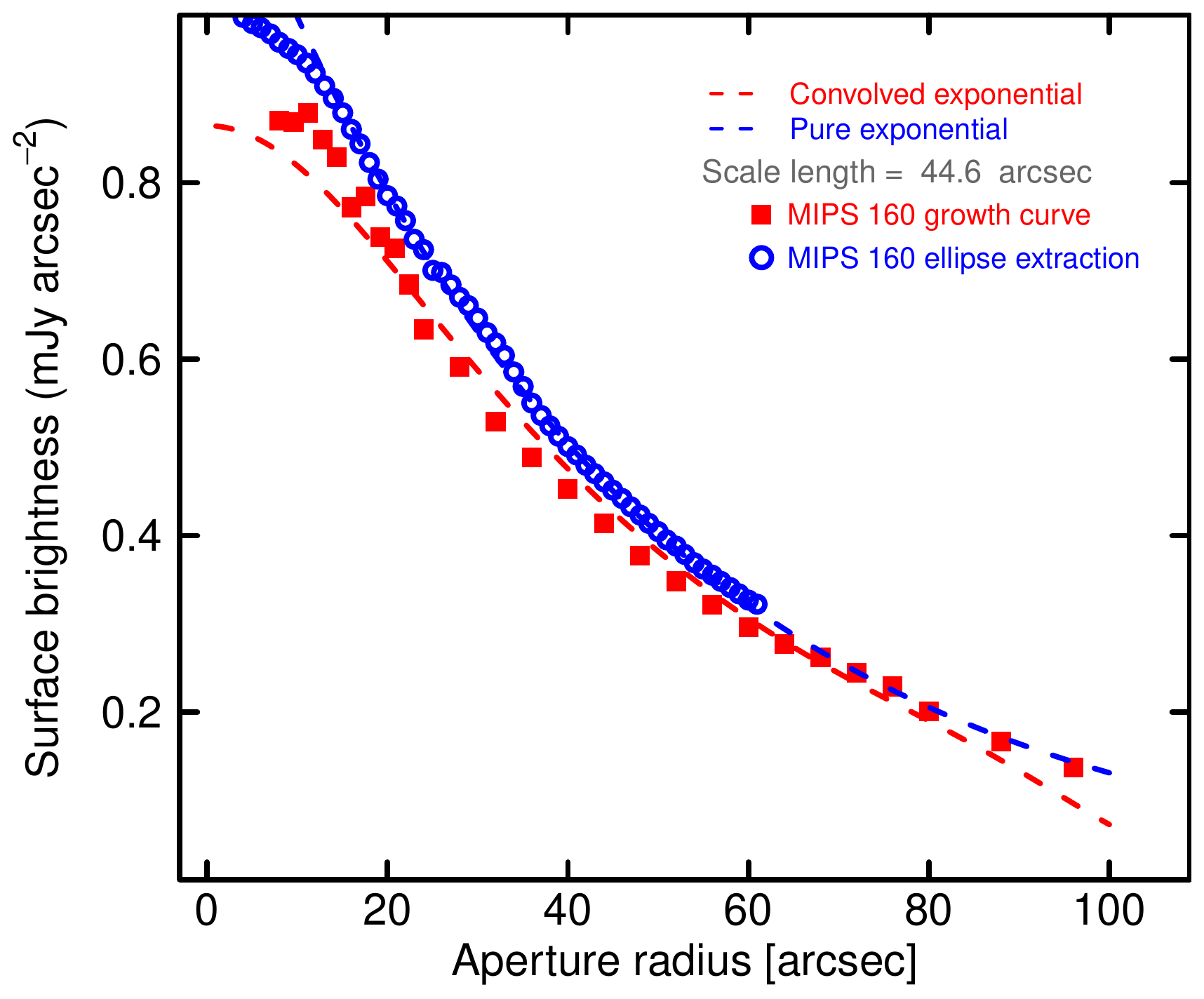}
\includegraphics[angle=0,width=0.45\linewidth]{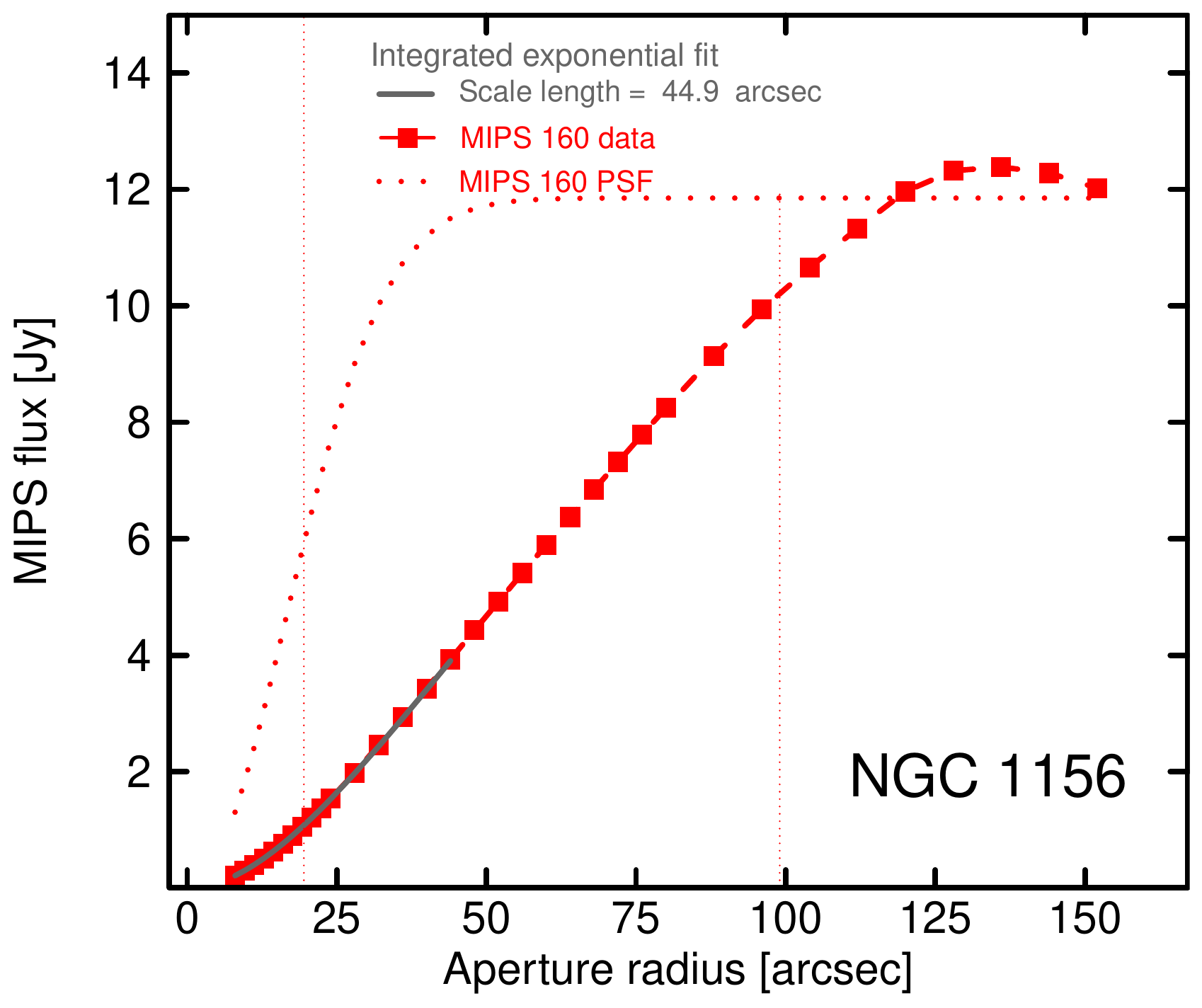}
}
}
\caption{{\it Left panel:} MIPS 160\,\micron\ radial surface brightness profile of NGC\,1156;
open (blue) circles give the profile obtained from averaging the image over fixed-position-angle ellipses,
and filled (red) squares the analogous profile obtained from the first derivative of the photometric
growth curve.
As in Fig. \ref{fig:beam_ngc1140},
the dashed curves show the exponential convolved with the MIPS 160\,\micron\ beam (lower),
and the upper the exponential without convolution; the best-fit scale length is 4\farcs0. 
{\it Right panel:} MIPS 160\,\micron\ growth curve centered on the brightness peak.
The left-most dotted vertical line gives the MIPS beam radius ($\sim$19\arcsec) and the right-most one
the optical radius of NGC\,1156 taken from NED (99\arcsec).
The solid grey curve corresponds to the best-fit exponential integral as described in the text;
like NGC\,3353 and UM\,448, the growth-curve best fit scalelength for NGC\,1156 is slightly different from
the radial profile one and we have adopted the growth-curve value for beam corrections.
}
\label{fig:beam_ngc1156}
\end{figure*}

PACS 160\,\micron\ images are available from the \hers\ archive for NGC\,1140, NGC\,3353, and UM\,448,
but for NGC\,1156, we had to use the MIPS 160\,\micron\ image from \spit.
In all cases, the 160\,\micron\ PACS fluxes reported by \citet{remy13} are beyond the peak of the dust emission,
implying that in these dwarf galaxies the 160\,\micron\ is tracing cool dust.
We obtained surface brightness profiles by
performing an azimuthal extraction centered on the brightness peak, and growth curves by
measuring photometric flux in increasingly large circular apertures.
The 160\,\micron\ surface brightness profiles are shown in the left panels 
and the growth curves in the right panels of Figs. 
\ref{fig:beam_ngc1140},
\ref{fig:beam_ngc3353},
\ref{fig:beam_um448},
and \ref{fig:beam_ngc1156}.
The dust distribution is clearly exponential (rather than Gaussian)
in all cases examined.
For the two remaining galaxies, Mrk\,996 and NGC\,7077, MIPS 160\,\micron\ images are available,
but the galaxies are not resolved 
(even at 24 or 70\,\micron\ where we obtained clear profiles of a diffraction-limited Airy ring).

Taking the source distribution to be exponential, with folding length $r_s$, 
the integrated flux density within a beam with FWHM $\theta_b$ corresponds to the 
following integral (assuming azimuthal symmetry) in polar coordinates:
 
\begin{equation}
S_\nu\,=\,A_0\,\int_0^{2\pi}\int_0^{\theta_b/2} \exp(-r/r_s)\, r \, dr \ d\phi
\label{eqn:expintegral}
\end{equation}
\noindent
where $A_0$ is the brightness at the origin of the exponential,
and $r$ corresponds to the angular distance from the origin.
Eqn. \ref{eqn:expintegral} can be solved analytically to obtain:

\begin{equation}
%N_{\rm cl}\,=\,2\,\pi\,N_0\,\left[ r_2^2\, \left( 1 - \exp \left( -\frac{\theta_b}{2\,r_s} \right) - \frac{\theta_b\,r_s}{2} \exp \left( -\frac{\theta_b}{2\,r_s} \right) \right]
S_\nu\,=\,2\,\pi\,A_0\,\left \{ r_s^2\, \left[ 1 - \exp \left( -\frac{\theta_b}{2\,r_s} \right)\right] - \frac{\theta_b\,r_s}{2} \exp \left( -\frac{\theta_b}{2\,r_s} \right) 
\right \}
\label{eqn:exparea}
\end{equation}
\noindent
The source FWHM $\theta_s$ is related to the exponential folding length by
$\theta_s = 2 r_s \ln(2)$,
obtained by calculating the diameter (i.e., twice the radius) at which the exponential distribution is equal to $A_0/2$.

However, Eqn. (\ref{eqn:expintegral}) is formally incorrect because of the need to convolve the pure exponential with the 
Gaussian beam
when the exponential scale length $r_s$ is small relative to the beam radius $\theta_b/2$.
This convolution is given by the following, where we have taken advantage of the azimuthal symmetry and calculate the convolution
along a single radial ray here given by $x$:

\begin{equation}
D(x,0)\,=\,A_0\,e^{-x^2/\sigma_b^2}\ \int_0^\infty \int_0^{2\pi} e^{-\frac{r_0}{r_s}}\,e^{\left( -\frac{r_0^2 - 2\,x\,r_0\,\cos\phi}{\sigma_b^2} \right)} r_0\,dr_0\,d\phi
\label{eqn:expconvolution}
\end{equation}
\noindent
where $\sigma_b$ corresponds to the Gaussian $\sigma$ assuming that the beam is
Gaussian with FWHM $\theta_b$ ($\theta_b\,=\,2\,\sigma_b \sqrt{2\,\ln\,2}$).

Equation (\ref{eqn:expconvolution}) can be integrated by separating the variables ($r_0$, $\phi$).
After integrating in $\phi$, we obtain:
\begin{equation}
D(x,0)\,=\,2\pi\,A_0\,e^{-x^2/\sigma_b^2}\ \int_0^\infty e^{-\frac{r_0}{r_s}}\,e^{-\frac{r_0^2 }{\sigma_b^2}} I_0\left( \frac{2xr_0}{\sigma_b^2}\right) r_0\,dr_0
\label{eqn:expconvolution_anal}
\end{equation}
\noindent
where $I_0$ is a modified Bessel function of the first kind.
Eqn. (\ref{eqn:expconvolution_anal}) can then be integrated numerically to obtain the surface
brightness profile of the convolved exponential, as
shown as the lower (red) dashed curves in Figs. 
\ref{fig:beam_ngc1140},
\ref{fig:beam_ngc3353},
\ref{fig:beam_um448},
and \ref{fig:beam_ngc1156}.

These figures show that for the four galaxies for which we were able to fit
surface-brightness profiles and growth curves
the pure exponential ($2\pi\,A_0\,\exp(-r/r_s)$) is equivalent to 
the integral given in Eqn. \ref{eqn:expconvolution_anal}
as long as the scale length $r_s$ is larger than $\theta_b/2$.
We thus fitted the growth curves with Eqn. (\ref{eqn:exparea}) in order to obtain the beam
corrections.
This is advantageous from a numerical point of view because otherwise it would be necessary to fit the growth curve
to the integral of Eqn. (\ref{eqn:expconvolution_anal}) which is in itself an integral.
%and thus becomes intractable.

To correct the fluxes (e.g., $S_\nu$ in Jy), we scaled the observed dust
fluxes to a common beam size of 22\arcsec\ by comparing
the growth-curve integrals in the different beam sizes using the best-fit of the analytical formula
in Eqn. (\ref{eqn:exparea}) to obtain $r_s$ and $A_0$. 
We thus derive a multiplicative factor
$AP_{\rm cor}$ as given in Col. (6) of Table \ref{tab:corrections}:
$S_{22}\,=\,S_{\rm orig}\, AP_{\rm cor}$.
We then apply the correction to velocity-integrated \tmb\ values
by noting that
$S/\Omega\,=\,2\,k\,\nu^2\,/c^2$\,\tmb,
where $\Omega$ is the beam solid angle,
$k$ is the Boltzmann constant, $\nu$ is the observed frequency,
and $c$ is the speed of light.
Thus:
\begin{equation}
T_{\rm mb}(22)\,=\,T_{\rm mb}({\rm orig})\ AP_{\rm cor}\ \frac{\Omega(\rm {orig})}{\Omega(22)}
\label{eqn:cor}
\end{equation}
\noindent
where ``orig" refers to the original values observed in the different beams.
The flux and \tmb\ corrections necessary to convert our observations to a 
common 
beam size of 22\arcsec\ are given in Table \ref{tab:corrections},
together with the corrected velocity-integrated \tmb\ values that we use throughout the paper when we
are examining line ratios (\cotwo/\coone) or fitting physical models.
Corrected values are also reported in Tables \ref{tab:lines}, \ref{tab:linesagain}, and \ref{tab:n1140}
in the main text.

For Mrk\,996 and NGC\,7077, the only available cool dust images are from
MIPS and, as mentioned above, the galaxies are not resolved.
Thus, for these we assumed that the disk scale length $r_s$ is equal
to 0.2\,\ropt\ (optical radius from NED), consistently with 
\citet{young95,leroy09,kuno07} who all found that CO exponential
scale lengths are, in the mean, equal to 1/5 the optical radius.
This is also roughly consistent with the other four galaxies for which
the mean $r_s$/\ropt\ given by the best fits of the growth curves is 0.27$\pm$0.16.

\begin{center}
\begin{table}
      \caption[]{Flux corrections to a common beam size$^{\mathrm a}$} 
\label{tab:corrections}
\resizebox{\linewidth}{!}{
% 28/2/2016
% see ~/statistics/moleculesBCDs_PaperII/mkcorrections
{%\small
%\tiny
\begin{tabular}{lllccccl}
\hline
\multicolumn{1}{c}{Galaxy} &
\multicolumn{1}{c}{$r_s$} &
\multicolumn{1}{c}{Transition} & 
\multicolumn{1}{c}{Original beam} & 
\multicolumn{1}{c}{Common beam} & 
\multicolumn{1}{c}{Aperture} & 
\multicolumn{1}{c}{\tmb} &
\multicolumn{1}{c}{$I_{CO}$}  \\ 
&
\multicolumn{1}{c}{(arcsec)} &
& 
\multicolumn{1}{c}{(arcsec)} & 
\multicolumn{1}{c}{(arcsec)} & 
\multicolumn{1}{c}{correction} & 
\multicolumn{1}{c}{correction} &
\multicolumn{1}{c}{corrected} \\ 
& & & & & 
\multicolumn{1}{c}{$AP_{\rm cor}$} & &
\multicolumn{1}{c}{(\kkms)} \\ 
\hline
Mrk\,0996  & 3.6 & \coone\          & 21.4 & 22.0 &  1.014 &  0.964 & 0.235\\
Mrk\,0996  & 3.6 & \cotwo\ (IRAM)   & 10.7 & 22.0 &  1.843 &  0.438 & 0.128\\
Mrk\,0996  & 3.6 & \cotwo\ (APEX)   & 27.2 & 22.0 &  0.908 &  1.389 & 0.938 \\
Mrk\,0996  & 3.6 & \cothree\        & 18.1 & 22.0 &  1.13  &  0.77 & $<$0.161\\
\\
%NGC\,1140  & 6.4 & \coone\         & 21.4 & 22.0 &  1.03 &  0.97 & 0.563\\
%NGC\,1140  & 6.4 & \cotwo\         & 10.7 & 22.0 &  2.51 &  0.59 & 1.179\\
%NGC\,1140  & 6.4 & \cofour\        & 13.6 & 22.0 &  1.78 &  0.68 & 0.582\\
% 2/3/2017 use the value fitted by *.pl, different because of roundoff
% these above are taken from R
% not commented values are the same
NGC\,1140  & 6.4 & \coone\         & 21.4 & 22.0 &  1.027 &  0.976 & 0.564\\
NGC\,1140  & 6.4 & \cotwo\         & 10.7 & 22.0 &  2.502 &  0.595 & 1.180\\
NGC\,1140  & 6.4 & \cothree\       & 18.1 & 22.0 &  1.238 &  0.841 & 0.972\\
NGC\,1140  & 6.4 & \cofour\        & 13.6 & 22.0 &  1.784 &  0.682 & 0.582\\
NGC\,1140  & 6.4 & \thirteencoone\ & 22.4 & 22.0 &  0.980 &  1.019 & 0.053\\
NGC\,1140  & 6.4 & \thirteencotwo\ & 11.2 & 22.0 &  2.342 &  0.609 & 0.105\\
%NGC\,1140  & 6.4 & \ci\            & 12.7 & 22.0 &  1.96 &  0.65 & 0.368\\
NGC\,1140  & 6.4 & \cione\         & 12.7 & 22.0 &  1.952 &  0.655 & 0.369\\
\\
NGC\,1156  & 44.9 & \coone\    & 21.4 & 22.0 &  1.055 &  0.995 & 1.841\\
NGC\,1156  & 44.9 & \cotwo\    & 10.7 & 22.0 &  3.903 &  0.921 & 0.668\\
\\
NGC\,3353  & 5.4 & \coone\    & 21.4 & 22.0 &  1.025 &  0.971 & 2.100\\
NGC\,3353  & 5.4 & \cotwo\    & 10.7 & 22.0 &  2.314 &  0.548 & 1.876\\
\\
NGC\,7077  & 4.8 & \coone\    & 21.4 & 22.0 &  1.022 &  0.969 & 0.474 \\
NGC\,7077  & 4.8 & \cotwo\    & 10.7 & 22.0 &  2.175 &  0.516 & 0.548\\
NGC\,7077  & 4.8 & \cothree\  & 18.1 & 22.0 &  1.186 &  0.804 & 0.431\\
\\
UM\,448    & 4.4 & \coone\          & 21.4 & 22.0 &  1.009 &  0.985 & 1.035\\
UM\,448    & 4.4 & \cotwo\ (IRAM)   & 10.7 & 22.0 &  2.036 &  0.497 & 1.237\\
UM\,448    & 4.4 & \cotwo\ (APEX)   & 27.6 & 22.0 &  0.869 &  1.365 & 2.106 \\
UM\,448    & 4.4 & \cothree\        & 18.1 & 22.0 &  1.154 &  0.806 & 0.705\\
\hline
\end{tabular}
}
}
\vspace{0.5\baselineskip}
\begin{description}
\item
$^{\mathrm a}$~Assuming that the CO is distributed in an exponential disk like the dust.
\end{description}
\end{table}
\end{center}

\section{Gaussian beam size correction}
\label{sec:appendix_gaussian}

To ensure that our results do not depend on the specific formulation of the correction
for beam dilution, here we investigate a different assumption for the distribution
of the CO emission, namely a Gaussian.
For a Gaussian source distribution with a FWHM $\theta_s$
(corresponding to a Gaussian $\sigma_s\,=\,\theta_s/ [2\,\sqrt{2\,\ln(2)}]$),
the integrated flux density within a beam with FWHM $\theta_b$ is given by the 
following integral (assuming azimuthal symmetry) in polar coordinates:
 
\begin{equation}
S_\nu\,=\,A_0\,\int_0^{2\pi}\int_0^{\theta_b/2} \exp\left( \frac{-r}{2\,\sigma_s^2} \right)\, r \, dr \ d\phi
\label{eqn:expintegral_gauss}
\end{equation}
\noindent
where $A_0$ is the normalization constant, 
and $r$ corresponds to the angular distance from the origin.
Eqn. \ref{eqn:expintegral_gauss} can be solved analytically to obtain:

\begin{equation}
S_\nu\,=\,2\,\pi\,A_0\,\sigma_s^2\, \left[ 1 - \exp \left( -\frac{\theta_b^2}{8\,\sigma_s^2} \right)\right] 
\label{eqn:exparea_gauss}
\end{equation}
%\noindent
%The source FWHM $\theta_s$ is related to the Gaussian $\sigma$ by
%$\sigma_s\,=\,\theta_s/ (2\,\sqrt{2\,\ln(2)})$).

However, like Eqn. (\ref{eqn:expintegral}), Eqn. (\ref{eqn:expintegral_gauss}) is formally incorrect because of the need to convolve the source Gaussian
distribution with the Gaussian beam.
This convolution is given by the following, where we have taken advantage of the azimuthal symmetry and calculate the convolution
in Cartesian coordinates:

\begin{equation}
D(x,y)\,=\,A_0\ \int_{-\infty}^\infty \int_{-\infty}^\infty e^{-\frac{-(x_0^2 + y_0^2)}{\sigma_s^2}}
 e^{-\frac{-(x-x_0)^2 - (y-y_0)^2}{\sigma_b^2}}\,dx_0\,dy_0
\label{eqn:expconvolution_gauss}
\end{equation}
\noindent
where $\sigma_b$ corresponds to the Gaussian $\sigma$ assuming that the beam is
Gaussian with FWHM $\theta_b$ ($=\,2\,\sigma_b \sqrt{2\,\ln\,2}$).

It is straightforward to integrate Eqn. (\ref{eqn:expconvolution_gauss}) analytically to obtain:
\begin{equation}
D(r)\,=\,A^\prime_0\,\frac{\sigma_s^2}{(\sigma_b^2 + \sigma_s^2)}\,\exp \left[ \frac{-r^2}{(\sigma_b^2 + \sigma_s^2)} \right] 
\label{eqn:expconvolution_anal_gauss}
\end{equation}
\noindent
where $A_0$ has been substituted by the inclusion of the normalization constant of a two-dimensional Gaussian with integral unity,
$A_0\,=\,A^\prime_0/(\pi\,\sigma_b^2)$;
the independent Cartesian coordinates $x$ and $y$ have now been substituted with
$r\,=\sqrt{x^2 + y^2}$.
Unlike Eqn. (\ref{eqn:expconvolution_anal}) for the exponential source distribution,
Eqn. (\ref{eqn:expconvolution_anal_gauss}) can be easily integrated to a circular aperture radius $\theta_b/2$
to obtain the aperture correction:
\begin{eqnarray}
S_\nu & = & A^\prime_0\,\frac{\sigma_s^2}{(\sigma_b^2 + \sigma_s^2)}\ \int_0^{2\pi}\int_0^{\theta_b/2} \exp\left( \frac{-r^2}{(\sigma_b^2 + \sigma_s^2)} \right)\, r \, dr \ d\phi \nonumber \\
 & = & A^\prime_0\,\pi\,\sigma_s^2\ \left[ 1 - \exp\left( {\frac{-\theta_b^2}{4\,(\sigma_b^2 + \sigma_s^2)}} \right) \right] 
\label{eqn:expintegral_convolvedgauss}
\end{eqnarray}

\begin{figure*}[h!]
\vspace{\baselineskip}
\hbox{
\centerline{
\includegraphics[angle=0,width=0.48\linewidth]{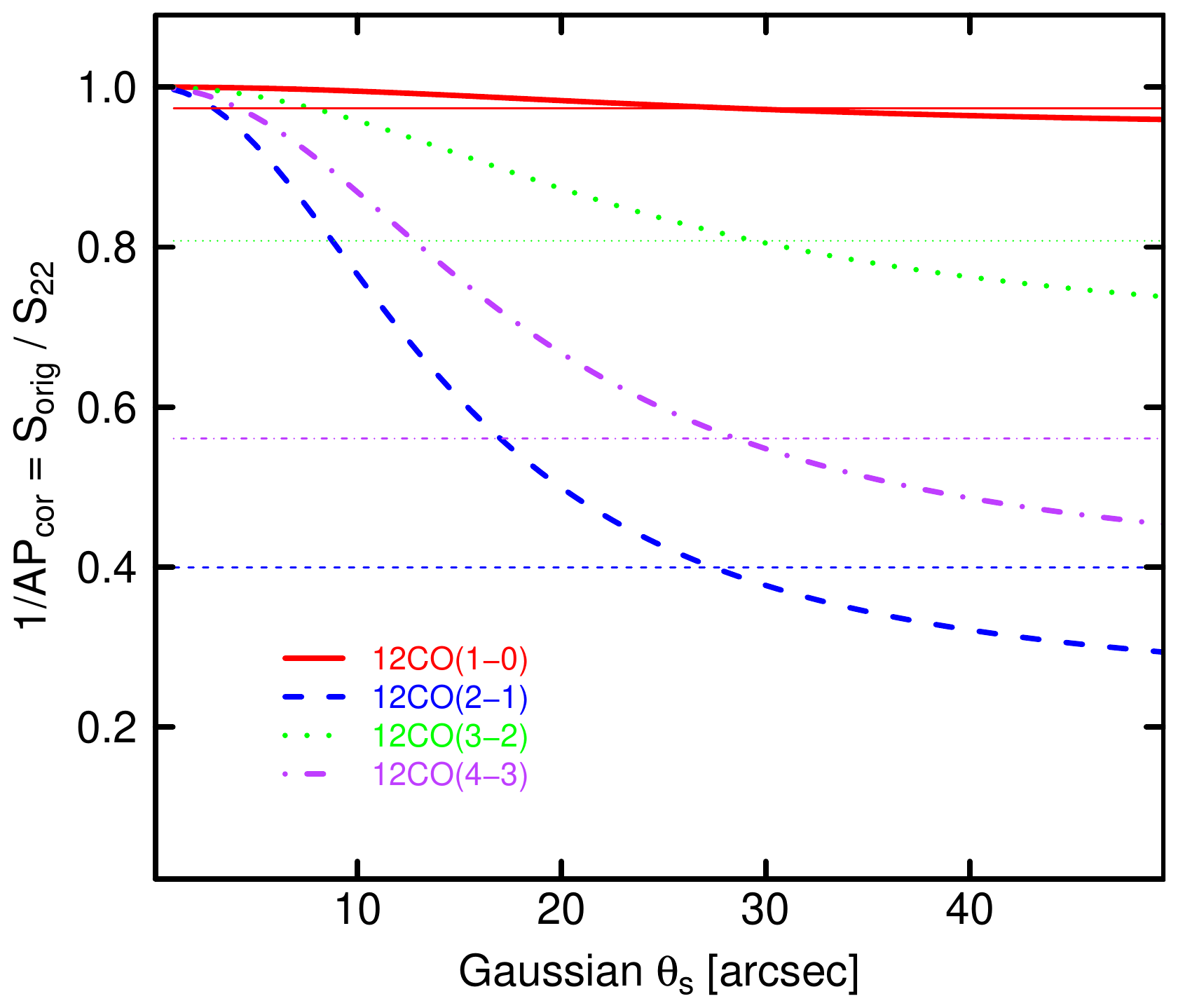}
\hspace{0.04\linewidth}
\includegraphics[angle=0,width=0.48\linewidth]{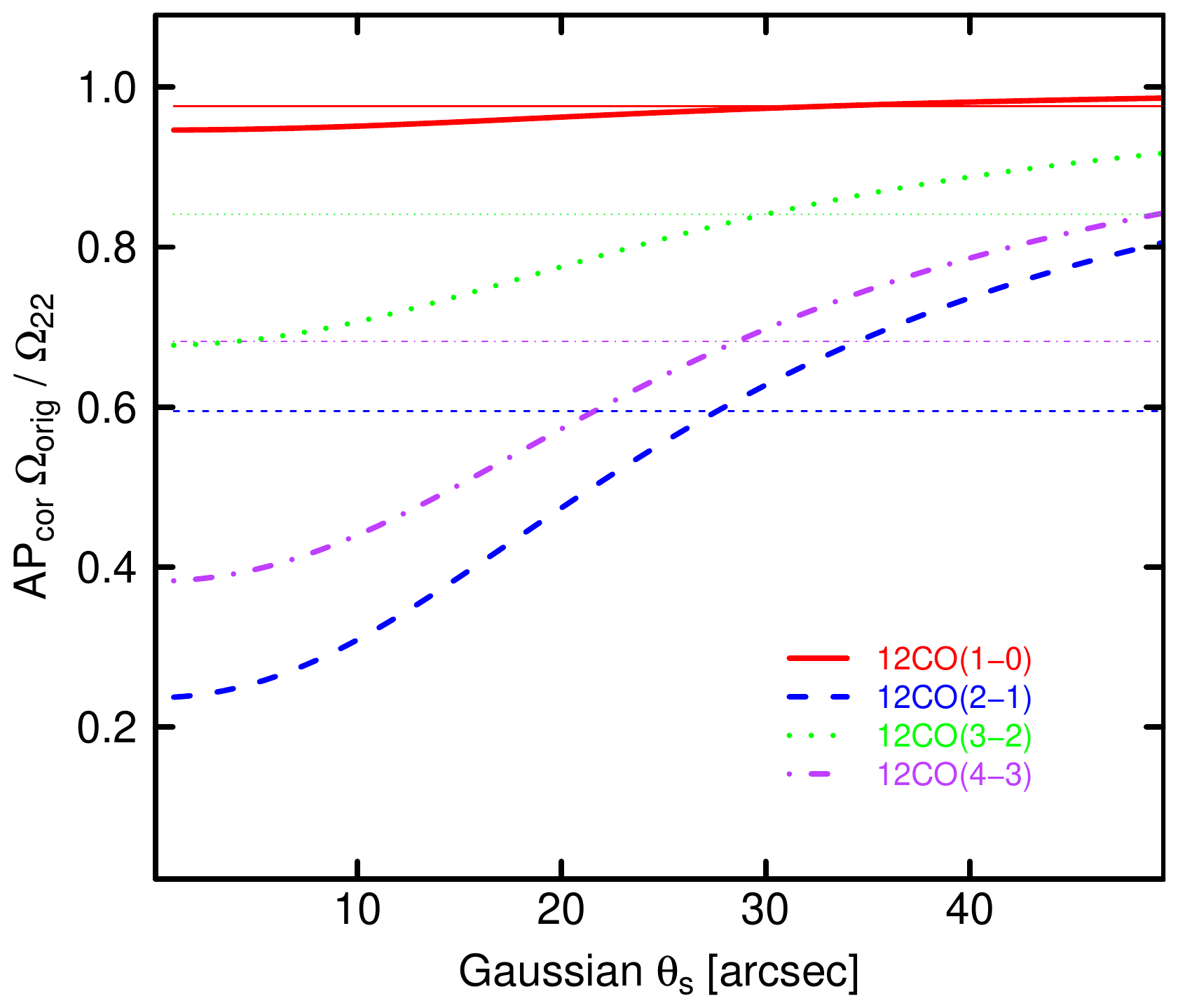}
}
}
\caption{{\it Left panel:}~(Inverse of) Aperture corrections as given in Eqn. (\ref{eqn:expintegral_convolvedgauss})
plotted against Gaussian source FWHM $\theta_s$.
{\it Right panel:}~\tmb\ corrections (AP$_{\rm cor}$plotted against $\theta_s$.
In both panels, different \twelveco\ beams are shown by different colors as line types as described in the legend:
\coone: solid (red);
\cotwo: dashed (blue); 
\cothree: dotted (green);
\cofour: dot-dashed (purple).
The horizontal (light-weighted) lines of the same type correspond to the exponentially-derived
quantities for aperture (left panel) and \tmb\ corrections (right) for NGC\,1140.
They converge roughly to a common Gaussian beam size of $\theta_s\sim$\,26\arcsec\
(see main text for more details).
}
\label{fig:beam_gauss}
\end{figure*}

For the case of a Gaussian source distribution,
to correct the fluxes (e.g., $S_\nu$ in Jy) to a common beam size, we scaled the observed CO
fluxes %to a common beam size of 22\arcsec\ 
by comparing the growth-curve integrals in the different beam sizes using the analytical formula
in Eqn. (\ref{eqn:expintegral_convolvedgauss}).
Figure \ref{fig:beam_gauss} shows the aperture and the \tmb\ corrections
for different transitions in the case of a Gaussian source distribution.
Also shown are the exponentially-derived beam corrections
for NGC\,1140 (see Appendix \ref{sec:appendix_exponential}). 

In the best-fit procedure described in the main text, we sample source FWHM $\theta_s$ ranging from 2\arcsec\ to
40\arcsec\ with steps of 1\arcsec, and for each of these
derive a multiplicative factor $AP_{\rm cor}$ $S_{22}\,=\,S_{\rm orig}\, AP_{\rm cor}$
[see Eqn. (\ref{eqn:cor})].
We then apply this correction to velocity-integrated \tmb\ values, 
and consider the best fit as the one with the lowest $\chi^2$ value.
Corrected values for NGC\,1140 are given in Table \ref{tab:corrections_gauss}.

It is interesting that the Gaussian beam corrections
found by the $\chi^2$ minimization technique 
are the {\it same} as the independently-derived exponential beam corrections.
It seems unlikely that this is a fortuitous result, but rather that it is
confirming the validity of either approach to beam corrections. 

\begin{center}
\begin{table}
      \caption[]{Flux corrections to a common beam size$^{\mathrm b}$} 
\label{tab:corrections_gauss}
\resizebox{\linewidth}{!}{
% 28/2/2016
% see ~/statistics/moleculesBCDs_PaperII/mkcorrections
{%\small
%\tiny
\begin{tabular}{llccccl}
\hline
\multicolumn{1}{c}{Galaxy} &
\multicolumn{1}{c}{Transition} & 
\multicolumn{1}{c}{Original beam} & 
\multicolumn{1}{c}{Common beam} & 
\multicolumn{1}{c}{Aperture} & 
\multicolumn{1}{c}{\tmb} &
\multicolumn{1}{c}{$I_{CO}$}  \\ 
&
& 
\multicolumn{1}{c}{(arcsec)} & 
\multicolumn{1}{c}{(arcsec)} & 
\multicolumn{1}{c}{correction} & 
\multicolumn{1}{c}{correction} &
\multicolumn{1}{c}{corrected} \\ 
& & & & 
\multicolumn{1}{c}{$AP_{\rm cor}$} & &
\multicolumn{1}{c}{(\kkms)} \\ 
\hline
\\
NGC\,1140  & \coone\         & 21.4 & 22.0 &  1.022 &  0.972 & 0.562\\
NGC\,1140  & \cotwo\         & 10.7 & 22.0 &  2.406 &  0.572 & 1.135\\
NGC\,1140  & \cothree\       & 18.1 & 22.0 &  1.204 &  0.818 & 0.945\\
NGC\,1140  & \cofour\        & 13.6 & 22.0 &  1.705 &  0.652 & 0.556\\
NGC\,1140  & \thirteencoone\ & 22.4 & 22.0 &  0.983 &  1.023 & 0.053\\
NGC\,1140  & \thirteencotwo\ & 11.2 & 22.0 &  2.246 &  0.584 & 0.100\\
NGC\,1140  & \cione\         & 12.7 & 22.0 &  1.865 &  0.626 & 0.352\\
\\
\hline
\end{tabular}
}
}
\vspace{0.5\baselineskip}
\begin{description}
\item
$^{\mathrm a}$~Assuming that the CO is distributed as a Gaussian with the best-fit $\theta_s$ = 26\arcsec.
\end{description}
\end{table}
\end{center}

\end{document}